\documentclass[11pt,a4paper]{article}
\pdfoutput=1

\usepackage{booktabs}
\usepackage{siunitx}

\usepackage{jheppub}
\usepackage{latexsym}
\usepackage{multirow}
\usepackage{color}
\usepackage[usenames,dvipsnames,table]{xcolor}
\usepackage{graphicx}
\usepackage{epsfig} 
\usepackage{epsf}   
\usepackage{dcolumn}
\usepackage{bm}
\usepackage{dcolumn}
\usepackage{textcomp}
\usepackage{float}
\usepackage{subfig}
\usepackage{hypcap}
\usepackage[]{hyperref}
\usepackage{makecell}
\usepackage{color}
\usepackage{pifont}
\usepackage{appendix}
\usepackage{amsmath}
\usepackage{multirow,bigdelim}
\usepackage{comment}
\usepackage{tabularx}
\usepackage{array}

\usepackage{lineno}
\usepackage{enumitem}
\usepackage[normalem]{ulem}
\usepackage{orcidlink}
\graphicspath{{fig/}}
\allowdisplaybreaks

\usepackage{tikz,xcolor,hyperref}

\definecolor{lime}{HTML}{A6CE39}
\DeclareRobustCommand{\orcidicon}{\hspace{-4pt}
	\begin{tikzpicture}
		\draw[lime, fill=lime] (0,0) 
		circle [radius=0.16] 
		node[white] {\hspace{0.1mm}{\fontfamily{qag}\selectfont \tiny ID}};
		\draw[white, fill=white] (-0.07,0.1) 
		circle [radius=0.01];
	\end{tikzpicture}
	\hspace{-3.2mm}
}

\foreach \x in {A, ..., Z}{\expandafter\xdef\csname 
	orcid\x\endcsname{\noexpand\href{https://orcid.org/\csname orcidauthor\x\endcsname}
		{\noexpand\orcidicon}}
}

\usepackage{amssymb}

\hypersetup{
	unicode=false,          
	pdftoolbar=true,        
	pdfmenubar=true,        
	pdffitwindow=true,   
	pdfstartview={FitH},    
	pdfsubject={LIV},   
	pdfnewwindow=true,      
	pdfcreator={RevTeX},
	colorlinks=true,     
	linkcolor=red,         
	citecolor=blue,        
	filecolor=black,     
	urlcolor=blue,           
}

\preprint{}

\title{Interplay of CPT-Violating and CPT-Conserving Lorentz Invariance Violation at DUNE}

\author[a]{Priyankush Deka\orcidA{},}
\author[b]{Arnab Sarker\orcidB{},}
\author[a]{Moon Moon Devi\orcidC{},}
\author[c]{Sushant K. Raut\orcidD{}}

\affiliation[a]{Department of Physics, Tezpur University, Napaam, Sonitpur, Assam, 784028, India}
\affiliation[b]{School of Physical Sciences, Indian Association for the	Cultivation of Science, Jadavpur, Kolkata, 700032, India}
\affiliation[c]{Division of Sciences, Krea University, Sri City, 517646, India}

\emailAdd{pdekaph@tezu.ernet.in}
\emailAdd{spsas3453@iacs.res.in}
\emailAdd{devimm@tezu.ernet.in}
\emailAdd{sushant.raut@krea.edu.in}

\date{\today}

\abstract{
We present a study of Lorentz invariance violation (LIV) in neutrino oscillations, with primary emphasis on the interplay between CPT-violating and CPT-conserving Standard Model Extension (SME) coefficients. We find that $a_{ee}$ and $a_{\tau\tau}$ dominate among the diagonal LIV coefficients, whereas $a_{e\mu}$ and $a_{e\tau}$ provide the most significant off-diagonal contributions. In contrast, the corresponding $c_{\alpha\beta}$ coefficients have comparatively sub-leading effects. Taking DUNE as a representative long-baseline experiment, we show that the sensitivity to CP violation is modified by LIV in a parameter-specific manner. We find correlations between off-diagonal LIV parameters, and non-trivial dependence on the new phases. In absence of LIV, DUNE is expected to establish CP violation at $5\sigma$ for a limited fraction of $\delta_{CP}$ values. We find that the presence of $a_{ee}$, $a_{\tau\tau}$, $a_{e\mu}$ and $a_{e\tau}$ weakens the CP discovery potential, reducing the achievable significance to below $3\sigma$ over a substantial fraction of $\delta_{\rm CP}$. The independent impact of $c_{\alpha\beta}$ terms also results in suppression, but in combination with $a_{\alpha\beta}$, they introduce degeneracies, complicating the extraction of $\delta_{CP}$. This generally results in a deterioration of CP violation sensitivities below $5\sigma$. These findings emphasize the importance of incorporating LIV effects in precision oscillation studies at upcoming long-baseline experiments.}


\begin{document}
\maketitle

\section{Introduction}\label{sec:intro}
Neutrinos ($\nu$) have emerged as one of the most powerful probes of physics beyond the Standard Model (BSM)~\cite{Gonzalez-Garcia:2007dlo}. The discovery of neutrino oscillations~\cite{Super-Kamiokande:1998kpq, SNO:2001kpb, K2K:2002icj, KamLAND:2002uet, T2K:2011ypd} has firmly established that the Standard Model (SM), which treats neutrinos as massless, is incomplete. This breakthrough has not only reshaped our understanding of fundamental particles but has also opened new pathways to explore deeper symmetries of nature that may break down at high energy scales. The standard 3-flavor $\nu$-oscillations depend on three mixing angles ($\theta_{12}$, $\theta_{13}$, $\theta_{23}$), two mass-squared differences ($\Delta m_{21}^2 \equiv m_{2}^2 - m_{1}^2$ and $\Delta m_{31}^2 \equiv m_{3}^2 - m_{1}^2$) and one Dirac CP-phase ($\rm \delta_{CP}$). With the phenomenon of $\nu$-oscillations firmly established~\cite{Super-Kamiokande:1998kpq, SNO:2001kpb}, current and future experiments focus on the precision measurement of these oscillation parameters. Among these six parameters, the solar mixing angle $\theta_{12}$, the solar mass-squared difference $\Delta m_{21}^2$, and the reactor mixing angle $\theta_{13}$ have been measured with high precision~\cite{KamLAND:2008dgz, KamLAND:2010fvi, KamLAND:2013rgu, DayaBay:2012fng}. However, parameters related to the atmospheric sector remain relatively less constrained. In particular, the mass ordering (i.e., the sign of $\rm\Delta m_{31}^2$), the octant of atmospheric mixing angle $\theta_{23}$, and the exact value of CP-violating phase $\rm \delta_{CP}$ are still under investigation \cite{ParticleDataGroup:2024cfk}. A value  of $\rm \delta_{CP}$ different from $0^{\circ}$ or $180^{\circ}$ would indicate CP violation (CPV) in the leptonic sector~\cite{Burguet-Castell:2001ppm}.

Several experiments including Borexino~\cite{BOREXINO:2014pcl}, SNO \cite{SNO:2011hxd}, Super-K \cite{Super-Kamiokande:2019gzr}, IceCube-DeepCore \cite{IceCube:2017lak}, DayaBay \cite{DayaBay:2018yms}, RENO \cite{RENO:2018dro}, DoubleChooz \cite{DoubleChooz:2019qbj}, T2K \cite{T2K:2019bcf}, NO$\nu$A \cite{NOvA:2019cyt}, Miniboone \cite{MiniBooNE:2020pnu}, MINOS \cite{MINOS:2013utc}, JUNO \cite{JUNO:2015zny} have played key roles in advancing the precision measurements of oscillation parameters. Among the currently running long-baseline (LBL) experiments, T2K and NO$\nu$A have provided important hints about mass ordering and CP violation. The recent T2K data favors normal mass ordering (NO) and the higher octant (HO) of $\theta_{23}$, with $\sin^2 \theta_{23}=0.561^{+0.021}_{-0.032}$ and a nearly maximal CP-violating phase $\rm \delta_{CP}=-1.97^{+0.97}_{-0.70}$. CP-conserving values ($\rm \delta_{CP} = 0,\pi$) are excluded at over $90\%$ confidence level (CL)~\cite{T2K:2023smv}. Similarly, the results from NO$\nu$A hint towards normal mass ordering and higher octant, with $\sin^2 \theta_{23}=0.57^{+0.03}_{-0.04}$. They disfavor inverted ordering with $\rm \delta_{CP}=\pi/2$ at more than $3\sigma$ and normal ordering with $\rm \delta_{CP}=3\pi/2$ at $2\sigma$ \cite{NOvA:2021nfi}. Despite the overall preference for normal mass ordering, T2K and NO$\nu$A exhibit a notable tension in their preferred values of the CP-violating phase $\delta_{CP}$. Recently, JUNO reported its first high-precision measurements with 59.1 days of data collection, obtaining $\sin^2\theta_{12}=0.3092 \pm 0.0087$ and $\rm \Delta m^2_{21} = (7.50 \pm 0.12)\times10^{-5} eV^2$ for the normal-ordering scenario, improving the previous global precision by a factor of 1.6 \cite{JUNO:2025gmd}. As current experiments continue to accumulate data, their sensitivity is expected to improve further. Next-generation LBL neutrino experiments such as the Deep Underground Neutrino Experiments (DUNE) \cite{DUNE:2020jqi, DUNE:2020lwj, DUNE:2021tad, DUNE:2021hwx, DUNE:2024wvj} and Tokai to Hyper-Kamiokande (T2HK) \cite{Hyper-Kamiokande:2016srs, Hyper-Kamiokande:2018ofw, Hyper-Kamiokande:2025fci} can significantly enhance the precision of these measurements. In particular, DUNE, with its 1300 km baseline and broadband beam, promises unprecedented statistics and sensitivity to mass ordering, CP violation, and the $\theta_{23}$ octant. It may also help resolve the existing tensions between NO$\nu$A and T2K observations \cite{Rahaman:2021zzm}.

While the Standard Model has excelled in describing particle interactions at energies accessible to current experiments, it is widely regarded as a low-energy effective gauge theory, valid only up to a certain scale. Theoretical arguments and models suggest that at the Planck scale $(\rm M_P \sim 10^{19}\, GeV)$, the structure of spacetime and the symmetries that govern physical laws, particularly Lorentz and CPT invariance, may be modified or broken \cite{Kostelecky:1988zi, Kostelecky:1989jp, Kostelecky:1991ak}. Notably, any violation of CPT symmetry necessarily implies Lorentz invariance violation, though the converse does not always hold \cite{Greenberg:2002uu}. In the observable low-energy limit, these possible departures from fundamental symmetries can be systematically explored within the framework of effective field theories, the most general of which is the Standard Model Extension (SME) \cite{Colladay:1998fq}. The SME allows for coordinate-independent Lorentz invariance violation (LIV), and includes both CPT-conserving and CPT-violating operators. The minimal SME framework, consisting of only renormalizable operators, serves as the standard and well-defined basis for most phenomenological studies of Lorentz and CPT violation \cite{Colladay:1996iz, Kostelecky:2003fs}.

In this framework, $\nu$-oscillations provide a particularly sensitive probe of LIV effects. Apart from measuring the standard oscillation parameters, LBL experiments are particularly well suited to probe the effects of LIV. Several experiments have studied the possible effects of LIV parameters, including MINOS \cite{MINOS:2010kat, MINOS:2012ozn}, MiniBooNE \cite{MiniBooNE:2011pix}, Super-Kamiokande \cite{Super-Kamiokande:2014exs}, Double Chooz \cite{DoubleChooz:2012eiq}, IceCube \cite{IceCube:2017qyp}, T2K \cite{T2K:2017ega}, and KM3NeT \cite{KM3NeT:2025mfl, KM3NeT:2026kuj}. Among these, Super-Kamiokande provides some of the most stringent bounds on off-diagonal LIV parameters at 95\% CL~\cite{Super-Kamiokande:2014exs}. The experimental bounds on the LIV parameters in the $\mu - \tau$ sector are given by IceCube at $90\%$ CL~\cite{IceCube:2017qyp}. The KM3NeT collaboration has recently reported constraints on isotropic LIV parameters using 1.4 years of atmospheric neutrino data~\cite{KM3NeT:2026kuj}. In addition, numerous phenomenological studies have explored the implications of LIV on $\nu$-oscillations, aiming to place bounds on LIV parameters and investigate experimental sensitivities~\cite{Agarwalla:2023wft, Sarker:2023mlz, Pan:2023qln, Majhi:2019tfi, Fiza:2022xfw, Rahaman:2021leu, Delgadillo:2024vqu, Mishra:2023tdj, Barenboim:2018ctx, Giarnetti:2024mdt, Cordero:2024hjr, Araya-Santander:2025jfd, Hilding-Norkjaer:2026fhp, Brdar:2026jbu}. Some of these efforts also focus on distinguishing LIV effects from other BSM effects such as non-standard interactions (NSI)~\cite{Majhi:2022fed, Sahoo:2022nbu}. While majority of these studies have specifically focused on the CPT-violating components of LIV, a few have examined CPT-conserving contributions and placed constraints on their magnitudes~\cite{Raikwal:2023lzk, Agarwalla:2023wft, Mishra:2023tdj}. However, a combined and comparative analysis incorporating both CPT-violating and CPT-conserving LIV components remains largely unexplored. Such a study is particularly important, as the interplay between these terms can lead to distinct modifications in oscillation probabilities, with different energy dependences and characteristic signatures. Disentangling these effects is essential not only for identifying potential signals of LIV but also for separating genuine CP-violating effects from new physics contributions. In this context, $\nu$-oscillations provide a sensitive framework to probe such scenarios, where even small LIV effects can induce observable deviations.

In this work, we particularly explore the interplay of CPT-violating and CPT-conserving isotropic LIV parameters in $\nu$-oscillations within the long-baseline sector. We study their individual as well as combined impacts to understand the distinct ways in which each component, and their interplay, can modify the standard 3-flavor oscillation paradigm. For the analysis, we adopt DUNE as a representative case study, owing to its long baseline, broad energy coverage, and strong potential to identify fundamental neutrino parameters. We begin by deriving analytical expressions for the oscillation probabilities in the presence of each type of LIV component, treating them separately to clearly identify their individual effects. Building on these analytical insights, we then extend our investigation by exploring the possible interplay and correlations between the CPT-conserving and CPT-violating LIV contributions through exact numerical simulations. This includes examining how the presence of one component may enhance or obscure the effects of the other. Finally, we assess the phenomenological consequences of these LIV terms at the level of experimental sensitivities. Specifically, we evaluate how the inclusion of such combined LIV effects influences the experiment's sensitivity towards CP violation. Through this comprehensive approach, we aim to provide a clearer understanding of how potential violations of Lorentz and CPT symmetries could affect the interpretation of results from next-generation $\nu$-oscillation experiments like DUNE.

This paper is organized as follows. In section~\ref{sec:tf}, we introduce the theoretical framework for incorporating LIV effects in $\nu$-oscillations. We present the analytically calculated probability expressions for the individual LIV components in section~\ref{sec_analexp}. The experimental setup and simulation methodology, including the $\chi^2$ analysis, are described in section~\ref{sec_exconfig}. In section~\ref{sec:PmueE}, we discuss the impact of LIV on the $\nu$-oscillation probabilities, including the bi-probability plots. In section~\ref{sec:CPV_sens}, we obtain the projected 2D constraints on the LIV parameters from DUNE, followed by the results on DUNE's sensitivity to CP violation in the presence of LIV. Finally, we summarize our findings and observations in section~\ref{sec_clsn}.

\section{Theoretical framework}   \label{sec:tf}
Lorentz invariance is a foundational symmetry of both the Standard Model of particle physics as well as general theory of relativity. However, as a low-energy manifestation of Planck-scale physics $\rm (M_{P} \sim 10^{19} \,GeV)$, potential violations can be studied using effective field theories that extend the standard relativistic framework. LIV can be incorporated into the Standard Model framework as a small perturbation to the Lagrangian. The effective Lagrangian that includes LIV effects takes the form~\cite{Kostelecky:2003cr, Kostelecky:2011gq}
\begin{equation}
    \mathcal{L} = \frac{1}{2} \bar\psi_{A} \left(i\gamma^{\mu}\delta_{\mu}\delta_{AB} - M_{AB} + \hat{\mathcal{Q}}_{AB} \right)\psi_{B} + \text{h.c.}
    \label{eq:lagrange}
\end{equation}
where $\psi_{A(B)}$ is a $2N$-dimensional spinor that includes both the spinor fields $\psi_{\alpha(\beta)}$ (where $\alpha, \beta$ index over $N$ flavors) and their charge conjugates $\psi_{\alpha(\beta)}^{C} = C\,\bar{\psi}_{\alpha(\beta)}^{T}$. This structure can be written compactly as $\psi_{A(B)} = \left(\psi_{\alpha(\beta)} \quad \psi^{C}_{\alpha(\beta)}\right)^T$. In eq.~\ref{eq:lagrange}, the three terms correspond respectively to the kinetic term, the mass term, and the LIV correction involving the operator $\hat{\mathcal{Q}}_{AB}$. The description of LIV in $\nu$-oscillations has been approached through several variants of the Standard Model Extension (SME), as extensively explored in literature%
~\cite{MINOS:2008fnv, Kostelecky:2011gq, MINOS:2010kat, MiniBooNE:2011pix,
DoubleChooz:2012eiq, IceCube:2017qyp, Kostelecky:2003cr, T2K:2017ega, Super-Kamiokande:2014exs, Bora:2025xfj}. We have adopted the minimal SME framework, which includes only renormalizable operators with mass dimension $\leq 4$, as these dominate at low energies. In this framework, the LIV contribution to the Lagrangian density for $\nu$ is
\begin{equation}
    \mathcal{L}_{\rm LIV} = -\frac{1}{2} \left[p_{\alpha\beta}^{\mu} \bar{\psi}_{\alpha} \gamma_{\mu} \psi_{\beta} + q^{\mu}_{\alpha\beta} \bar{\psi}_{\alpha}\gamma_{5}\gamma_{\mu} \psi_{\beta} - i r_{\alpha\beta}^{\mu\nu} \bar{\psi}_{\alpha} \gamma_{\mu}\delta_{\nu} \psi_{\beta} - i s_{\alpha\beta}^{\mu\nu} \bar{\psi}_{\alpha} \gamma_{5}\gamma_{\mu}\delta_{\nu} \psi_{\beta} \right] + \text{h.c.}
\end{equation}
Here, \( p_{\alpha\beta}^{\mu}, ~q_{\alpha\beta}^{\mu}, ~r_{\alpha\beta}^{\mu\nu} \) and \( s_{\alpha\beta}^{\mu\nu} \) are flavor-dependent LIV coefficients. Focusing on left-handed neutrinos, these contributions can be recast using
\begin{equation}
    (a_{L})^{\mu}_{\alpha\beta} = (p+q)^{\mu}_{\alpha\beta}, \qquad (c_{L})^{\mu\nu}_{\alpha\beta} = (r + s)^{\mu\nu}_{\alpha\beta}
\end{equation}
Both \( a_L \) and \( c_L \) are Hermitian matrices in flavor space. The parameter \( a_{L} \) represents CPT-violating terms, while \( c_{L} \) corresponds to CPT-conserving terms. In this study, we have considered the isotropic limit, which assumes rotational invariance and simplifies the LIV contributions to their time-like components~ \( \mu = \nu = 0 \). The commonly used Sun-centered celestial equatorial frame is employed here. Accordingly, we define
\begin{equation}
    a_{\alpha\beta} \equiv (a_{L})^0_{\alpha\beta}, \quad c_{\alpha\beta} \equiv (c_{L})^{00}_{\alpha\beta}
\end{equation}
By incorporating these LIV contributions, the effective Hamiltonian for $\nu$-oscillations can be expressed as
\begin{equation}
    H_{\rm eff} = H_{\rm vac} + H_{\rm mat} + H_{\rm LIV}
\end{equation}
The terms are defined as follows
\begin{equation}
    H_{\rm vac} = \frac{1}{2E} U \begin{pmatrix}
    m_1^2 & 0 & 0 \\
    0 & m_2^2 & 0 \\
    0 & 0 & m_3^2
    \end{pmatrix} U^{\dagger}, \quad
    H_{\rm mat} = \sqrt{2} G_{F} N_{e} \begin{pmatrix}
    1 & 0 & 0 \\
    0 & 0 & 0 \\
    0 & 0 & 0
    \end{pmatrix}
\end{equation}
\begin{equation}
    H_{\rm LIV} = 
    \begin{pmatrix}
    a_{ee} & a_{e\mu} & a_{e\tau} \\
    a_{e\mu}^* & a_{\mu\mu} & a_{\mu\tau} \\
    a_{e\tau}^* & a_{\mu\tau}^* & a_{\tau\tau}
    \end{pmatrix}
    - \frac{4}{3}E
    \begin{pmatrix}
    c_{ee} & c_{e\mu} & c_{e\tau} \\
    c_{e\mu}^* & c_{\mu\mu} & c_{\mu\tau} \\
    c_{e\tau}^* & c_{\mu\tau}^* & c_{\tau\tau}
    \end{pmatrix}
\end{equation}

\noindent where \( U \) is the  Pontecorvo–Maki–Nakagawa–Sakata matrix (PMNS matrix), \( G_F \) is the Fermi coupling constant, $E$ is the neutrino energy, and \( N_e \) is the electron number density. The matrix with \( a_{\alpha\beta} \) elements introduces CPT-violating effects, while \( c_{\alpha\beta} \) introduces CPT-conserving effects. The diagonal elements (\( \alpha = \beta \)) are real, while the off-diagonal elements (\( \alpha \neq \beta \)) are complex and can be parameterized as
\begin{equation}
    a_{\alpha\beta} = |a_{\alpha\beta}| e^{i\phi^a_{\alpha\beta}}, \quad
    c_{\alpha\beta} = |c_{\alpha\beta}| e^{i\phi^c_{\alpha\beta}}
\end{equation}
where $\phi_{\alpha\beta}^{a}$ and $\phi_{\alpha\beta}^{c}$ are the associated phases. The factor \( -\frac{4}{3}E \) accompanying \( c_{\alpha\beta} \) arises due to the isotropic assumption in the chosen reference frame~\cite{Sahoo:2021dit}. For antineutrinos ($\bar{\nu}$), the transformations are:
\begin{equation}
U \rightarrow U^*, \quad
\sqrt{2} G_F N_e \rightarrow -\sqrt{2} G_F N_e, \quad
a_{\alpha\beta} \rightarrow -a^*_{\alpha\beta}, \quad
c_{\alpha\beta} \rightarrow c^*_{\alpha\beta}
\label{eq:anticonversion}
\end{equation}
It is noteworthy that neutral current non-standard interactions arising from various neutrino mass models yield terms structurally similar to the LIV contributions~\cite{Miranda:2015dra, Farzan:2017xzy, Medhi:2021wxj, Sarker:2024ytu}. A formal correspondence can be established as
\begin{equation}
    \epsilon_{\alpha\beta} \longleftrightarrow \frac{a_{\alpha\beta}}{\sqrt{2} G_F N_e}    \quad  \text{and} \quad \epsilon_{\alpha\beta} \longleftrightarrow -\frac{4\,E\,c_{\alpha\beta}}{3\,\sqrt{2} G_F N_e}
\end{equation}
where $\epsilon_{\alpha\beta}$ represents the NSI elements. Although this relation illustrates how NSI and LIV enter the Hamiltonian in an analogous mathematical form, their physical origins and phenomenological implications remain fundamentally distinct. LIV is an intrinsic, frame-dependent modification of the neutrino propagation that persists even in vacuum and stems from possible violations of fundamental spacetime symmetries. In contrast, NSI arise from effective matter-induced interactions, and their effects vanish in the absence of background matter. In next section, we explore the analytical probability expressions in presence of LIV elements using a perturbative approach.


\section{Analytic formulation of $P_{\mu e}$ in presence of LIV}\label{sec_analexp}
The analytical study of $\nu$-oscillation probabilities will provide valuable insights into how different parameters contribute to the overall transition probability, and how correlations among these parameters arise. A perturbative approach is one of the most effective methods for deriving relatively simple analytical expressions, as discussed in several previous works~\cite{Cervera:2000kp, Freund:2001pn, Akhmedov:2004ny, Denton:2016wmg, Chattopadhyay:2022ftv, Bezboruah:2024yhk}. In the standard 3-flavor scenario, the oscillation probability can be expanded perturbatively in terms of the small parameters \( s_{13} \,(\equiv \sin\theta_{13}) \) and \( \alpha \, (\equiv \Delta m^2_{21}/\Delta m^2_{31}) \), since \( s_{13} \sim 0.14 \) and \( \alpha \sim 0.03 \). To maintain a consistent order in the perturbative expansion, an auxiliary parameter \( \lambda \sim 0.2 \) is introduced such that $ s_{13} \sim \mathcal{O}(\lambda) $ and $ \alpha \sim \mathcal{O}(\lambda^2) $. The transition amplitude for neutrinos propagating through matter with Hamiltonian \( H \) is given by  
\begin{equation}
    A = e^{-i H L}
    \label{eq:Ampstd}
\end{equation}
and the corresponding transition probability from flavor state $i$ to $j$ is\footnote{
Throughout this work, the indices $i,j \in \{e,\mu,\tau\}$ denote neutrino flavor labels used for transition probabilities $P_{ij}$. In contrast, Greek indices $\alpha,\beta$ are reserved exclusively for LIV coefficients.
} 
\begin{equation}
    P_{ij} = |A_{ji}|^2.
    \label{eq:PinA}
\end{equation}
The transition amplitude $A_{ji}$ can be evaluated using the Cayley--Hamilton formalism~\cite{Akhmedov:2004ny}, in which the evolution operator is expressed as a finite polynomial in the effective Hamiltonian. The standard oscillation probability is then obtained from the amplitude through eq.~\ref{eq:PinA}. To study the impact of LIV, we modify the interaction Hamiltonian by including additional terms representing the CPT-violating (\( a_{\alpha\beta} \)) and CPT-conserving (\( c_{\alpha\beta} \)) contributions. These are incorporated through the effective Hamiltonian \( H_{\text{eff}} \). For simplicity, the effects of CPT-violating and CPT-conserving components are evaluated independently for analytical calculations. The best-fit magnitudes of \( a_{\alpha\beta} \) and \( c_{\alpha\beta} \) are typically of $\mathcal{O}(10^{-23}~\mathrm{GeV})$ and $\mathcal{O}(10^{-24})$, respectively, which are small compared to the matter potential \( V_{CC} \sim 10^{-22}~\mathrm{GeV} \). Therefore, it is convenient to consider the dimensionless parameters $a_{\alpha\beta}/V_{CC}$ and $c_{\alpha\beta} E/V_{CC}$, so that both are of order comparable to \( s_{13} \), i.e. of $\mathcal{O}(\lambda)$. Hence, the modified Hamiltonian can be expanded perturbatively in terms of \( s_{13} \), \( \alpha \), $a_{\alpha\beta}/V_{CC}$ and $c_{\alpha\beta} E/V_{CC}$. The inclusion of LIV modifies the eigenvalues and consequently the transition probabilities. The eigenvalues of the effective Hamiltonian can be expressed as  
\begin{equation}
    E^{\rm eff} = E^{\rm SI} + E^{\rm LIV},
\end{equation}
where \( E^{\rm SI} \) represents the standard contribution and \( E^{\rm LIV} \) denotes the LIV-induced correction. The corresponding oscillation probability can then be written as  
\begin{equation}
    P_{ij} = P_{ij}^{\rm SI} + P_{ij}^{\rm LIV}.
\end{equation}
Here, $P_{ij}^{\rm SI}$ denotes the standard 3-flavor oscillation
probability in matter, while
$P_{ij}^{\rm LIV}$ represents the contribution arising from LIV parameters $a_{\alpha\beta}$ and $c_{\alpha\beta}$ (including the interference among standard and LIV parameters). In this work, we focus on the appearance channel $\nu_\mu \to \nu_e$, as it is
particularly sensitive to matter effects, $\delta_{CP}$ and potential new physics contributions. In LBL-experiments, this channel is governed by interference among multiple oscillation amplitudes, making it especially responsive to small perturbations of the effective Hamiltonian, including those induced by LIV. The frequently used dimensionless parameters appearing throughout this analysis
are defined as
\begin{equation}
A = \frac{2 E V_{CC}}{\Delta m^2_{31}}, \quad
\Delta = \frac{\Delta m^2_{31} L}{4 E},
\end{equation}
where $L$ is the baseline, and $V_{CC}$ denotes the matter potential. Using these definitions, the standard oscillation probability in matter in the $\nu_{\mu} \to \nu_{e}$ channel can be expressed as
\begin{equation}
\begin{aligned}
P_{\mu e}^{\rm SI} =\;&
4 s_{13}^{2} s_{23}^{2}
\frac{\sin^{2}\!\left[(A - 1)\Delta\right]}{(A - 1)^{2}} \\
&+ 2 \alpha s_{13} \sin(2\theta_{12}) \sin(2\theta_{23})
\cos(\delta_{\mathrm{CP}} + \Delta)
\frac{\sin\!\left[(A - 1)\Delta\right]}{(A - 1)}
\frac{\sin(A \Delta)}{A}.
\end{aligned}
\end{equation}

\noindent Using the Cayley--Hamilton formalism, the additional terms contributing to \( P_{\mu e} \) in presence of LIV can then be derived explicitly as 
\begin{equation}
\begin{split}
P^{\rm LIV}_{\mu e} = {} K^{(1)}_{\alpha\beta}\,\Big[a_{\alpha\beta} - \frac{4}{3}E\,c_{\alpha\beta}\Big] + K^{(2)}_{\alpha\beta}\,\Big[a^2_{\alpha\beta} &+ \frac{16}{9} E^2\,c^2_{\alpha\beta} \\ 
& - \frac{8}{3} E\,\cos{(\phi_{\alpha\beta}^{a}-\phi_{\alpha\beta}^c)}\, a_{\alpha\beta}c_{\alpha\beta}\Big]
\end{split}
\label{eq:Pac}
\end{equation}

\noindent where $\rm K^{(1)}_{\alpha\beta}$ and $\rm K^{(2)}_{\alpha\beta}$ are the 1$^{st}$ and 2$^{nd}$ order LIV contributions. Note that, for the diagonal LIV elements ($\alpha = \beta$), the last term reduces to $- \frac{8}{3} E\, a_{\alpha\alpha}c_{\alpha\alpha}$. The expressions for the first-order coefficients $K^{(1)}_{\alpha\beta}$ are

\begin{subequations}\label{eq:K1}
\begin{align}
K^{(1)}_{ee} =\ & \frac{2\,E\,s_{13}}{A\,(A-1)^3\,\Delta m_{31}^2} \Bigg[ 
8 A \sin[(A-1)\Delta]
\Big( (A-1)\Delta \cos[(A-1)\Delta] + \sin[(1-A)\Delta] \Big)
s_{13} s_{23}^2 \nonumber \\
& + \alpha (A-1)\sin[A\Delta]
\Big(
2 (A-1)\Delta \cos[A\Delta + \delta_{\text{CP}}]
+ 2 \cos[\Delta + \delta_{\text{CP}}]\sin[(1-A)\Delta]
\Big) \nonumber \\
& \times
\sin(2\theta_{12}) \sin(2\theta_{23})
\Bigg],
\label{eq:K1a}
\\[1.5ex]
K^{(1)}_{\mu\mu} =\ & \frac{E \, s_{13}}{2 A (A-1)^3\,\Delta m_{31}^2}
\Bigg[
\frac{(A-1)}{A}\alpha c_{23}s_{23}\sin(2\theta_{12})
\bigg(
-2 \sin[2A\Delta] \nonumber \\
& \quad \times
\Big[
-2(A-1)^2 c_{23}^2 \sin\delta_{\text{CP}}
+ A\big(A+(A-2)\cos2\theta_{23}\big)\sin[2\Delta+\delta_{\text{CP}}] \nonumber \\
& \quad
+ (1-2A+\cos2\theta_{23})\sin[2A\Delta+\delta_{\text{CP}}]
+ 4A\Delta(A-1)\cos[2\Delta+\delta_{\text{CP}}] s_{23}^2
\Big] \nonumber \\
& \quad
+ 4 \sin^2[A\Delta]
\Big[
\cos[2A\Delta+\delta_{\text{CP}}](1-2A+\cos2\theta_{23}) + A\cos[2\Delta+\delta_{\text{CP}}] \nonumber \\
& \quad
\times \big(A+(A-2)\cos2\theta_{23}\big)
- 2(A-1)\Big(2A\Delta\sin[2\Delta+\delta_{\text{CP}}] s_{23}^2 \nonumber \\
& \quad
+ (A-1)c_{23}^2 \cos\delta_{\text{CP}}\Big)
\Big]
\bigg) + 4 \sin[(A-1)\Delta] s_{13}
\bigg(
2\big(-1 + A(A+2)  \nonumber \\
& \quad + (A(A-2)-1)\cos2\theta_{23}\big)
\sin[(A-1)\Delta] s_{23}^2 + (A-1)
\Big[
-8A\Delta\cos[(A-1)\Delta] s_{23}^4 \nonumber \\
& \quad - (A-1)\sin[(A+1)\Delta]\sin^2(2\theta_{23})
\Big]
\bigg)
\Bigg],
\label{eq:K1b}
\\[1.5ex]
K^{(1)}_{\tau\tau} = & \frac{E\,s_{13}}{A (A-1)^3\,\Delta m_{31}^2} \Bigg[ \frac{8 \alpha c^3_{23} (A-1) \sin[A \Delta]}{A} \Big(-A(A-1) \Delta \cos[(2 - A)\Delta + \delta_{\text{CP}}] \nonumber \\
& \quad + (A-2) A \cos[(1 - A)\Delta + \delta_{\text{CP}}] \sin\Delta + \cos\delta_{\text{CP}} \sin[A\Delta] \Big) \sin(2\theta_{12}) s_{23} \nonumber \\
& \quad + 4 \sin[(A-1)\Delta] \Big( -A(A-1) \Delta \cos[(A-1)\Delta]+ A(A-2) \cos[A \Delta] \sin\Delta \nonumber \\
& \quad + \cos\Delta \sin[A \Delta] \Big) s_{13} \sin^2(2\theta_{23}) \Bigg],
\label{eq:K1c}
\\[1.5ex]
|K^{(1)}_{e\mu}|e^{i\phi^{a/c}_{e\mu}} &= \frac{2\,E}{A(A-1)^2\,\Delta m_{31}^2} \Bigg[ \frac{(A - 1)}{A} \alpha c_{23} \sin[A\Delta] \sin(2\theta_{12}) \bigg( \Big[3A -2 + (A-2)\cos(2\theta_{23})\Big] \nonumber \\
&\quad \times \cos\phi^{a/c}_{e\mu} \sin[A\Delta] + 2A s^2_{23} \Big[-\cos[A\Delta] \sin\phi^{a/c}_{e\mu} + \sin[(A-2)\Delta + \phi^{a/c}_{e\mu}]\Big] \bigg) \nonumber \\
&\quad + 2 s_{13} s_{23} \sin[(A-1)\Delta] \Big(-2A s^2_{23} \sin[(1- A)\Delta - (\delta_{\rm CP} + \phi^{a/c}_{e\mu})] \nonumber \\ &\quad + \big(1 - 2A + \cos(2\theta_{23})\big)\sin[(1- A)\,\Delta + (\delta_{\rm CP} + \phi^{a/c}_{e\mu})] \Big) \nonumber \\
&\quad + 4(A-1)\,c^2_{23} s_{13} s_{23} \sin[(A-1)\Delta] \sin[(1 + A)\,\Delta + (\delta_{\rm CP} + \phi^{a/c}_{e\mu})] \Bigg],
\label{eq:K1d}
\\[1.5ex]
|K^{(1)}_{e\tau}|e^{i\phi^{a/c}_{e\tau}} &= \frac{E \sin{2\theta_{23}}}{A(A - 1)^2\,\Delta m_{31}^2} \Bigg[ 
    4 \sin[(A -1)\Delta] \Big(
        -A \cos[A \, \Delta + \delta_{\text{CP}} + \phi^{a/c}_{e\tau}] \sin\Delta \nonumber \\
        &\quad + \cos[\Delta + \delta_{\text{CP}} + \phi^{a/c}_{e\tau}] \sin[A \,\Delta]
    \Big) s_{13} s_{23}  + \frac{\alpha(A -1) c_{23} \sin[A\, \Delta] \sin(2\theta_{12})}{A} \nonumber \\
    &\quad \times \Big(
        \sin[A\,\Delta - \phi^{a/c}_{e\tau}] + A \sin[(A -2)\Delta + \phi^{a/c}_{e\tau}] - (A -1) \sin[A\,\Delta + \phi^{a/c}_{e\tau}]
    \Big) \Bigg],
    \label{eq:K1e}
    \\[1.5ex]
|K^{(1)}_{\mu \tau}|e^{i\phi^{a/c}_{\mu \tau}}
&= -\frac{4Es_{13}}{A^2(A-1)^3 \Delta m_{31}^2}
    \Bigg[ 2 \alpha A(A-2)(A-1)
    \cos[(1-A)\Delta+\delta_{\text{CP}}] c^2_{23} \cos(2\theta_{23})
    \cos\phi_{\mu\tau}^{a/c}
    \nonumber\\
    &\quad \times
    \sin\Delta \sin[A\Delta]
    \sin(2\theta_{12}) +2\alpha (A-1) c^2_{23}
    \cos\delta_{\text{CP}}
    \cos(2\theta_{23})
    \cos\phi_{\mu\tau}^{a/c}
    \sin^2[A\Delta]
    \sin(2\theta_{12})
    \nonumber\\
    &\quad
    +A\cos\phi_{\mu\tau}^{a/c}
    \bigg(
    2\alpha (A-1)
    \cos[\Delta+\delta_{\text{CP}}]
    c^2_{23}
    \sin[A\Delta]
    \sin[(1-A)\Delta]
    \sin(2\theta_{12})
    \nonumber\\
    &\quad
    +4A(A-1)\Delta
    c_{23}
    \sin[2(A-1)\Delta]
    s_{13} s^3_{23}
    +\sin(2\theta_{23})
    \Big[
    -2A\sin^2[(A-1)\Delta]s_{13}
    \nonumber\\
    &\quad
    +\alpha(A-1)^2\Delta
    \cos[(2-A)\Delta+\delta_{\text{CP}}]
    \sin[A\Delta]
    \sin(2\theta_{12})
    \sin(2\theta_{23})
    \Big]
    \nonumber\\
    &\quad
    +\sin[(A-1)\Delta]
    \Big[
    A(A-2)\cos[A\Delta]\sin\Delta
    +\cos\Delta\sin[A\Delta]
    \Big] s_{13}
    \sin(4\theta_{23})
    \bigg)
    \nonumber\\
    &\quad
    +(A-1)^2\sin[A\Delta]
    \bigg(\alpha \Big[
    -(A-1)\cos[A\Delta-\delta_{\text{CP}}] +A\cos[(2-A)\Delta+\delta_{\text{CP}}]
    \nonumber\\
    &\quad
    -\cos[A\Delta+\delta_{\text{CP}}]
    \Big] c^2_{23}
    \sin(2\theta_{12}) +2A\sin\Delta
    \sin[(A-1)\Delta] s_{13}
    \sin(2\theta_{23})
    \bigg) \sin\phi_{\mu\tau}^{a/c}
    \Bigg].
\label{eq:K1f}
\end{align}
\end{subequations}
\noindent The second–order contributions will bring relatively smaller impact on the oscillation probabilities, but they can become relevant at higher energies. For completeness, the expressions for $K^{(2)}_{\alpha\beta}$ are provided in appendix~\ref{sec_app}.

These analytical expressions are useful for understanding and interpreting the effect of standard and LIV parameters on the oscillation probabilities, and the correlations among them. However, note that, our analyses shown in section 5 and beyond make use of exact numerical simulations as discussed in the following section.

\section{Experimental configuration \& simulation}\label{sec_exconfig}
We have focused on the DUNE as the long-baseline setup for our simulations~\cite{DUNE:2020jqi, DUNE:2020lwj, DUNE:2021tad, DUNE:2021hwx, DUNE:2024wvj}. DUNE features a baseline of 1300 km, extending from the Fermi National Accelerator Laboratory (Fermilab) in Illinois to the Sanford Underground Research Facility in South Dakota. Its far detector consists of a 40 kiloton (kt) liquid argon time projection chamber (LArTPC), designed to observe neutrino interactions with high spatial and energy resolution. The experiment is designed to operate with a 1.07 MW proton beam that corresponds to a total exposure of approximately 300 kt·MW·yr. We perform the simulation using GLoBES~\cite{Huber:2004ka, Kopp:2006wp, Huber:2007ji} with DUNE configuration inputs, including neutrino fluxes, interaction cross-sections, energy reconstruction matrices, and detector efficiencies, as provided by the DUNE collaboration~\cite{DUNE:2016ymp}. We incorporate our own probability engine within GLoBES to calculate the exact numerical oscillation probabilities in the presence of LIV. We assume a total runtime of 7 years with 3.5 years in $\nu$-mode and 3.5 years in $\bar{\nu}$-mode, corresponding to an integrated flux of $1.47 \times 10^{21}$ protons on target (POT) per year. Throughout this analysis, we assume normal mass ordering with the values of standard oscillation parameters as listed in table~\ref{tab:bestfit}. Following existing experimental bounds and recent phenomenological studies in literature \cite{Super-Kamiokande:2014exs, IceCube:2017qyp, KM3NeT:2026kuj, Raikwal:2023lzk, Agarwalla:2023wft, Mishra:2023tdj}, we fix the diagonal LIV parameters as $a_{\alpha\alpha} \sim \mathcal{O}(10^{-22}\,\mathrm{GeV})$ and $c_{\alpha\alpha} \sim \mathcal{O}(10^{-23})$, while the off-diagonal elements are taken to be one order of magnitude smaller, with $a_{\alpha\beta} \sim \mathcal{O}(10^{-23}\,\mathrm{GeV})$ and $c_{\alpha\beta} \sim \mathcal{O}(10^{-24})$.

\begin{table}[!b]
\renewcommand{\arraystretch}{1}
\centering
\begin{tabular}{lcc}
\toprule
Parameter & True values & Marginalization range \\
\midrule
$\theta_{12}$ & $33.68^\circ$ & -- \\
$\theta_{13}$ & $8.52^\circ$  & -- \\
$\theta_{23}$ & $48.5^\circ$  & $[40.5^\circ,\,50.5^\circ]$ \\
$\Delta m_{21}^{2}$ & $7.49\times10^{-5}\,\mathrm{eV}^2$ & -- \\
$\Delta m_{31}^{2}$  & $2.534\times10^{-3}\,\mathrm{eV}^2$ 
& $[2.45,\,2.60]\times10^{-3}\,\mathrm{eV}^2$ \\
$\delta_{CP}$  & $-90^{\circ}$ 
& $[-180^\circ,\,180^\circ]$ \\
\bottomrule
\end{tabular}
\caption{Values of $\nu$-oscillation parameters are mostly taken from NuFit-6.0~\cite{Esteban:2024eli}.}
\label{tab:bestfit}
\end{table}

To estimate the statistical sensitivity of the experiment to CP violation, we define the Poissonian $\chi^2$ function appropriate for event-counting
experiments as
\begin{align}
\chi^2 \equiv
\underset{\{\eta_k\}}{\text{min}}
\left[
2 \sum_{i} \sum_{j}
\left(
N_{\text{test}}^{i,j} - N_{\text{true}}^{i,j}
+ N_{\text{true}}^{i,j}
\ln\frac{N_{\text{true}}^{i,j}}{N_{\text{test}}^{i,j}}
\right)
+ \sum_k \left(\frac{\eta_k}{\sigma_k}\right)^2
\right],
\label{eq:chi2_poisson}
\end{align}
where $N_{\text{true}}^{i,j}$ and $N_{\text{test}}^{i,j}$ denote the number of events in the $\{i,j\}^{\text{th}}$ energy and channel bin for the true and test hypotheses, respectively. The index $k$ runs over all sources of systematic uncertainties, each characterized by a pull parameter $\eta_k$ and its corresponding $1\sigma$ uncertainty $\sigma_k$. The effect of systematic uncertainties is incorporated through the pull method, where the test event rates are modified as
\begin{align}
N_{\text{test}}^{i,j}(\{\eta_k\})
=
N_{\text{test}}^{i,j}(0)
\left(
1 + \sum_k \pi_k^{\,i,j}\,\eta_k
\right),
\end{align}
\noindent with $\pi_k^{\,i,j}$ denoting the fractional change in the event rate of the $\{i,j\}^{\text{th}}$ bin due to the $k^{\text{th}}$ systematic uncertainty. The systematic uncertainties considered in this analysis are summarized in table~\ref{tab:systematics}. The statistical sensitivity is obtained by minimizing the $\chi^2$ function over all the considered systematic uncertainties using the pull method~\cite{Gonzalez-Garcia:2004pka, Fogli:2002pt, Fogli:2003th}. In the next section, we examine the impact of LIV elements at the probability level, building upon the simulation details described here.

\begin{table}[!h]
\centering
\renewcommand{\arraystretch}{1.1}
\begin{tabular}{lcc}
\toprule
Channel & Signal uncertainty (\%) & Background uncertainty (\%) \\
\midrule
$\nu_e\,(\bar{\nu}_e)$ appearance     & $2\,(2)$ & $5\,(5)$ \\
$\nu_\mu\,(\bar{\nu}_\mu)$ disappearance & $5\,(5)$ & $5\,(5)$ \\
\bottomrule
\end{tabular}
\caption{Normalization uncertainties for signal/background for each analysis channel~\cite{DUNE:2016ymp}.}
\label{tab:systematics}
\end{table}
\section{Signatures of CPT-violating \& CPT-conserving LIV}  \label{sec:PmueE}
We show here, employing exact numerical computations, the impact of LIV on the appearance probability ($\nu_\mu \to \nu_e$) at the baseline of DUNE (1300~km). Throughout this section, we assume normal ordering with the oscillation parameters listed in table \ref{tab:bestfit}. In section~\ref{subsec:Pmue}, we explore the energy dependence of the oscillation probabilities ($P_{\mu e}$/$\bar{P}_{\mu e}$ ) in the neutrino/anti-neutrino channels. We also examine the $\delta_{CP}$ dependence in section~\ref{sec:dcp_PPbar}, which provides insight into potential CP-violating effects induced by LIV. Section \ref{sec_biprob} further explores this through CP trajectories in the bi--probability plane, where distortions of the standard elliptical structure provide a geometric visualization of LIV--induced modifications. In section~\ref{subsec:parscan}, we perform a parameter scan of the LIV and SI parameters, visualized through 2D heatmaps of the probabilities. These scans include pairwise comparisons of $(a'_{\alpha\beta}, \delta_{\rm CP})$, $(c'_{\alpha\beta}, \delta_{\rm CP}$),
$(a'_{\alpha\beta}$, $c'_{\alpha\beta})$, and $(\phi^{a/c}_{\alpha\beta}, \delta_{\rm CP})$. Such analyses may reveal correlations among the parameters as well as regions of degeneracy.
\subsection{Variation of $P_{\mu e}\, (\bar{P}_{\mu e})$ in presence of $a_{\alpha\beta}$ and $c_{\alpha\beta}$}
\label{subsec:Pmue}
In figure~\ref{fig:Pmue_E}, we show the variation of $P_{\mu e}\, (\bar{P}_{\mu e})$ as a function of neutrino energy. The top panel shows the impact of CPT-violating and CPT-conserving LIV parameters in the $\nu$-channel. The bottom panel shows the corresponding effects for the $\bar{\nu}$-channel. The solid black line represents the probability for the SI case ($a'_{\alpha\beta}=c'_{\alpha\beta}=0$), whereas other solid (dashed) lines represent the effect of $a_{\alpha\beta}\,(c_{\alpha\beta})$ elements. As a representative choice, we fix the CP phase to $\delta_{\rm CP} = -90^\circ$. For the off-diagonal elements, we consider the case where $\phi_{\alpha\beta}^{a}$ and $\phi_{\alpha\beta}^{c}$ are set to $-90^{\circ}$. In all the figures, we adopt the following two scaled terms for the ease of representation.
\begin{align}
    a'_{\alpha \beta} &= \frac{a_{\alpha \beta}}{\rm 10^{-23}\,GeV} \quad &
    c'_{\alpha \beta} &= \frac{c_{\alpha \beta}}{10^{-23}}
\end{align}
\noindent However, we will primarily refer to the original \( a_{\alpha\beta} \) and \( c_{\alpha\beta} \) in the discussion since they only differ by a scaling factor. Our observations from figure~\ref{fig:Pmue_E} are as follows:
\begin{figure}[!t]
\includegraphics[width=0.48\linewidth]{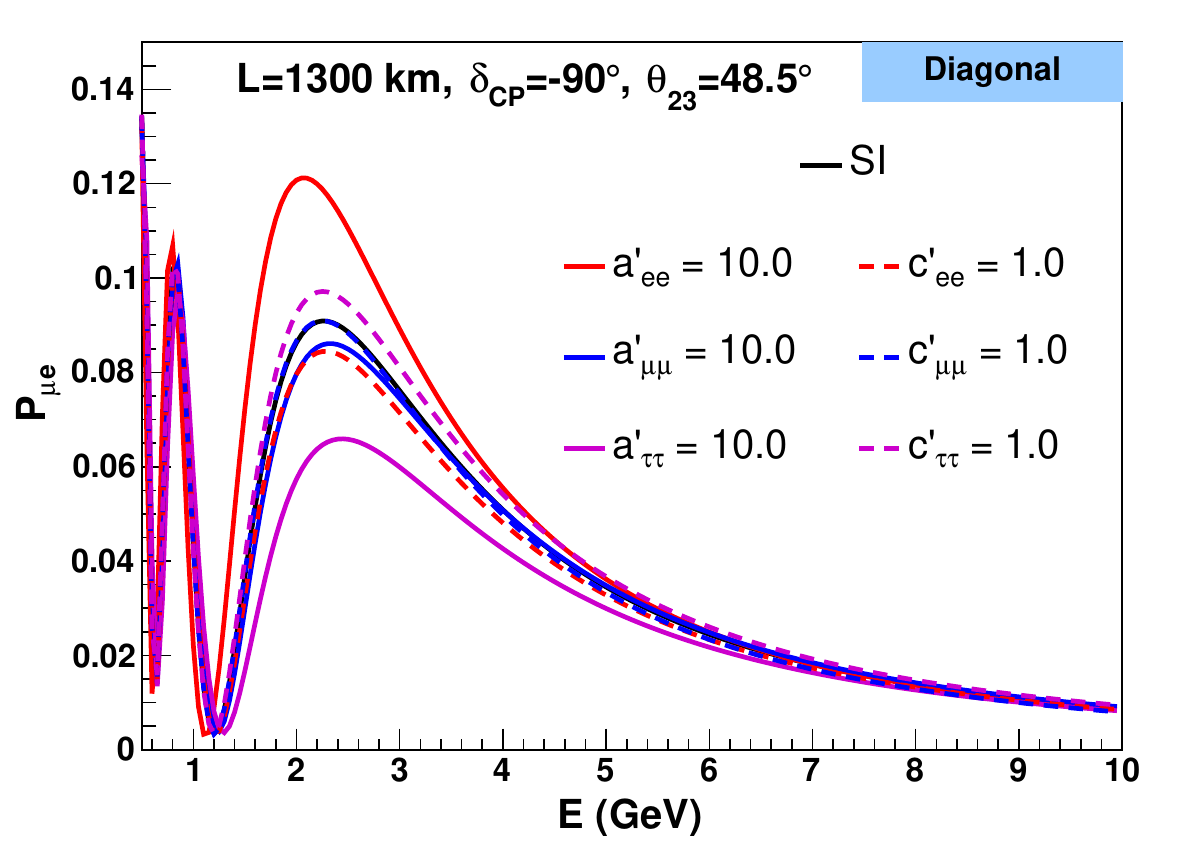}
\includegraphics[width=0.48\linewidth]{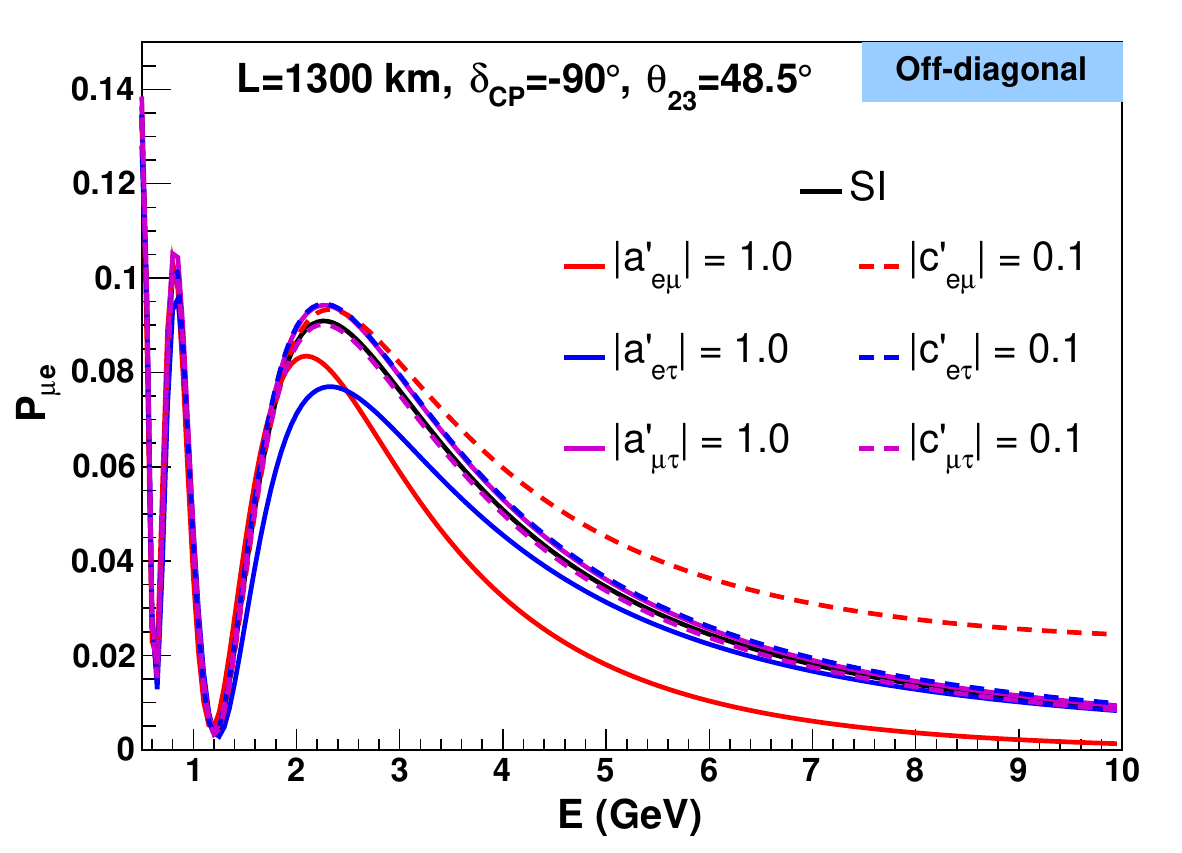}
\includegraphics[width=0.48\linewidth]{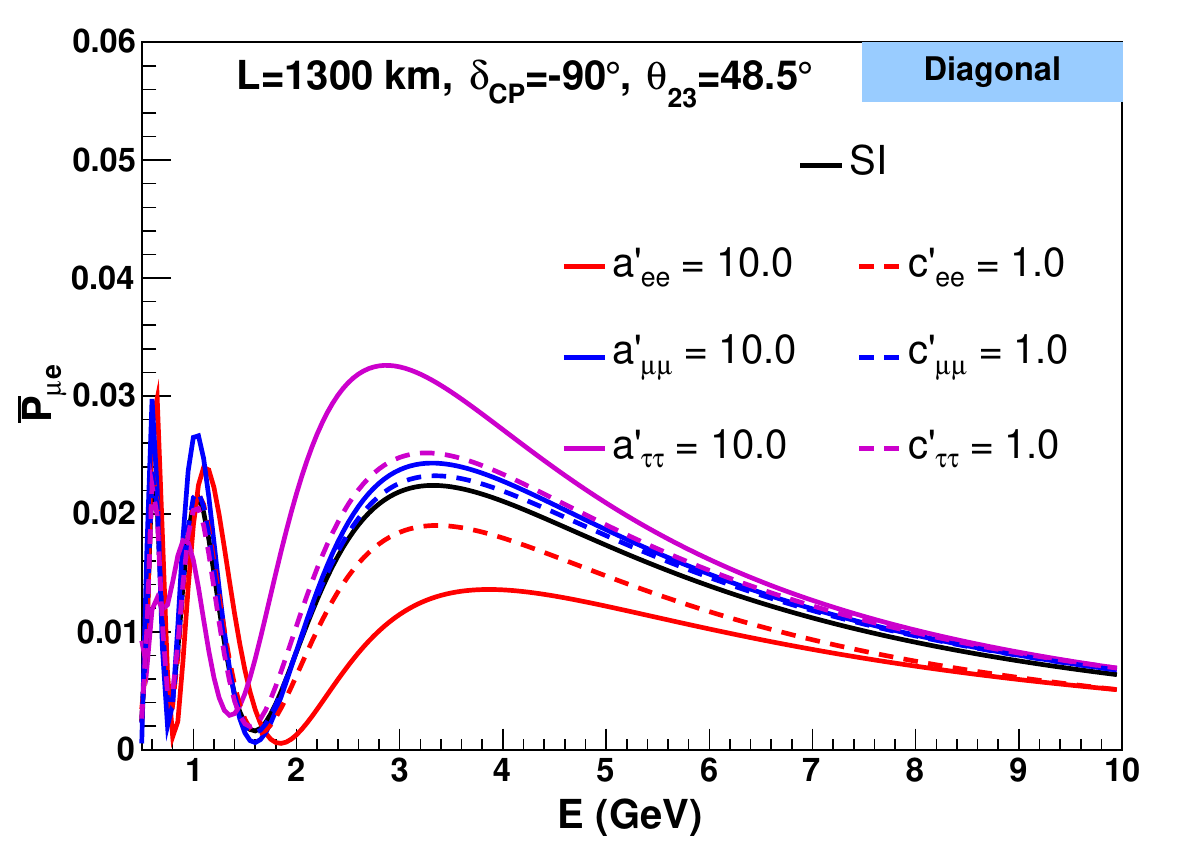}
\hfill
\includegraphics[width=0.48\linewidth]{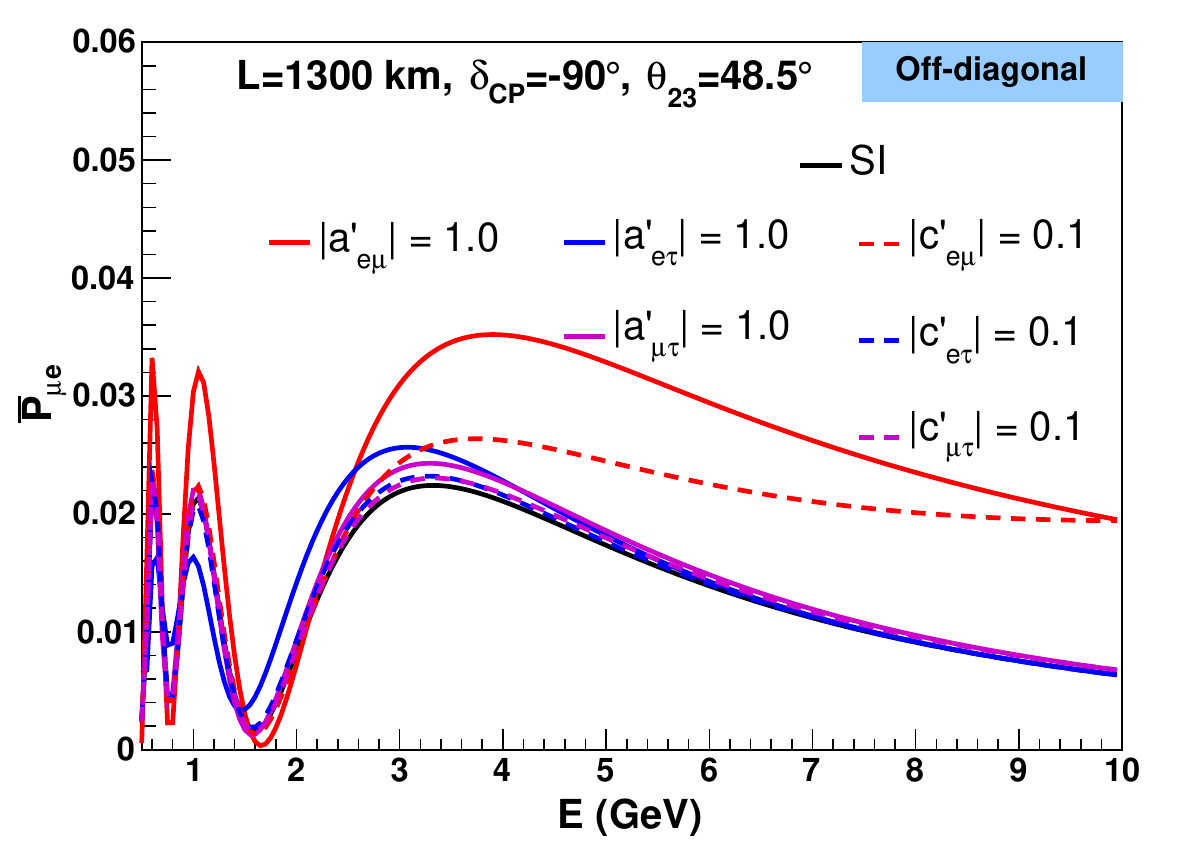}
\caption{Neutrino (top panel) and anti-neutrino (bottom panel) oscillation probabilities as a function of energy at DUNE baseline. The black solid curve represents the SI scenario. The solid (dashed) colored curves correspond to the $a'_{\alpha\beta}$ ($c'_{\alpha\beta}$) LIV contributions. For the off-diagonal elements, we consider $\phi_{\alpha\beta}^{a} = -90^{\circ}$ and $\phi_{\alpha\beta}^{c} = -90^{\circ}$.}
\label{fig:Pmue_E}
\end{figure}

\vspace{0.3cm}
\emph{For neutrino channel (top panel):}
\begin{itemize}
    \item The SI probability exhibits its first oscillation maximum near 2.5~GeV for DUNE. The presence of \( a_{ee} \) significantly enhances the probability compared to the SI case around this peak energy, whereas \( c_{ee} \) causes a relatively smaller suppression of \( P_{\mu e} \). The \( a_{\mu\mu} \) contribution slightly suppresses the probability, while \( c_{\mu\mu} \) has a minimal effect. The presence of \( a_{\tau\tau}\,(c_{\tau\tau}) \) shows a suppression (enhancement) of the standard probability. At higher energies, diagonal elements exhibit only a nominal impact.
    \item The off-diagonal elements have a noticeable impact on the SI probability for energies \( E \gtrsim 1.5 \, \mathrm{GeV} \). The presence of \( a_{e\mu}\,(c_{e\mu}) \) leads to suppression (enhancement) of probability across this energy range. For the \( a_{e\tau}\,(c_{e\tau}) \) element, the probability shows suppression (enhancement) near the peak energy, but it gradually decreases at high energies. The presence of \( a_{\mu\tau} \) element has only a marginal enhancement near the peak energy, and \( c_{\mu\tau} \) exhibits no noticeable effect.
\end{itemize}

\emph{For anti-neutrino channel (bottom panel):}
\begin{itemize}  
\item In SI scenario, the $\bar{\nu}_\mu \to \bar{\nu}_e$ appearance probability attains a smaller oscillation maximum of $\bar{P}_{\mu e} \simeq 0.022$, compared to the $\nu$-channel, with the peak shifted to a higher energy of approximately
$3.3~\mathrm{GeV}$. In \(\bar{\nu}\)-channel, both the CPT-violating and the corresponding CPT-conserving elements exhibit similar behavior rather than opposite effects, in contrast to the \(\nu\)-channel.
\item The diagonal $a_{\alpha\alpha}$ elements influence the standard probability in the opposite manner relative to $\nu$-channel. The elements causing enhancements in $\nu$-channel cause suppression in the $\bar{\nu}$-mode, and vice versa. The overall influence of all diagonal LIV terms diminishes progressively with increasing energy. 
\item The presence of off-diagonal element \(a_{e\mu}\) exhibits a strong impact on \(\bar{P}_{\mu e}\), enhancing the probability by a significant margin compared to the SI scenario. The \(c_{e\mu}\) element also produces a significant enhancement, with its effect gradually increasing with energy. The \(a_{e\tau}\) contribution shows an enhancement near the peak energy, followed by a relatively small enhancement due to \(a_{\mu\tau}\). These two elements and their corresponding \(c_{\alpha\beta}\) elements show negligible effects at higher energies.
\end{itemize}

\noindent We can gain insight into the observed effects of LIV on oscillation probabilities by analyzing the analytical expressions for the probabilities. A key observation is that, for oscillations in the \(\nu\) channel, any CPT-violating element \(a_{\alpha\beta}\) induces an opposing and relatively smaller effect from the corresponding CPT-conserving element \(c_{\alpha\beta}\). This arises because the contributing terms involving \(a_{\alpha\beta}\) and \(c_{\alpha\beta}\) in eq.~\ref{eq:Pac} are structurally similar but appear with opposite signs. However, for the \(\bar{\nu}\) oscillation probability, both the \(a_{\alpha\beta}\) and corresponding \(c_{\alpha\beta}\) elements exhibit similar effects on the standard oscillation. This is understood from eq.~\ref{eq:anticonversion}, which shows how the parameters transform for the \(\bar{\nu}\) case. While \(c_{\alpha\beta}\) terms carry an additional energy dependence, this does not significantly amplify their impact in the energy range specific to DUNE. But it will play an important role while probing high-energy neutrinos. Moreover, the enhancement or suppression of the oscillation probability near the peak energy, due to either \(a_{\alpha\beta}\) or \(c_{\alpha\beta}\), is predominantly governed by their respective first-order contributions. The impact of diagonal LIV elements tends to diminish at higher energies. For off-diagonal elements $a_{\mu\tau}$ and $c_{\mu\tau}$, the LIV effects generally vanish at high energies. The analytical expressions show that, for each coefficient, the high-energy contributions from $a_{e\mu}$ and $a_{e\tau}$, as well as their corresponding $c_{\alpha\beta}$ terms, arise prominently from second-order effects. These contributions are modulated by specific trigonometric factors, as can be seen in appendix~\ref{sec_app}. We note that
\begin{subequations}
\begin{align}
P^{(2)}_{\mu e} 
&\propto s_{13}^2 s_{23}^2 \approx 0.012, 
\quad \text{for each of } a_{ee},\, c_{ee},\, a_{\mu\mu},\, c_{\mu\mu},\, a_{\mu\tau} \text{ and } c_{\mu\tau}
\label{P_sub1} \\
P^{(2)}_{\mu e} 
&\propto s_{13}^2 c_{23}^2 \approx 0.0096, 
\quad \text{for each of } a_{\tau\tau} \text{ and }  c_{\tau\tau}
\label{P_sub2} \\
P^{(2)}_{\mu e} 
&\propto c_{23}^2 s_{23}^2 \approx 0.246, 
\quad \text{for each of } a_{e\tau} \text{ and } c_{e\tau}
\label{P_sub3}
\end{align}
\end{subequations}
where we observe strong suppression in eqs.~\ref{P_sub1}-\ref{P_sub2} due to $s_{13}^2$, which leads to a negligible impact at high energies. In contrast, eq.~\ref{P_sub3} shows that the contribution from either \( a_{e\tau} \) or \( c_{e\tau} \) is relatively significant. The second-order contributions from $a_{e\mu}$ and $c_{e\mu}$ are not suppressed, so their effects remain significant even at high energies. Therefore, among all LIV parameters, $a_{e\mu}$ and $c_{e\mu}$ show the strongest effects at high energies.

\subsection{$\delta_{CP}$ dependence of $P_{\mu e}$ and $\bar{P}_{\mu e}$}\label{sec:dcp_PPbar}
\begin{figure}[!h]
    \centering
    \includegraphics[width=0.325\linewidth, height=5cm]{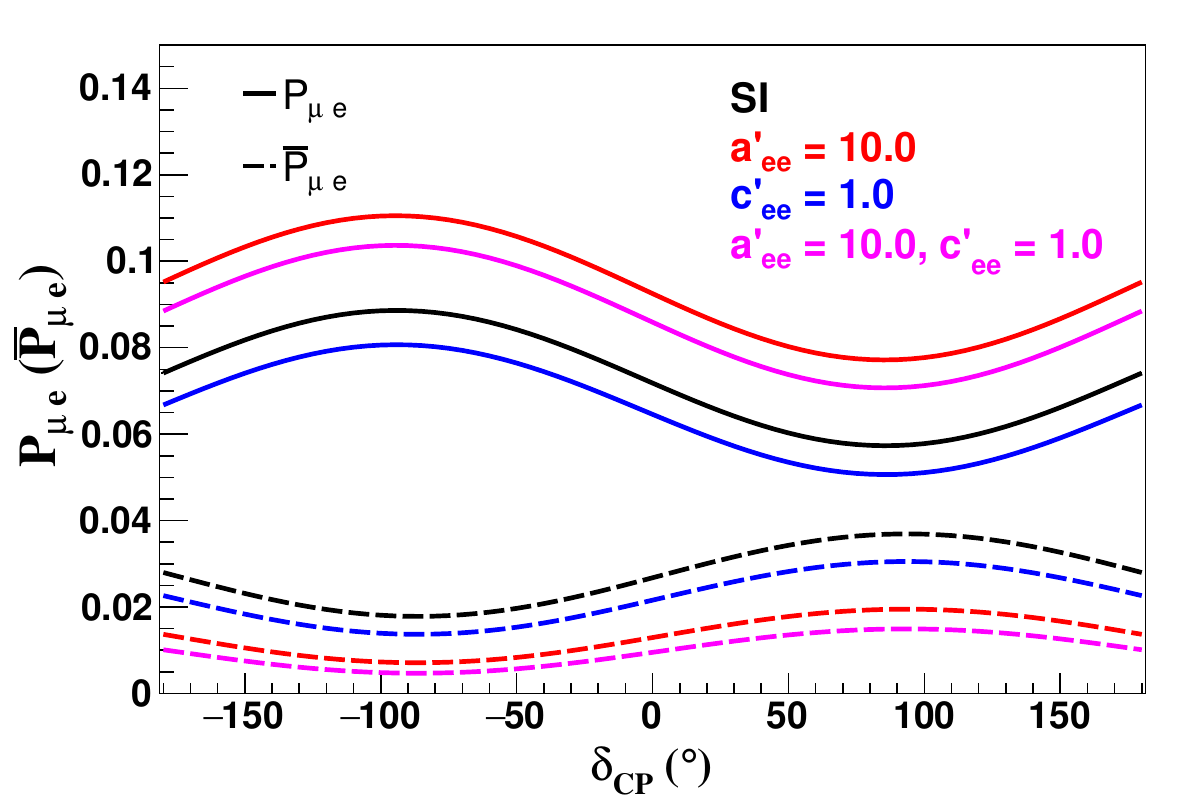}
    \includegraphics[width=0.325\linewidth, height=5cm]{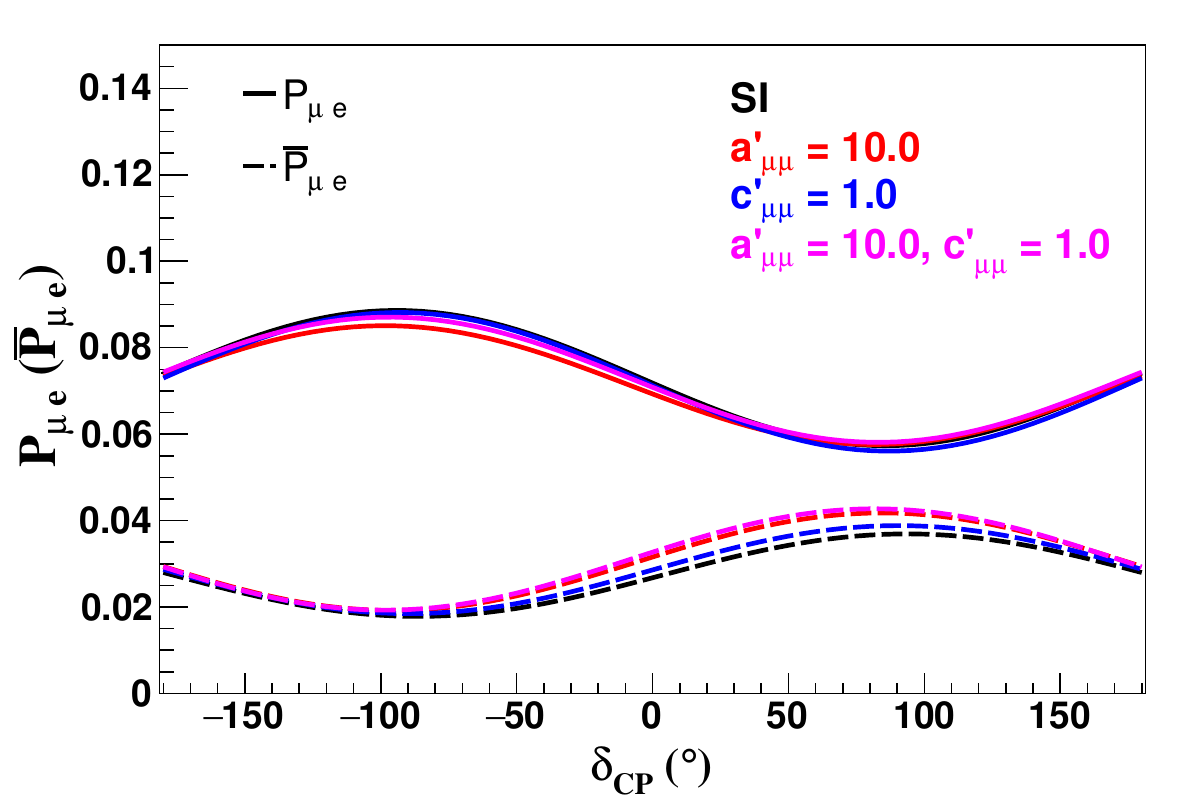}
    \includegraphics[width=0.325\linewidth, height=5cm]{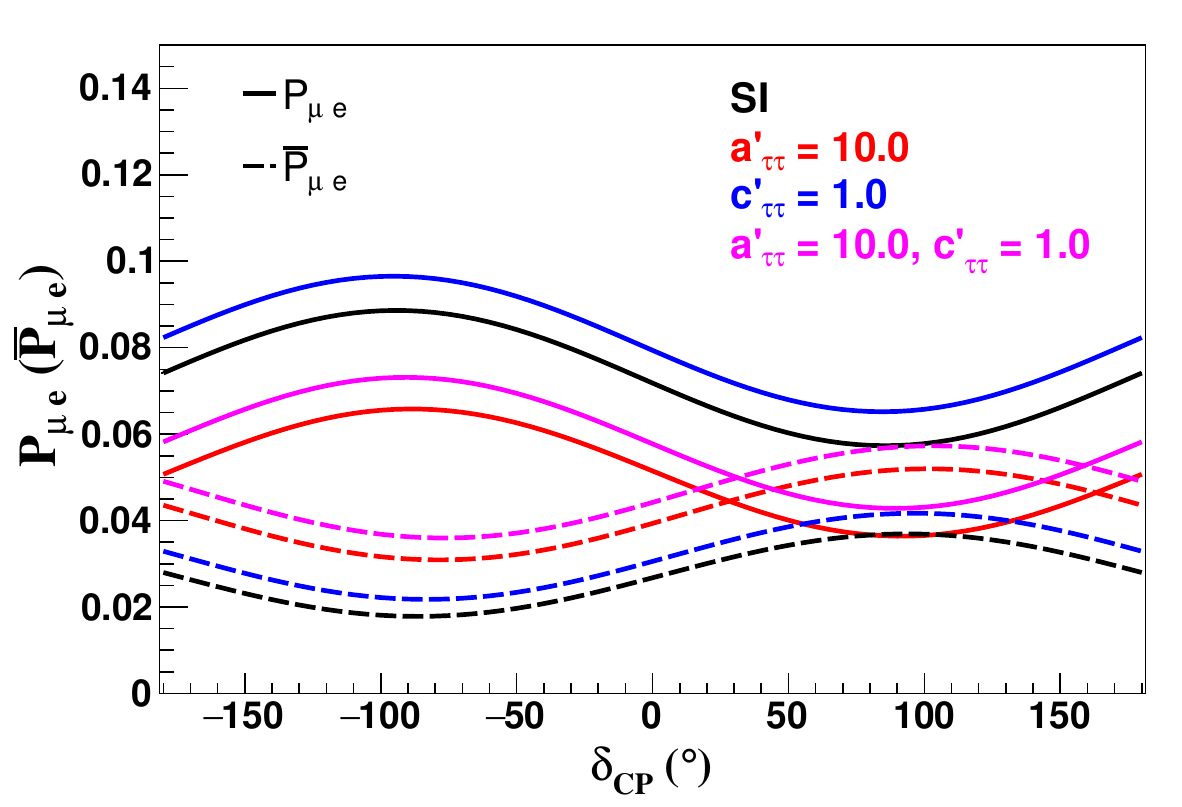}
    \includegraphics[width=0.325\linewidth, height=5cm]{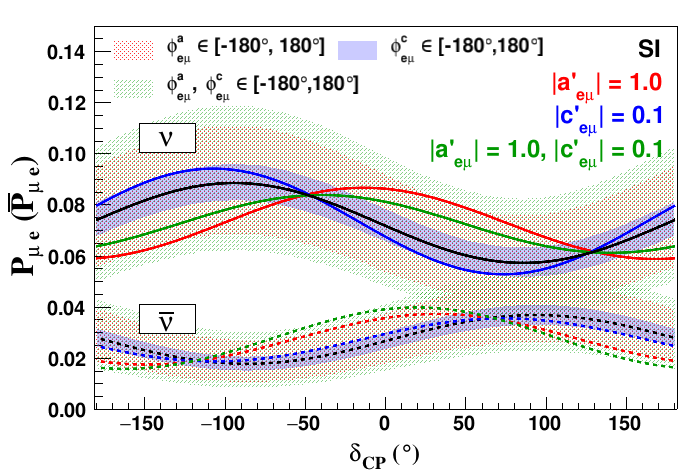}
    \includegraphics[width=0.325\linewidth, height=5cm]{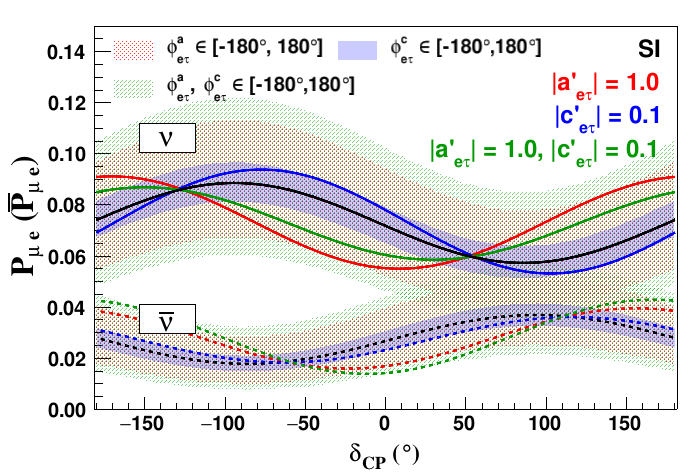}
    \includegraphics[width=0.325\linewidth, height=5cm]{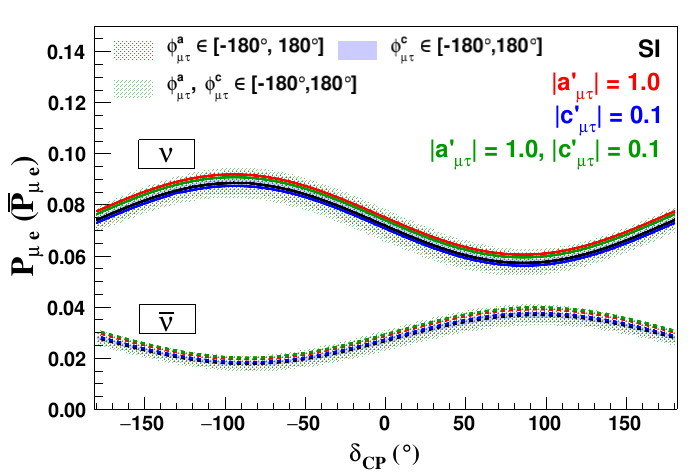}
    \caption{Neutrino (solid) and anti-neutrino (dashed) oscillation probabilities as a function of $\delta_{CP}$ at DUNE baseline and $E = 2.5~{\rm GeV}$. The top (bottom) panel corresponds to the diagonal (off-diagonal) LIV parameters. The shaded bands indicate the variation of off-diagonal phases ($\phi_{\alpha\beta}^{a}$, $\phi_{\alpha\beta}^{c}$), whereas the curves correspond to fixed $\phi_{\alpha\beta}^{a} = \phi_{\alpha\beta}^{c}= -90^\circ$.}
    \label{fig:PvsCP}
\end{figure}

Figure \ref{fig:PvsCP} shows $P_{\mu e}$ and $\bar{P}_{\mu e}$ as functions of $\delta_{CP}$ for a fixed energy of 2.5 GeV. The solid and dashed curves correspond to $P_{\mu e}$ and $\bar{P}_{\mu e}$, respectively. In all panels, the black curve represents the SI probability.

The top panel illustrates the effect of diagonal LIV elements. In presence of $a_{ee}$ (top-left panel), the probability curve shifts upward, leading to a significant enhancement in the oscillation probability, whereas $a_{\tau\tau}$ (top-right panel) causes a downward shift, resulting in a reduction. The effect of $a_{\mu\mu}$ is comparatively negligible. The corresponding $c_{\alpha\alpha}$ elements exhibit opposite trends, although with smaller magnitudes. Specifically, $c_{ee}$ reduces $P_{\mu e}$, while $c_{\tau\tau}$ enhances it, with $c_{\mu\mu}$ again producing only a marginal effect. When both $a_{\alpha\alpha}$ and $c_{\alpha\alpha}$ are present simultaneously, their contributions combine effectively, producing trends similar to those observed for the individual $a_{\alpha\alpha}$ elements, but with reduced overall magnitude.

For the off-diagonal scenarios shown in the bottom panel, the parameters $a_{e\mu}$ and $a_{e\tau}$, and similarly $c_{e\mu}$ and $c_{e\tau}$, exhibit significant effects. For the $a_{e\mu}$ and $c_{e\mu}$ curves with $\phi_{e\mu}^{a}=\phi_{e\mu}^{c}=-90^{\circ}$ (bottom-left panel), a clear phase shift relative to the SI curve is observed in both the $\nu$ and $\bar{\nu}$ probabilities. The red and blue bands correspond to varying $\phi_{e\mu}^{a}$ and $\phi_{e\mu}^{c}$ over the range $[-180^{\circ},180^{\circ})$, respectively. The green bands represent the combined effect obtained by simultaneously varying both phases over the same range, resulting in a broader band. The observations for $a_{e\tau}$--$c_{e\tau}$ pair in the bottom-middle panel are qualitatively similar. In contrast, $a_{\mu\tau}$--$c_{\mu\tau}$ pair exhibits a nominal deviation from the SI scenario with extremely narrow bands.

\subsection{CP trajectories in Bi-probability plane}
\label{sec_biprob}
A CP trajectory refers to the closed curve traced in the $(P_{\mu e},\, \bar{P}_{\mu e})$ plane when the CP--violating phase $\delta_{\mathrm{CP}}$ is varied over $[-180^{\circ}, 180^{\circ})$. For standard 3-flavor scenario, this curve takes the form of an ellipse. Each point on the ellipse corresponds to a particular value of 
$\delta_{\mathrm{CP}}$, and the overall size, tilt, and position of the ellipse encode how intrinsic CP violation, matter effects, and possible new physics influence the oscillation probabilities \cite{Minakata:2001qm, Hyde:2018tqt}.

With this interpretation, the bi–probability plane provides a natural geometric setting for visualizing how LIV parameters distort the CP trajectory. It thus serves as a useful tool for examining LIV–induced modifications to $\nu$ oscillations. The displacement of these ellipses from the vacuum symmetry line ($P_{\mu e} = \bar{P}_{\mu e}$) reflects the so-called fake CP asymmetry arising from matter effects \cite{Minakata:1998bf}, while deviations in their sizes or tilts may indicate the presence of LIV contributions. Thus, bi-probability plots offer a clear and intuitive way to disentangle standard and non-standard effects when probing the underlying symmetry properties of $\nu$ oscillations. For the standard oscillation case, the appearance probabilities for $\nu$ and $\bar{\nu}$ can be expressed as \cite{Minakata:2001qm}:

\begin{equation}
    P_{\mu e} = A \cos{\delta_{\mathrm{CP}}} + B \sin{\delta_{\mathrm{CP}}} + C,
    \label{eq:biprobnu}
\end{equation}
\begin{equation}
    \bar{P}_{\mu e} = \bar{A} \cos{\delta_{\mathrm{CP}}} - \bar{B} \sin{\delta_{\mathrm{CP}}} + \bar{C},
    \label{eq:biprobanti}
\end{equation}

\noindent where \(A\) and \(B\) are the respective coefficients of the CP--conserving and CP--violating terms, while \(C\) is the CP--independent contribution. \(\bar{A}\), \(\bar{B}\), and \(\bar{C}\) denote the corresponding $\bar{\nu}$ coefficients. In vacuum, where $\nu$ and $\bar{\nu}$ experience identical
propagation, one has
\begin{equation}
A = \bar{A}, \qquad
B = \bar{B}, \qquad
C = \bar{C},
\end{equation}
and the only difference between \( P_{\mu e} \) and
\( \bar{P}_{\mu e} \) arises from flipping of sign of the
\(\sin\delta_{\mathrm{CP}}\) term. In matter, however, the MSW potential breaks the symmetry between
\(\nu\) and \(\bar{\nu}\), leading to
\begin{equation}
A \neq \bar{A}, \qquad
B \neq \bar{B}, \qquad
C \neq \bar{C}.
\end{equation}

\noindent Nevertheless, eqs.~\ref{eq:biprobnu}--\ref{eq:biprobanti} remain useful, with the coefficients replaced by their effective matter-modified forms. Together they define a parametric curve in the $(P_{\mu e},\,\bar P_{\mu e})$ plane, which is generically an ellipse as $\delta_{\mathrm{CP}}$ is varied over $[-180^{\circ},180^{\circ})$. The extent and orientation of such ellipses are determined by the relative sizes and correlations of the CP-even and CP-odd interference contributions, encoded in the $\cos\delta_{\mathrm{CP}}$ and $\sin\delta_{\mathrm{CP}}$ terms, respectively.

\begin{figure}[h]
\centering
\includegraphics[width=0.35\linewidth, height=5cm]{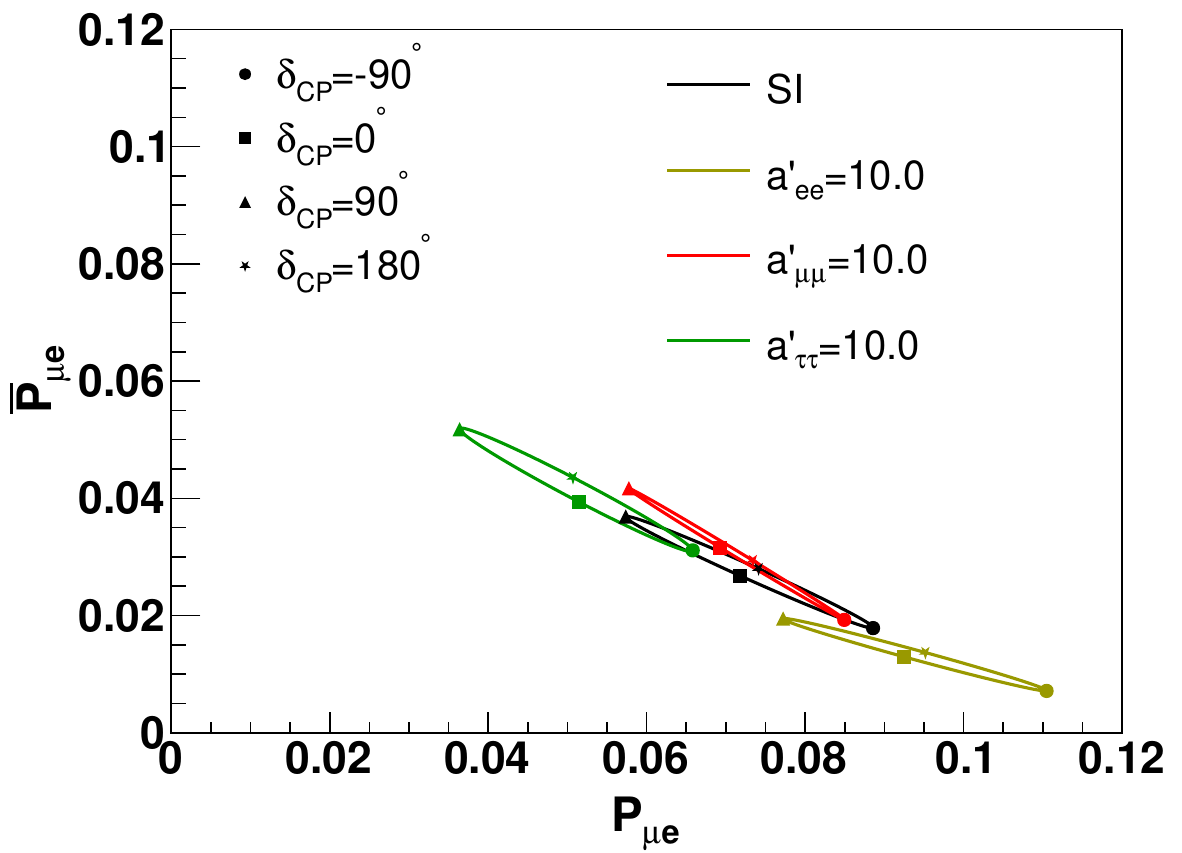}
\includegraphics[width=0.35\linewidth, height=5cm]{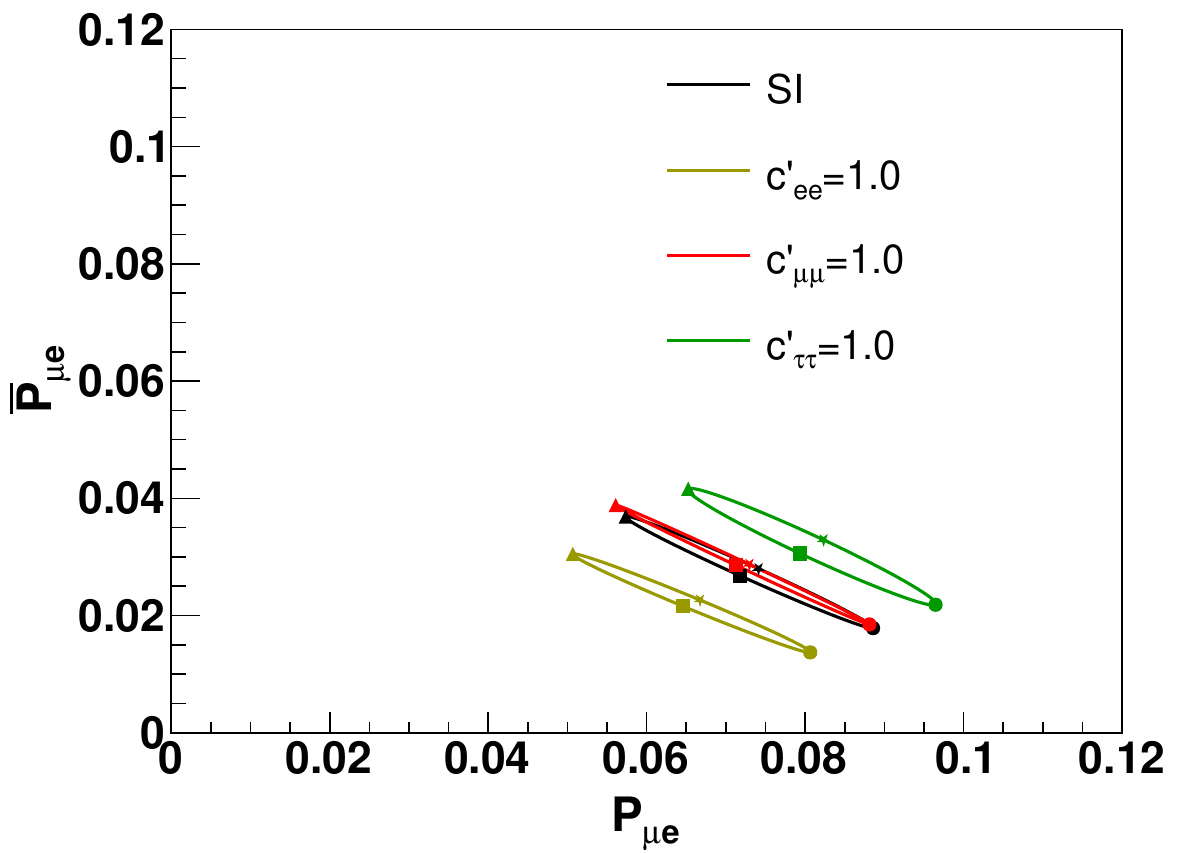}
\caption{Bi-probability plots ($P_{\mu e}$ vs $\bar{P}_{\mu e}$) at $E=2.5$ GeV. The left (right) panel shows the scenario with diagonal $a'_{\alpha\alpha}$ ($c'_{\alpha\alpha}$) elements.}
\label{fig:biprob_diag}
\end{figure}

In figure~\ref{fig:biprob_diag}, we show the CP trajectories in presence of diagonal LIV coefficients. The left (right) panel corresponds to the CPT-violating (CPT-conserving) cases. In both panels, the black ellipse represents the SI scenario. All bi-probability plots in this section are evaluated at the peak appearance energy of $\rm 2.5\,GeV$. At higher energies, the ellipses contract towards the origin, as the appearance probabilities for both $\nu$ and $\bar{\nu}$ decrease relative to the SI scenario.

In the left panel, we observe that introducing diagonal $a_{\alpha\alpha}$ coefficients does not significantly modify the ellipse shape relative to the SI scenario. Instead, the ellipses primarily undergo a displacement in the $(P_{\mu e},\,\bar P_{\mu e})$ plane along the direction of major axis. For instance, the ellipse corresponding to $a_{ee}$ (deep yellow) shifts towards slightly higher $P_{\mu e}$ and lower $\bar P_{\mu e}$, while the one with $a_{\tau\tau}$ (light green) moves oppositely. This behavior resembles a modified matter effect, since diagonal $a_{\alpha\alpha}$ terms enter the Hamiltonian in the same manner as the MSW potential \cite{Minakata:2001qm}:
\begin{equation}
H_{\rm eff}= H_{\rm SI} + {\rm diag}(a_{ee},a_{\mu\mu},a_{\tau\tau}).
\end{equation}

\begin{figure}[!t]
\centering
\includegraphics[width=0.325\linewidth, height=5cm]{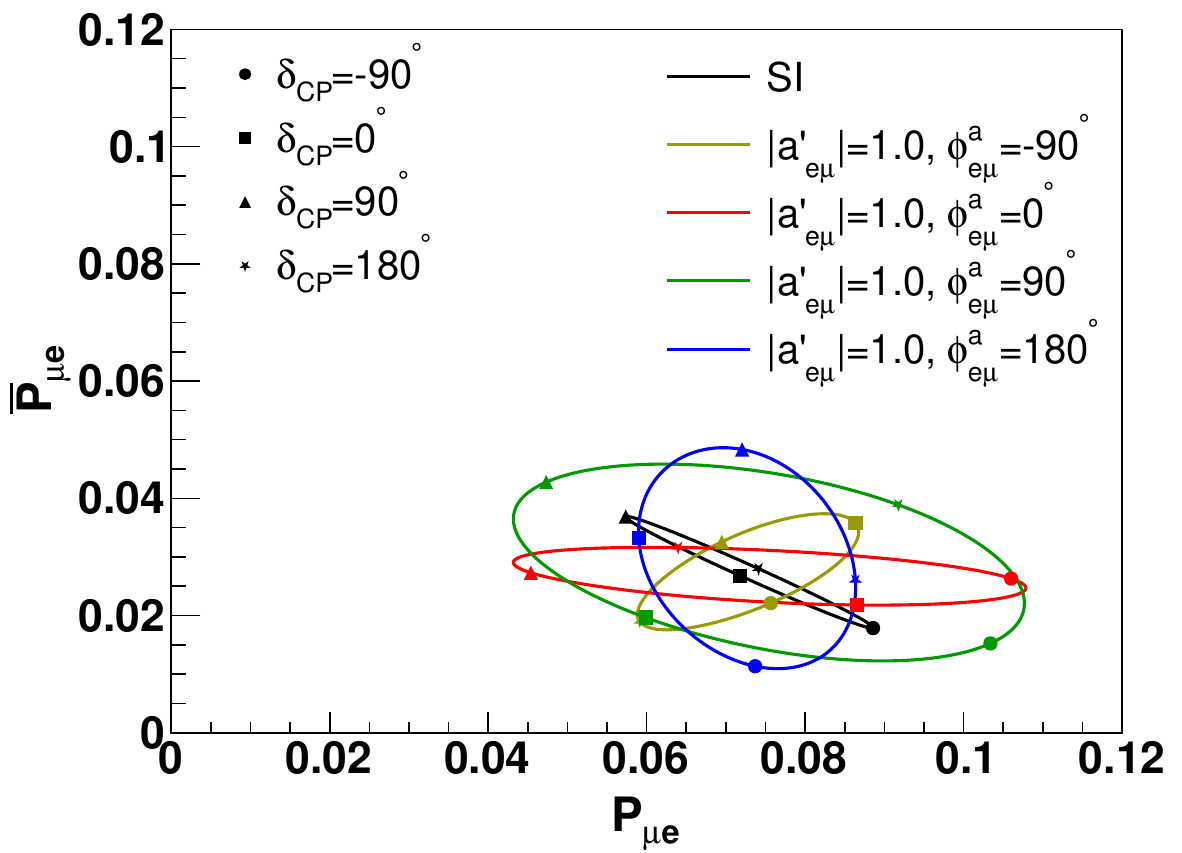}
\includegraphics[width=0.325\linewidth, height=5cm]{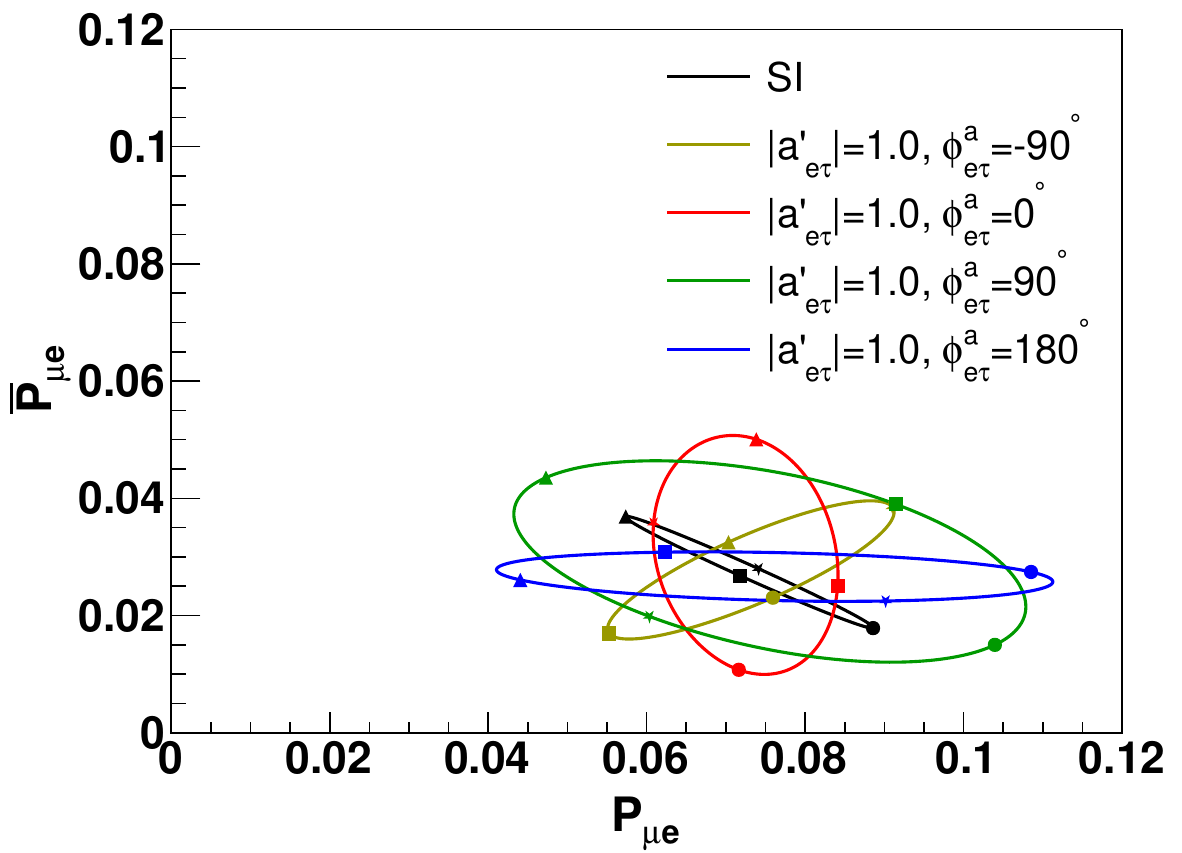}
\includegraphics[width=0.325\linewidth, height=5cm]{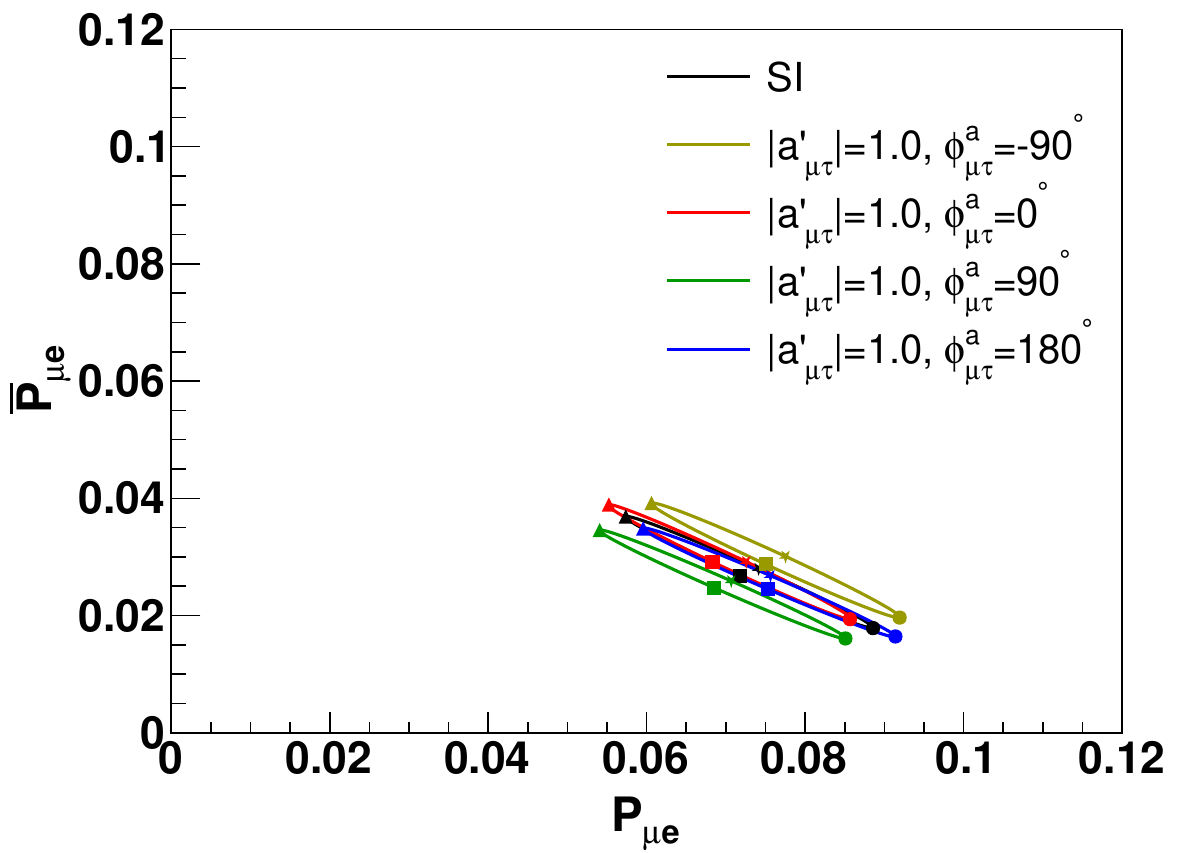}
\includegraphics[width=0.325\linewidth, height=5cm]{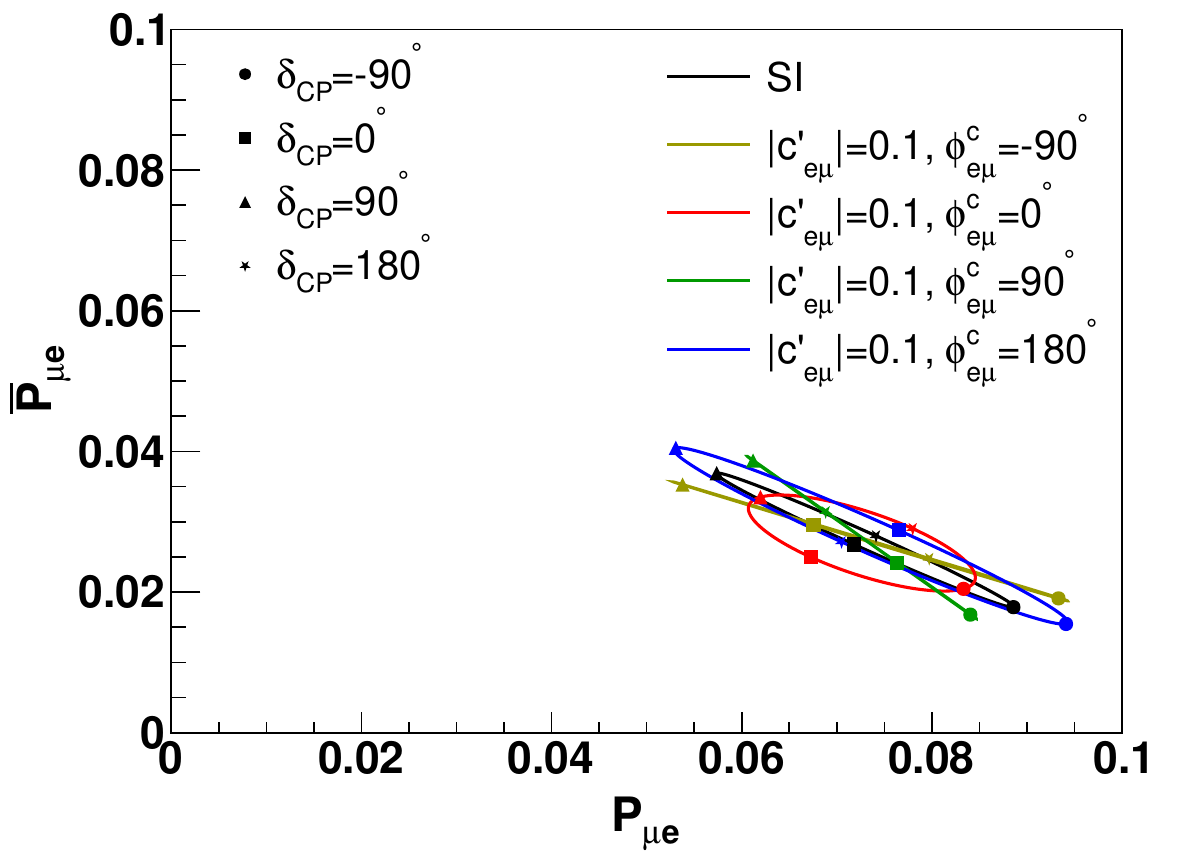}
\includegraphics[width=0.325\linewidth, height=5cm]{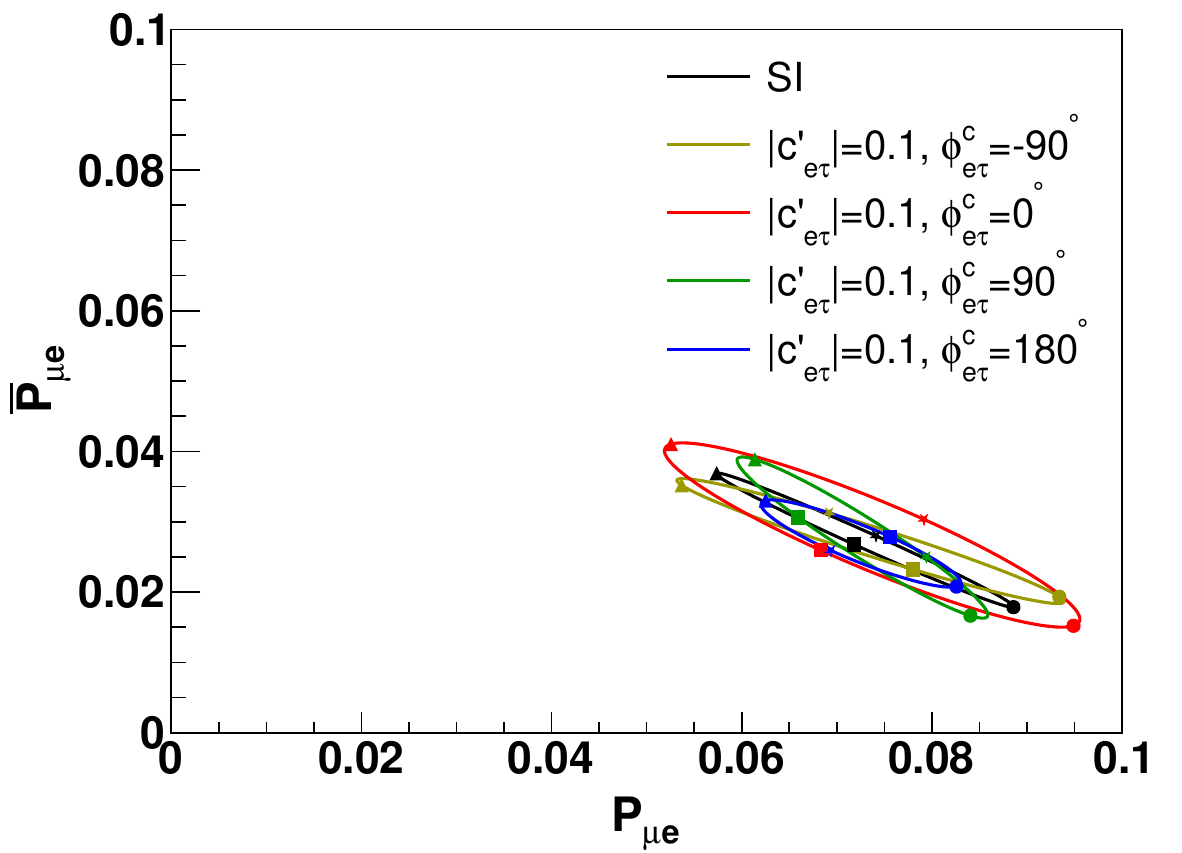}
\includegraphics[width=0.325\linewidth, height=5cm]{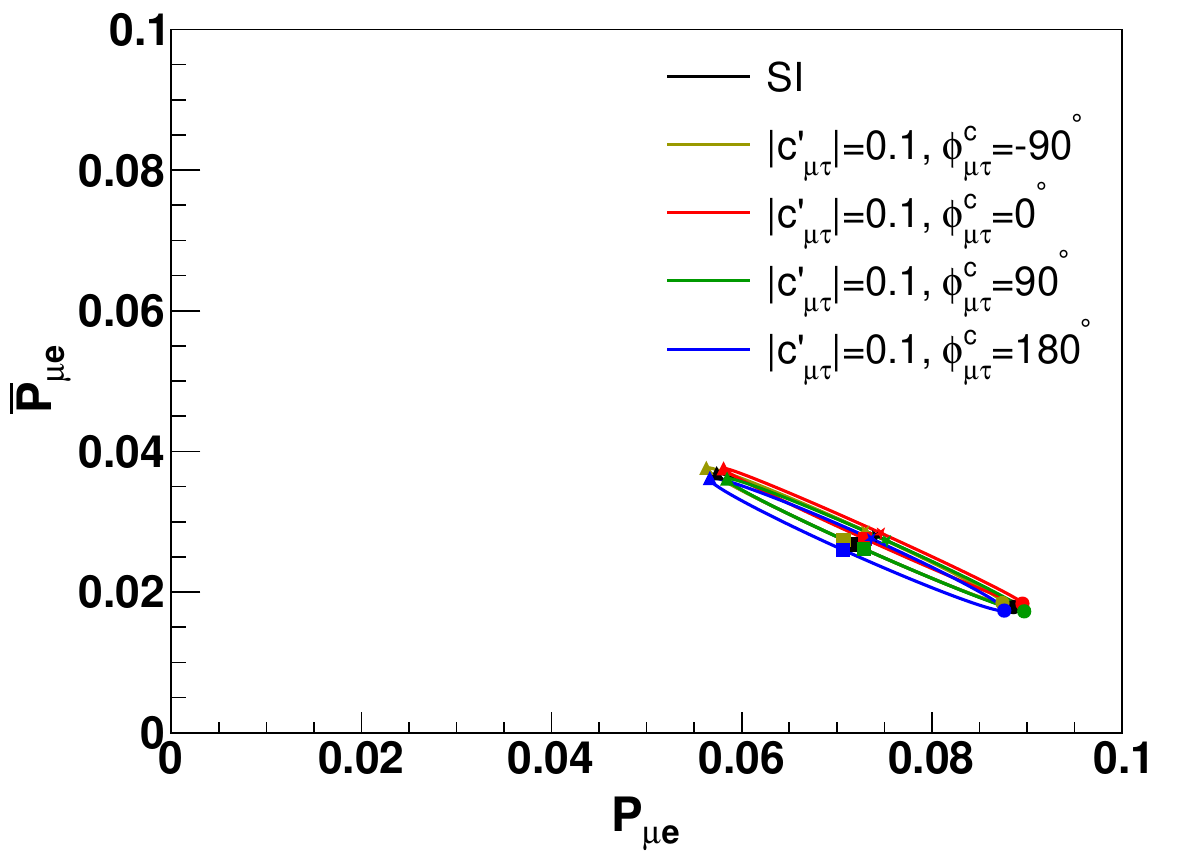}
\caption{Bi-probability plots ($P_{\mu e}$ vs $\bar{P}_{\mu e}$) at $E = 2.5~\mathrm{GeV}$. The top (bottom) panels correspond to the case of off-diagonal $a'_{\alpha\beta}$ ($c'_{\alpha\beta}$) elements. The black ellipses represent the SI case, whereas the coloured ellipses represent the effects of $a'_{\alpha\beta}$ and $c'_{\alpha\beta}$ for different values of $\phi^{a}_{\alpha\beta}$ and $\phi^{c}_{\alpha\beta}$, respectively.}
\label{fig:biprob_aoffdiag}
\end{figure}

In particular, for the $a_{ee}$ case, the effective potential becomes $V_{\rm eff}=V_{CC}+a_{ee}$, which alters the resonance condition and thus the oscillation probability. Because $a_{\alpha\alpha}$ is CPT-violating and flips sign between
$\nu$ and $\bar{\nu}$, this displacement increases the separation between
$P_{\mu e}$ and $\bar P_{\mu e}$ even when $\delta_{\rm CP}=0$, thereby generating a
fake CP asymmetry of matter-like origin rather than genuine CP violation. In contrast, the diagonal $c_{\alpha\alpha}$ coefficients also preserve the ellipse shape and orientation but induce a displacement nearly parallel to the SI ellipse. This stems from their CPT-even nature: unlike $a_{\alpha\alpha}$, the $c_{\alpha\alpha}$ terms do not flip sign between $\nu$ and $\bar{\nu}$ (see eq.~\ref{eq:anticonversion}), and therefore shift both probabilities in the same direction, resulting in a symmetric deformation rather than a matter-like fake CP asymmetry.

When off-diagonal $a_{\alpha\beta}$ and $c_{\alpha\beta}$ elements are introduced, as shown in figure~\ref{fig:biprob_aoffdiag}, the CP trajectories exhibit distinct features. For $a_{e\mu}$ ($c_{e\mu}$) and $a_{e\tau}$ ($c_{e\tau}$), the ellipses become enlarged and rotated, with their major and minor axes varying for different values of $\phi^{a/c}_{e\mu}$ and $\phi^{a/c}_{e\tau}$. The increase in the major-axis length indicates that these LIV elements can significantly influence the observable CP-violating effects. The rotation arises because the inclusion of such off-diagonal terms introduces additional $\sin(\delta_{\mathrm{CP}} \pm \phi^{a/c}_{\alpha\beta})$ and $\cos(\delta_{\mathrm{CP}} \pm \phi^{a/c}_{\alpha\beta})$ components in the probabilities (see analytical expressions in section~\ref{sec_analexp}). In the case of $a_{\mu\tau}$ or $c_{\mu\tau}$, these mixed $\delta_{\mathrm{CP}}$–$\phi_{\alpha\beta}^{a/c}$ terms appear only at sub-leading order, and thus no significant rotation is observed for these parameters.

We also note that, for certain values of $\phi^{a/c}_{e\mu}$ and $\phi^{a/c}_{e\tau}$, the corresponding LIV ellipses intersect the SI ellipse at four points. Such intersections indicate degeneracies between the LIV phase and the standard CP-violating phase. At these degeneracy points, identical $(P_{\mu e}, \bar P_{\mu e})$ values arise from distinct parameter combinations $(\delta_{\mathrm{CP}}, \phi^{a/c}_{\alpha\beta})$, reducing the experiment’s ability to unambiguously attribute observed CP asymmetries to standard or LIV origins. Consequently, the overall CP sensitivity is degraded.

\subsection{Exploration of $(a_{\alpha\beta}$ -- $c_{\alpha\beta})$ parameter space }
\label{subsec:parscan}
\begin{figure}[!b]
\centering     \includegraphics[width=0.6\linewidth]{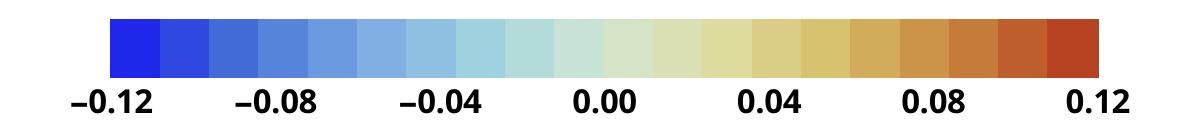}
\par
\includegraphics[width=0.325\linewidth, height=5cm]{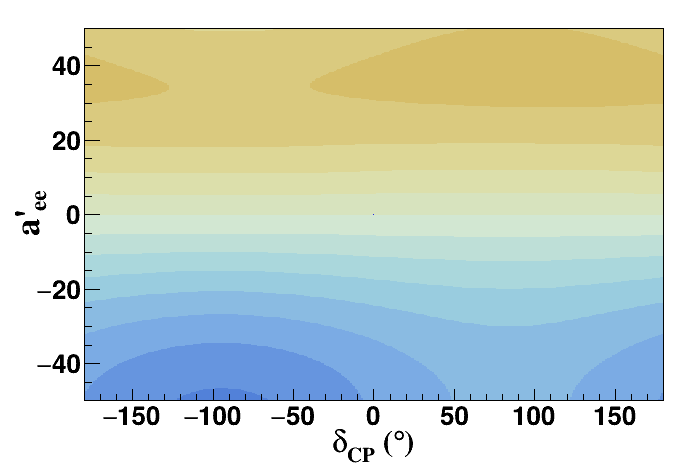}
\includegraphics[width=0.325\linewidth, height=5cm]{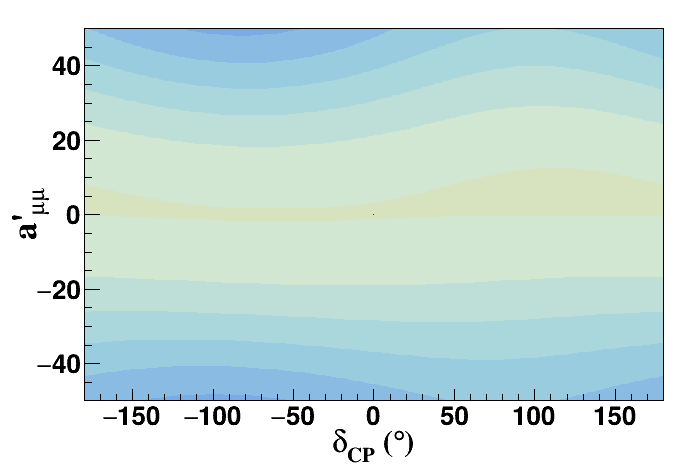}
\includegraphics[width=0.325\linewidth, height=5cm]{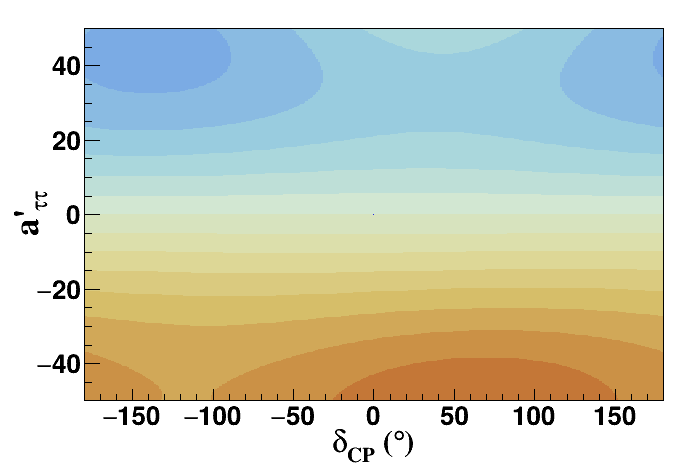}
\includegraphics[width=0.325\linewidth, height=5cm]{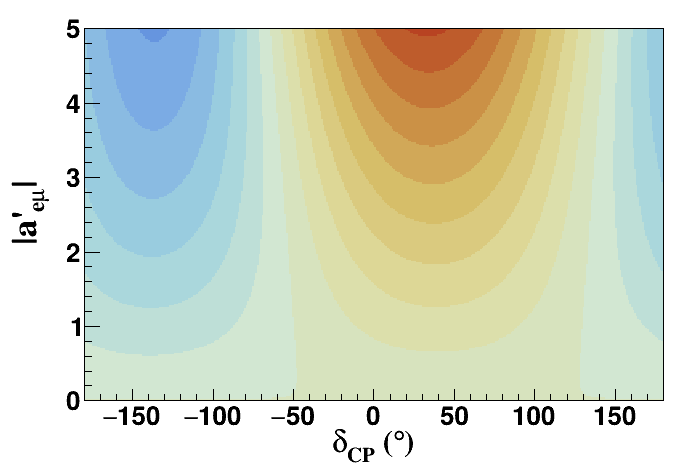}
\includegraphics[width=0.325\linewidth, height=5cm]{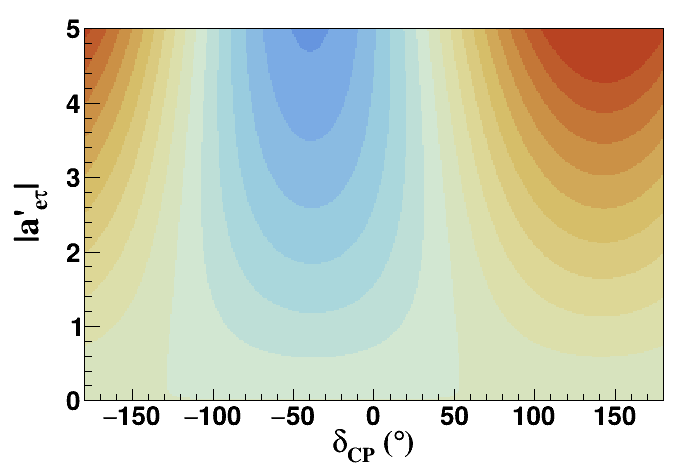}
\includegraphics[width=0.325\linewidth, height=5cm]{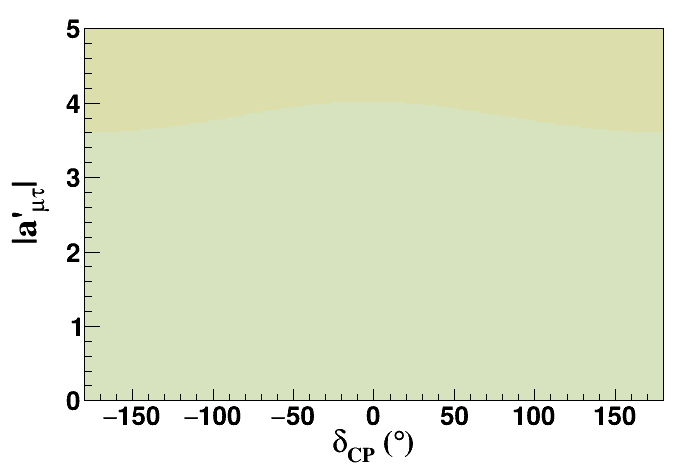}
\caption{2D heatmap of $\Delta P_{\mu e}$ 
on the $a'_{\alpha\beta}$--$\delta_{CP}$ plane at $\rm E = 2.5~\mathrm{GeV}$. The off-diagonal LIV phases are set to $\phi^{a}_{\alpha\beta} = -90^\circ$.}
\label{fig:a_vs_dcp2d}
\end{figure}

In the previous sections, we observed that both $a_{\alpha\beta}$ and $c_{\alpha\beta}$ elements can modify the standard oscillation probability, as well as the CP trajectories in the bi-probability planes. However, the analysis provides only a limited perspective, as it is considered for fixed parameter values. In order to obtain a more complete picture, we perform two-dimensional parameter scans that allow us to investigate correlations and possible degeneracies between different LIV parameters and $\delta_{CP}$. We produce the 2D-heatmaps of $\Delta P_{\mu e}$ which is defined as 
\begin{equation}
\Delta P_{\mu e} = P_{\mu e} - P_{\mu e}^{\rm SI}
\end{equation}
by varying two parameters at a time. The first part of the analysis examines the
$(a'_{\alpha\beta}$--$\delta_{\rm CP})$ and $(c'_{\alpha\beta}$--$\delta_{\rm CP})$ planes at a fixed energy of 2.5~GeV, which corresponds to the first oscillation maximum and lies within the high–flux region of the DUNE beam. We then study the $(a'_{\alpha\beta}$--$c'_{\alpha\beta})$ parameter space at $\delta_{\rm CP} = -90^\circ$, followed by an exploration of the $(\phi_{\alpha\beta}^{a/c}$--$\delta_{\rm CP})$ plane. We note our observations below

\begin{enumerate}[label={\arabic*.}]
\item \textit{For $(a'_{\alpha \beta}-\delta_{CP})$ plane:} In figure~\ref{fig:a_vs_dcp2d}, we examine the heatmaps of $\Delta P_{\mu e}$ in the ($a'_{\alpha\beta}$--$\delta_{CP}$) plane.  We observe that among the diagonal elements, for lower absolute magnitudes ($-20 \lesssim a'_{\alpha\alpha} \lesssim 20$) of $a_{ee}$ and $a_{\tau\tau}$, the variation of $\Delta P_{\mu e}$ is nearly independent of $\delta_{CP}$ and changes approximately in proportion to the value of $a'_{\alpha\alpha}$. For $a_{\mu\mu}$, such behavior is observed in the range $-30 \lesssim a'_{\mu\mu} \lesssim 0$. We identify two broad regions of approximate degeneracy for $a_{\mu\mu}$, located at $-15 \lesssim a'_{\mu\mu} \lesssim 0$ and $5 \lesssim a'_{\mu\mu} \lesssim 20$. As shown in figure~\ref{fig:Pmue_E} (top--left panel), the impact of $a_{\mu\mu}$ on $P_{\mu e}$ is very small, leading to weak $\delta_{CP}$ dependence and therefore broad degeneracy bands. For the off-diagonal elements, we fix $\phi^{a}_{\alpha\beta}$ at $-90^{\circ}$. In presence of either $a_{e\mu}$ or $a_{e\tau}$, $\Delta P_{\mu e}$ exhibits a sinusoidal dependence on $\delta_{CP}$ at any nonzero value of $|a'_{\alpha\beta}|$, and the amplitude of both the positive and negative peaks increases with increasing $|a'_{\alpha\beta}|$. The positive and negative regions are separated by narrow degenerate lines, which appear around $\delta_{CP} \sim -70^{\circ}, 150^{\circ}$ for $a_{e\mu}$, and $\delta_{CP} \sim -110^{\circ}, 40^{\circ}$ for $a_{e\tau}$.  We note that the impact of $a_{\mu\tau}$ is much smaller compared to $a_{e\mu}$ or $a_{e\tau}$. 
\begin{figure}[!h]
    \centering     \includegraphics[width=0.6\linewidth]{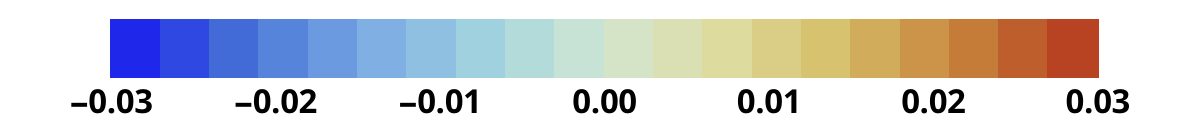}

\par
\includegraphics[width=0.325\linewidth, height=5cm]{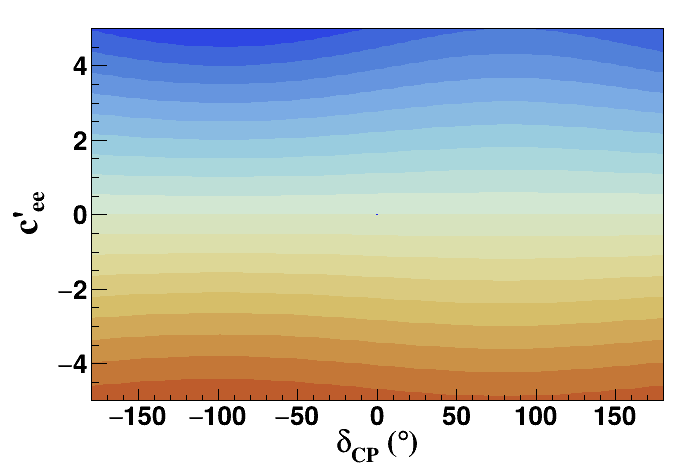}
\includegraphics[width=0.325\linewidth, height=5cm]{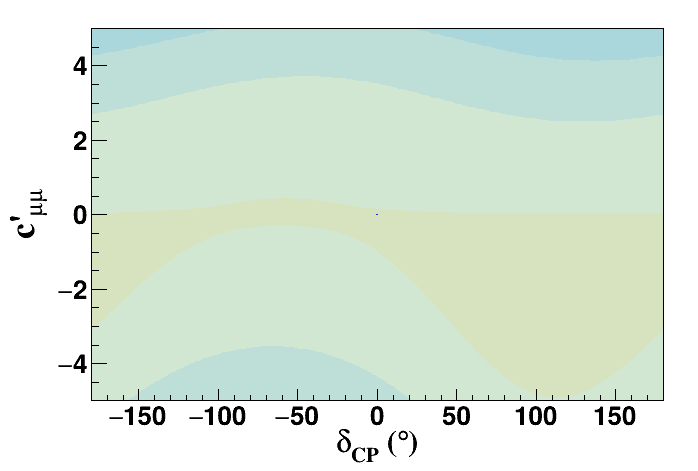}
\includegraphics[width=0.325\linewidth, height=5cm]{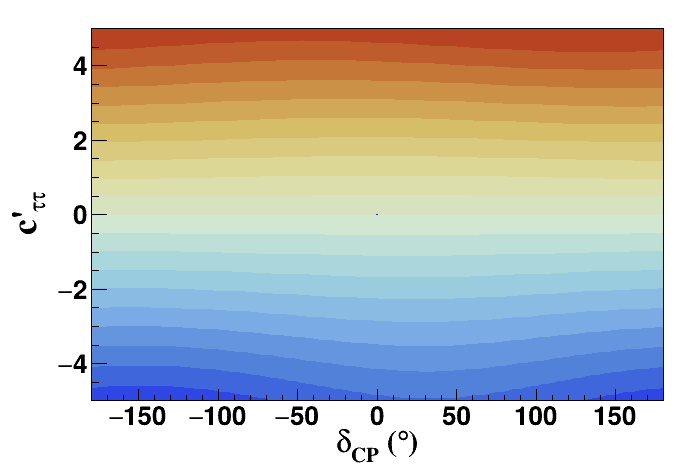}
\includegraphics[width=0.325\linewidth, height=5cm]{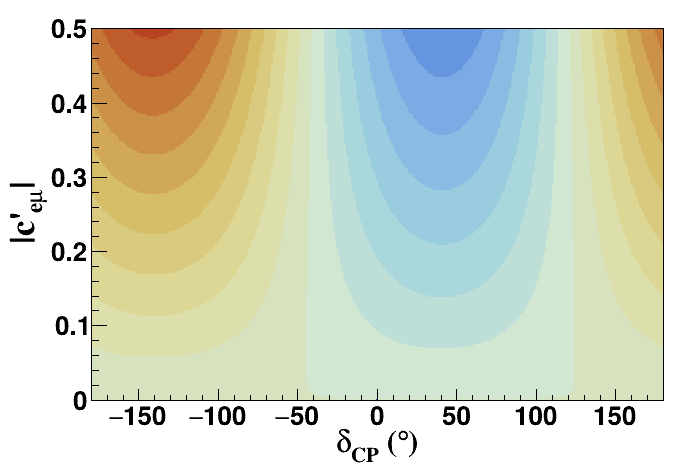}
\includegraphics[width=0.325\linewidth, height=5cm]{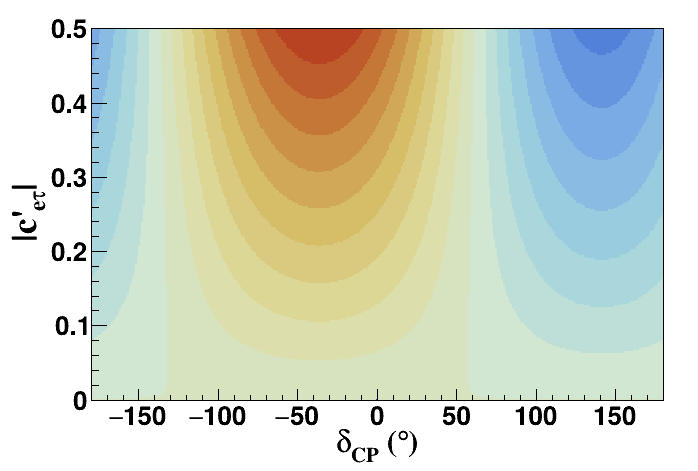}
\includegraphics[width=0.325\linewidth, height=5cm]{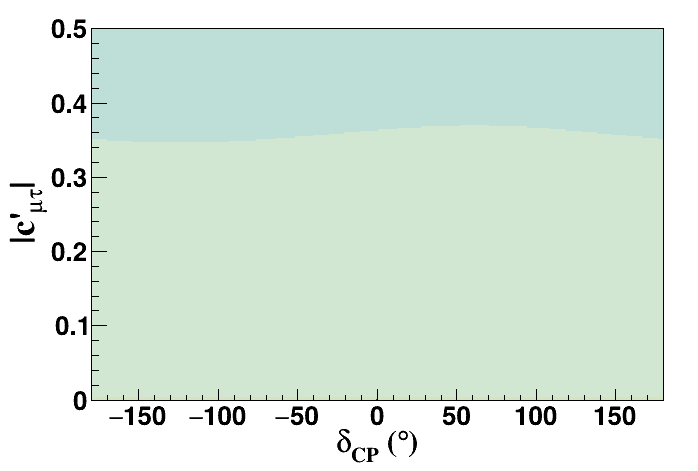}
\caption{2D heatmap of $\Delta P_{\mu e}$ 
on the $c'_{\alpha\beta}$--$\delta_{CP}$ plane at $\rm E = 2.5~\mathrm{GeV}$.  The off-diagonal LIV phases are set to $\phi^{c}_{\alpha\beta} = -90^\circ$.}
\label{fig:c_vs_dcp2d}
\end{figure}
\item \textit{For $(c'_{\alpha \beta}-\delta_{CP})$ plane:} In figure \ref{fig:c_vs_dcp2d}, we present heatmaps of $\Delta P_{\mu e}$ in the ($c'_{\alpha\beta}$–$\delta_{CP}$) plane. We see a clear distinction between the patterns for diagonal (top row) and off-diagonal (bottom row) elements. Among the diagonal elements, the impact of $c_{ee}$ and $c_{\tau\tau}$ are significant, and the pattern for $c_{\tau\tau}$ looks nearly opposite to that of $c_{ee}$ along the vertical axis. In contrast, the contribution from $c_{\mu\mu}$ is much weaker. The low variation in color at any particular $c_{\alpha\beta}$ suggests that the effect of $c_{ee}$ and $c_{\tau\tau}$ remains nearly constant across the full $\delta_{CP}$ range. As with the off-diagonal $a_{\alpha\beta}$ elements, we observe noticeable effects on $\Delta P_{\mu e}$ due to $c_{e\mu}$ and $c_{e\tau}$. In both cases, $\Delta P_{\mu e}$ displays a sinusoidal dependence on $\delta_{CP}$, with the amplitude of the oscillations increasing with $|c'_{\alpha\beta}|$. The positive and negative regions are separated by narrow degenerate lines. For $c_{e\mu}$, these lines appear around $\delta_{CP} \sim -40^\circ$ and $110^\circ$, while for $c_{e\tau}$, they appear around $\delta_{CP} \sim -140^\circ$ and $70^\circ$. The effect of $c_{\mu\tau}$ is negligible. From the individual analyses of $a_{\alpha\beta}$ and $c_{\alpha\beta}$ elements, it is evident that both can significantly alter the oscillation probability. This motivates us to investigate their combined effect, which we explore in the next part.
\begin{figure}[!h]
\centering
\includegraphics[width=0.6\linewidth]{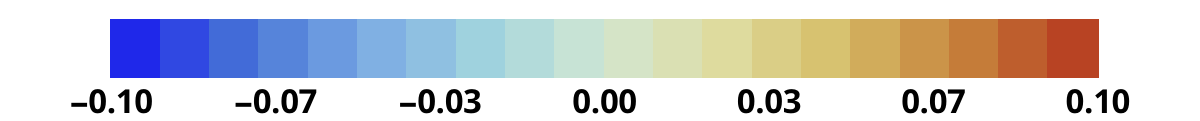}
\par
\includegraphics[width=0.325\linewidth, height=5cm]{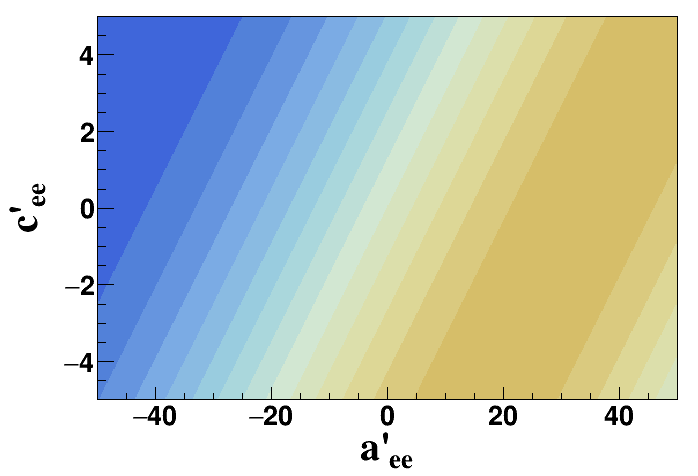}
\includegraphics[width=0.325\linewidth, height=5cm]{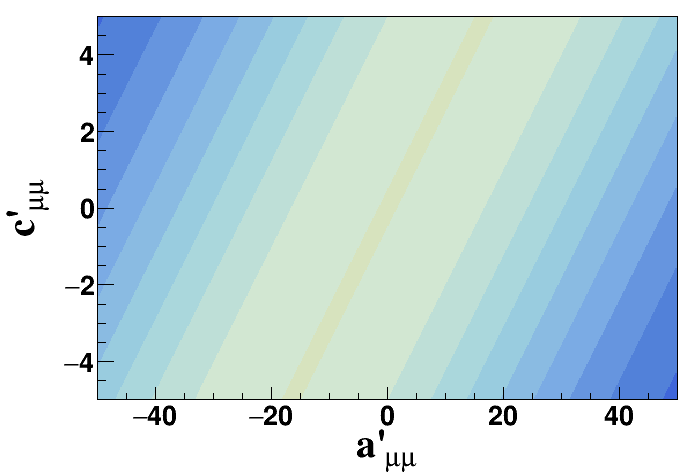}
\includegraphics[width=0.325\linewidth, height=5cm]{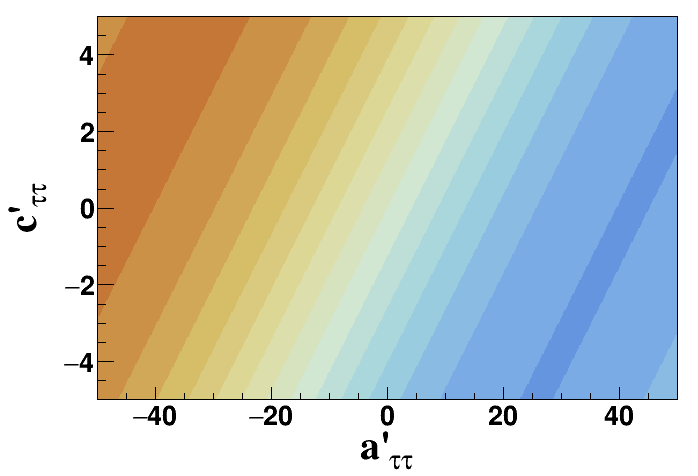}
\includegraphics[width=0.325\linewidth, height=5cm]{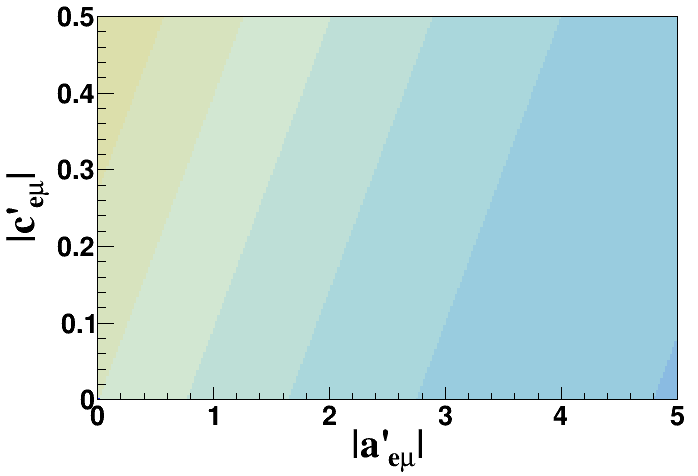}
\includegraphics[width=0.325\linewidth, height=5cm]{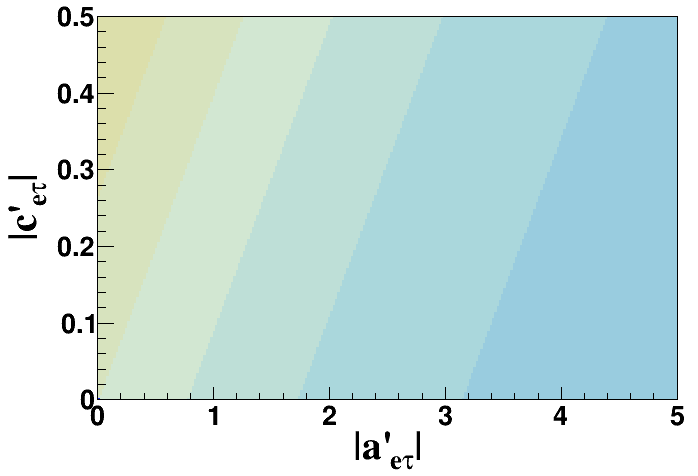}
\includegraphics[width=0.325\linewidth, height=5cm]{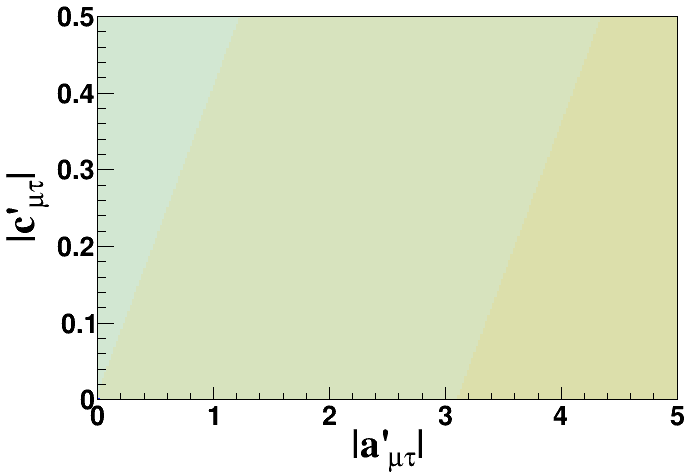}
\caption{2D heatmap of $\Delta P_{\mu e}$ on the $c'_{\alpha\beta}$--$a'_{\alpha\beta}$ plane at $E = 2.5~\mathrm{GeV}$ and $\delta_{CP}=-90^{\circ}$. The off-diagonal LIV phases are set to $\phi^{a}_{\alpha\beta} = -90^\circ$ and $\phi^{c}_{\alpha\beta} = -90^\circ$.}
\label{fig:c_vs_a2d}
\end{figure}
\item \textit{For $(c'_{\alpha \beta}-a'_{\alpha \beta})$ plane:} In figure \ref{fig:c_vs_a2d}, we examine the combined effect of $a_{\alpha\beta}$ and the corresponding $c_{\alpha\beta}$ elements on $\Delta P_{\mu e}$ . For all the pairs, the plots exhibit linear bands of varying width, slanting at a fixed angle with respect to the $a'_{\alpha\beta}$ axis. Each band corresponds to a nearly constant $\Delta P_{\mu e}$ value, implying a strong correlation between the $a_{\alpha\beta}$ and $c_{\alpha\beta}$ parameters. This indicates that multiple combinations of these LIV parameters can yield the same $\Delta P_{\mu e}$. We also observe bands corresponding to $\Delta P_{\mu e}\approx 0$, which indicates that the contributions from $a_{\alpha\beta}$ and $c_{\alpha\beta}$ can nearly cancel each other, leading to SI–LIV degeneracy. This degeneracy is most prominent for the $a_{\mu\tau}$–$c_{\mu\tau}$ pair due to their relatively smaller individual contributions. For the other LIV pairs, the degenerate regions appear but are narrower. The linear behavior can be understood from the first-order probability contributions discussed in eq.~\ref{eq:Pac}. The combined effect of any $a_{\alpha\beta}$--$c_{\alpha\beta}$ pair can be written as
\begin{equation}
    \Delta \mathrm{P}_{\mu e} \approx K^{(1)}_{\alpha\beta}\,a_{\alpha\beta} - \frac{4}{3} K^{(1)}_{\alpha\beta} E\, c_{\alpha\beta}\,,
    \label{eq:Pacslope}
\end{equation}
where $K^{(1)}_{\alpha\beta}$ is a constant for the chosen oscillation parameters. Since we evaluate at $E = 2.5$~GeV, the inclination of each linear band follows directly from eq.~\ref{eq:Pacslope} by imposing that $\Delta P_{\mu e}$ remains constant along the band. This results in a predicted inclination of $\approx 71.57^\circ$, which is in agreement with the observed inclination of $\approx 76^\circ$ in the figure.

\begin{figure}[!b]
\centering     \includegraphics[width=0.6\linewidth]{Figures/LIV_vs_dcp/klighttemp_colorbar_cvsdcp.pdf}
\par
\includegraphics[width=0.325\linewidth, height=5cm]{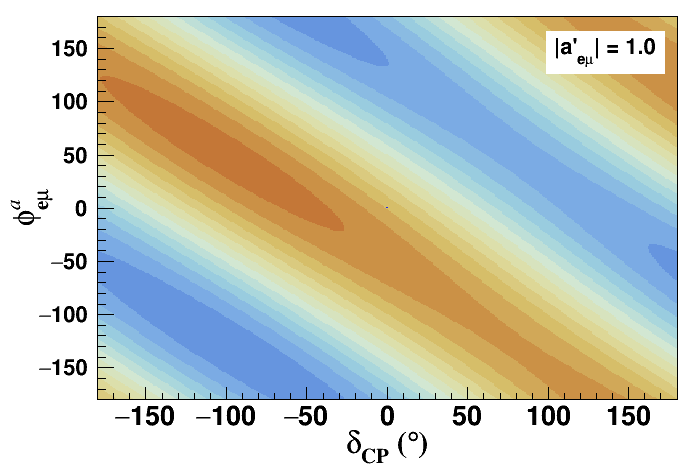}
\includegraphics[width=0.325\linewidth, height=5cm]{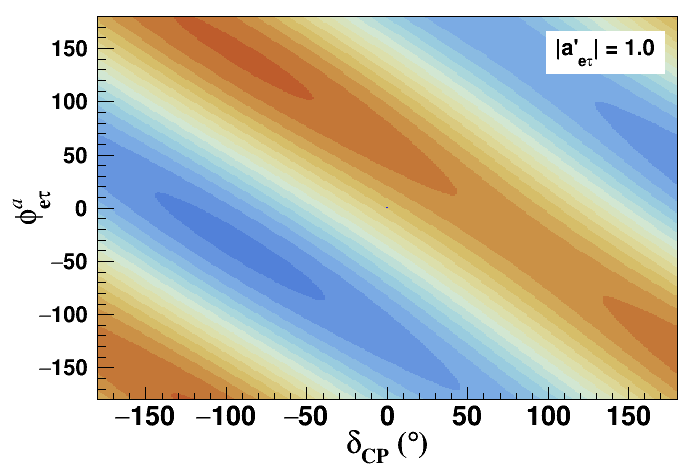}
\includegraphics[width=0.325\linewidth, height=5cm]{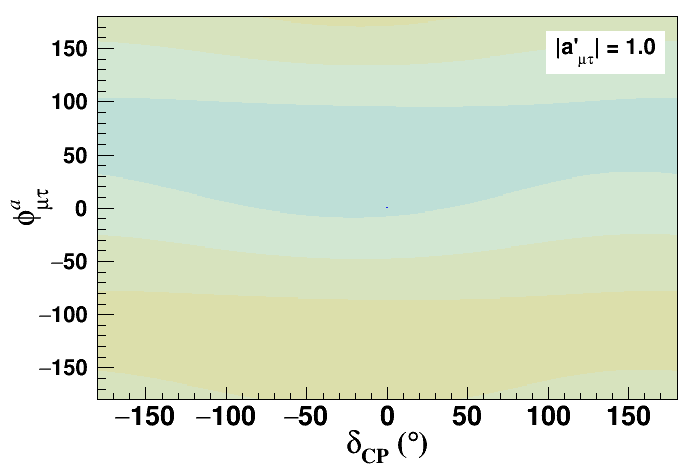}
\includegraphics[width=0.6\linewidth]{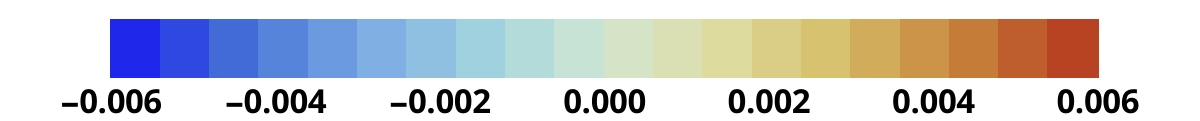}
\par
\includegraphics[width=0.325\linewidth, height=5cm]{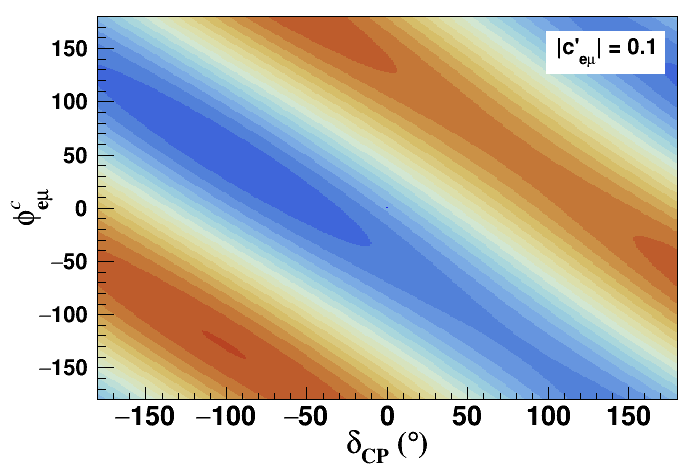}
\includegraphics[width=0.325\linewidth, height=5cm]{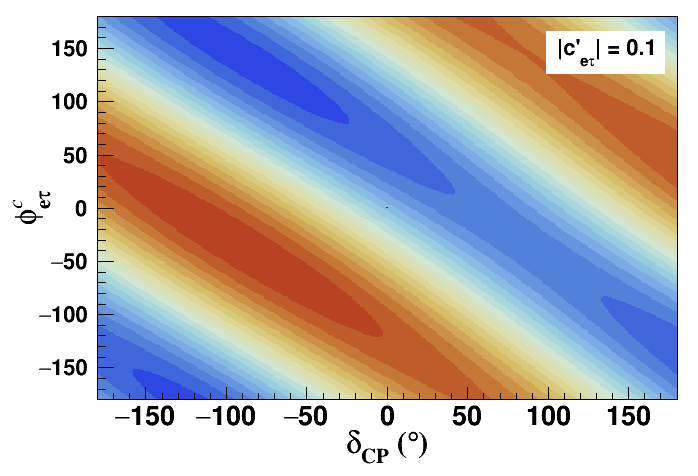}
\includegraphics[width=0.325\linewidth, height=5cm]{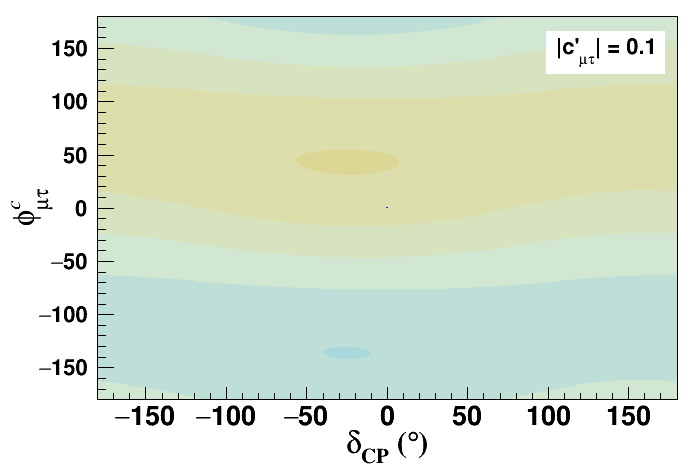}
\caption{Variation of $\Delta P_{\mu e}$ across the $\phi^{a}_{\alpha\beta}-\delta_{\rm CP}$ plane (top panel) and $\phi^{c}_{\alpha\beta}-\delta_{\rm CP}$ plane (bottom panel) at $E=2.5~\mathrm{GeV}$. The off-diagonal LIV parameters are fixed at $a'_{\alpha\beta}=1.0$ and $c'_{\alpha\beta}=0.1$.}
\label{fig:phivsdcp}
\end{figure}
\item \textit{For $(\phi_{\alpha \beta}^{a/c}-\delta_{CP})$ plane:} We consider the cases of non-zero LIV phases in our discussion. In figure \ref{fig:phivsdcp}, we depict the heatmap for $\Delta P_{\mu e}$ as a function of $\phi_{\alpha \beta}^{a/c}$ and $\delta_{CP}$ at the energy of 2.5 GeV.  The top (bottom) panel corresponds to the CPT-violating (conserving) off-diagonal LIV elements with an absolute magnitude of $|a'_{\alpha\beta}|=1.0 \,(|c'_{\alpha\beta}|=0.1)$. In the top panel, we observe a pattern of linear diagonal bands in $\Delta P_{\mu e}$ when either the $a_{e\mu}$ or $a_{e\tau}$ parameter is varied. Along each of these bands, $\Delta P_{\mu e}$ remains approximately constant, indicating a strong correlation between the standard CP phase $\delta_{\rm CP}$ and the LIV phase $\phi^{a}_{\alpha\beta}$. From the analytical expression of the $a_{e\mu}$ contribution to $P_{\mu e}$ given in eq.~\ref{eq:K1d}, we find that the dominant phase dependence arises from the terms $\sin\!\left[\Delta(1-A) - (\delta_{\rm CP} + \phi^{a}_{e\mu})\right] $ and $\sin\!\left[\Delta(1-A) + (\delta_{\rm CP} + \phi^{a}_{e\mu})\right]$. These two terms combine into a single effective sinusoidal dependence of the form -- $ R\,\sin\!\left(\delta_{\rm CP} + \phi^{a}_{e\mu} + \varphi_{e\mu}^{a}\right)$, where the amplitude $R$ and the phase shift $\varphi_{e\mu}$ are given by
\begin{align}
    R &= \sqrt{(C_-)^{\,2}\sin^2\!\big[\Delta(1-A)\big]
          + (C_+)^{\,2}\cos^2\!\big[\Delta(1-A)\big]}, \\
    \tan\varphi_{e\mu}^{a} &=
    \frac{(C_-)\,\sin\!\big[\Delta(1-A)\big]}
         {(C_+)\,\cos\!\big[\Delta(1-A)\big]},
\end{align}
\begin{align} 
    \text{with,} \hspace{0.5cm}
    C_- = 2\Big[(1-A) - (1+A)s_{23}^2\Big], \hspace{1cm}
    C_+ = 2(1-A)c_{23}^2 .
\end{align}
Here, $\varphi_{e\mu}^{a}$ represents a matter-dependent phase shift arising from the unequal weights of the $\cos(\delta_{\rm CP}+\phi^a_{e\mu})$ and
    $\sin(\delta_{\rm CP}+\phi^a_{e\mu})$ terms. As a result, the SI--LIV degeneracy occurs when the argument of the sine function remains approximately constant, leading to the condition
    \begin{align}
    \delta_{\rm CP} + \phi^{a}_{e\mu} + \varphi_{e\mu}^{a} \simeq n\pi,
    \qquad n = 0, \pm1, \pm2, \dots
    \label{eq:dcpphi_corr_aemu}
    \end{align}
This relation represents straight lines with slope $-1$ in the $(\delta_{\rm CP},\,\phi^{a}_{e\mu})$ plane, corresponding to an inclination angle of $135^\circ$. The separation between adjacent degenerate bands is $\pi$, while their intercepts are shifted by the matter-induced phase $\varphi_{e\mu}^{a}$. This behavior is in good agreement with the diagonal degeneracy bands.

\hspace{5mm} Similarly, for the $a_{e\tau}$ contribution to $P_{\mu e}$ given in
eq.~\ref{eq:K1f}, the dominant dependence on the CP phases arises through the combination $\delta_{\rm CP} + \phi^{a}_{e\tau}$. After combining the relevant trigonometric terms, the $a_{e\tau}$ contribution can be expressed in the form -- $ R_{e\tau}\,\sin\!\left(\delta_{\rm CP} + \phi^{a}_{e\tau} + \varphi_{e\tau}^{a}\right)$, where $R_{e\tau}$ is an overall amplitude and $\varphi_{e\tau}^{a}$ is an effective matter-dependent phase shift arising from the unequal weights of the contributing terms. Consequently, the SI--LIV degeneracy condition for $a_{e\tau}$ is given by
    \begin{align}
    \delta_{\rm CP} + \phi^{a}_{e\tau} + \varphi_{e\tau}^{a} \simeq n\pi,
    \qquad n = 0, \pm1, \pm2, \dots
    \label{eq:dcpphi_corr_aetau}
    \end{align}
This relation corresponds to straight lines with slope $-1$ in the $(\delta_{\rm CP},\,\phi^{a}_{e\tau})$ plane, leading to diagonal degeneracy bands separated by $\pi$, in good agreement with the numerical results.
    
\hspace{5mm} The effect of $a_{\mu \tau}$ on $\Delta \,P_{\mu e}$ is much less in comparison to the other off-diagonal elements, and we do not see such diagonal patterns. This behavior is expected since the dependence on $\delta_{CP}$ and $\phi_{\mu \tau}^{a}$ is complex, as seen in eq.~\ref{eq:K1f}. We note a similar effect for the corresponding $c_{\alpha\beta}$ elements, since the patterns are similar (bottom panel in figure \ref{fig:phivsdcp}). This is also supported by the analytical expressions.
\end{enumerate}
In the next section, we discuss the capability of DUNE to constrain both the CPT-violating and CPT-conserving LIV elements when considered simultaneously. We then explore how the combined LIV effects can affect DUNE's CP violation sensitivities. 

\section{Sensitivity studies of LIV and CP violation}\label{sec:CPV_sens}
The probability-level analyses presented in the previous sections show that both $a_{\alpha\beta}$ and $c_{\alpha\beta}$ can lead to significant deviations from the standard 3-flavor oscillation scenario. It also reveals nontrivial correlations and degeneracies in the presence of LIV, which can complicate the interpretation of experimental observables. To study these effects quantitatively, we first determine the allowed regions of the LIV parameter space and compare the resulting bounds with existing experimental constraints. We then examine the impact of LIV on the CPV sensitivity of DUNE. This allows us to assess whether LIV effects can mimic, enhance, or obscure genuine leptonic CP violation.

\subsection{Constraints and correlations of LIV parameters}
\begin{figure}[!h]
\includegraphics[width=0.325\linewidth, height=5cm]{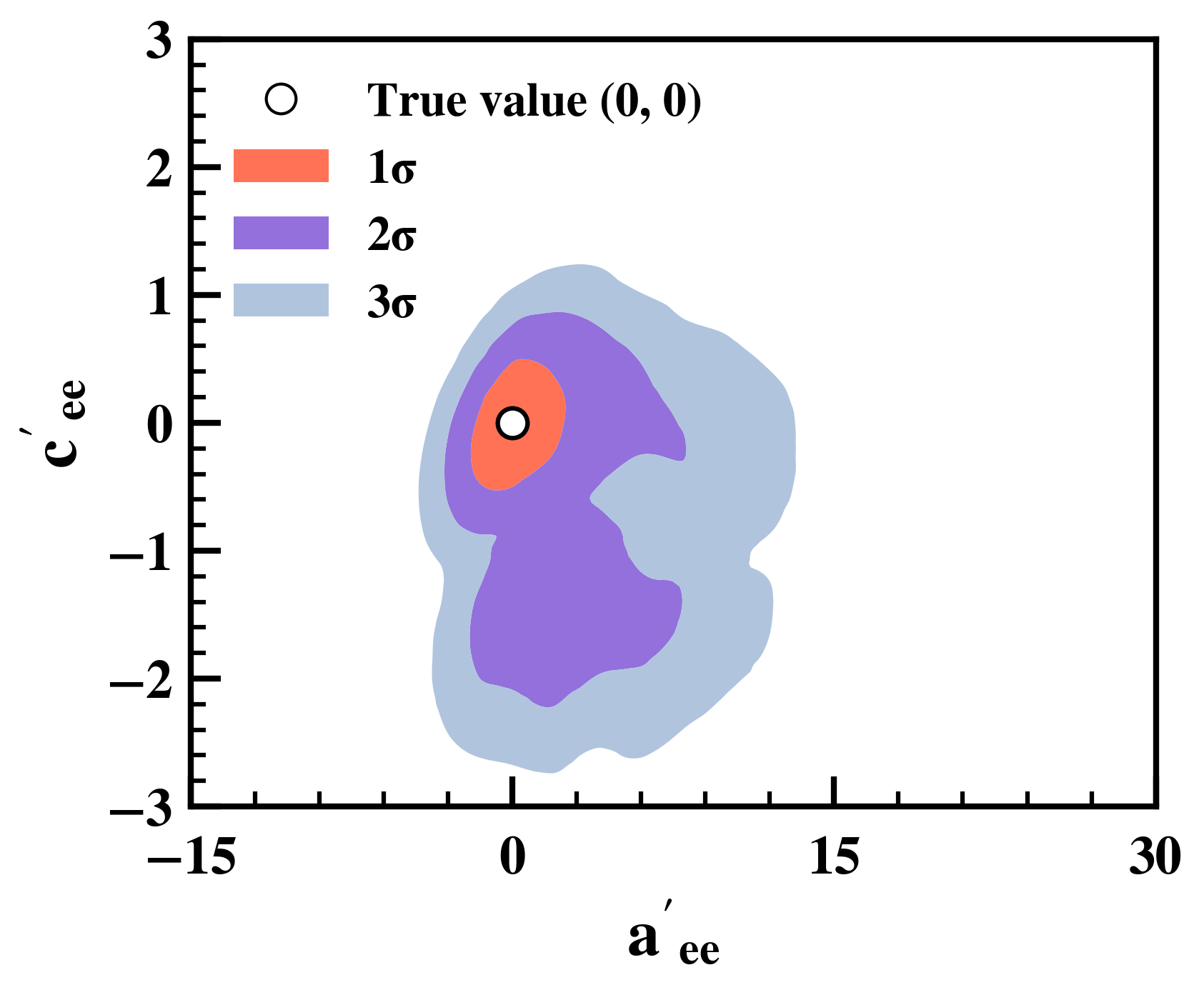}
\includegraphics[width=0.325\linewidth, height=5cm]{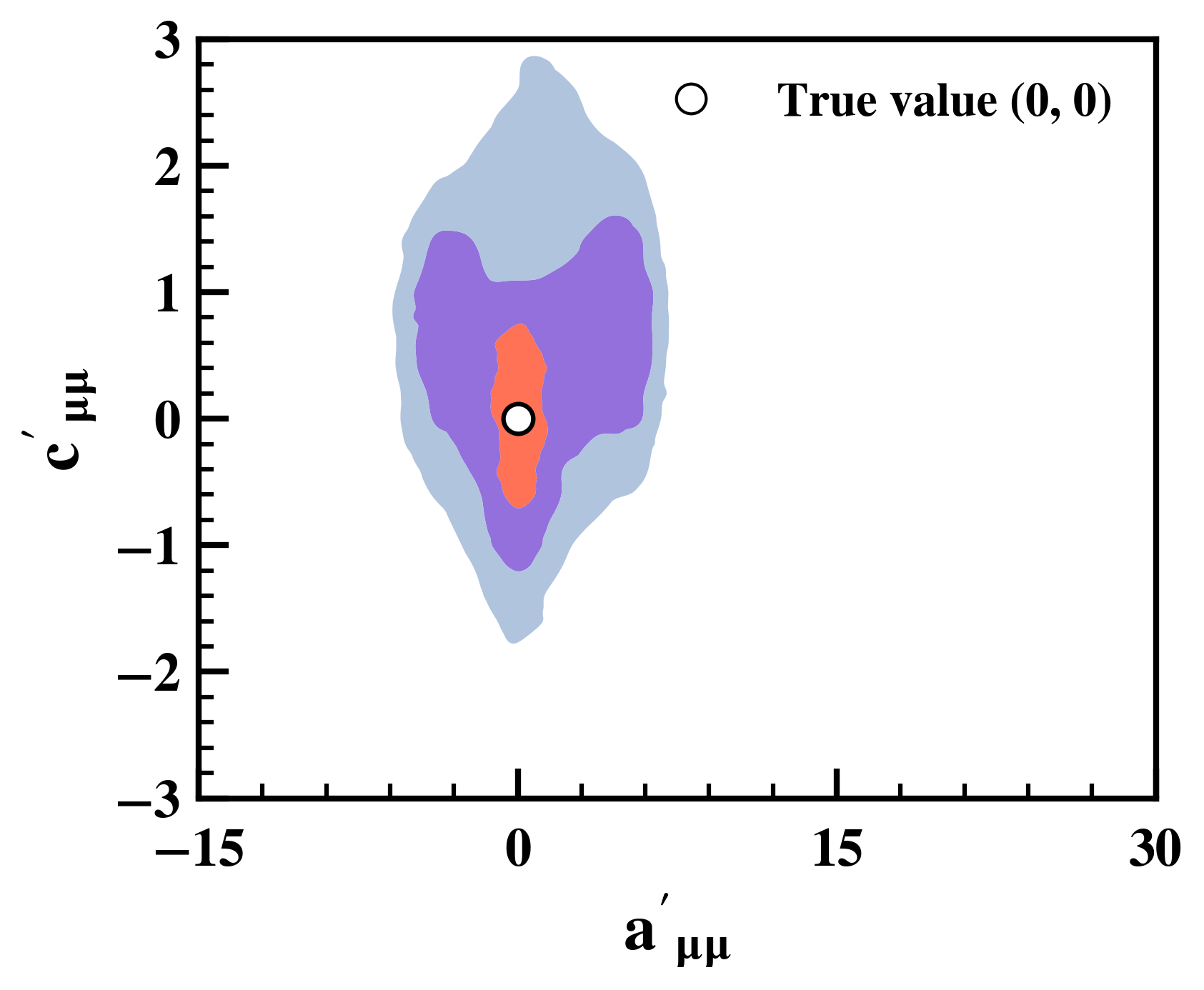}
\includegraphics[width=0.325\linewidth, height=5cm]{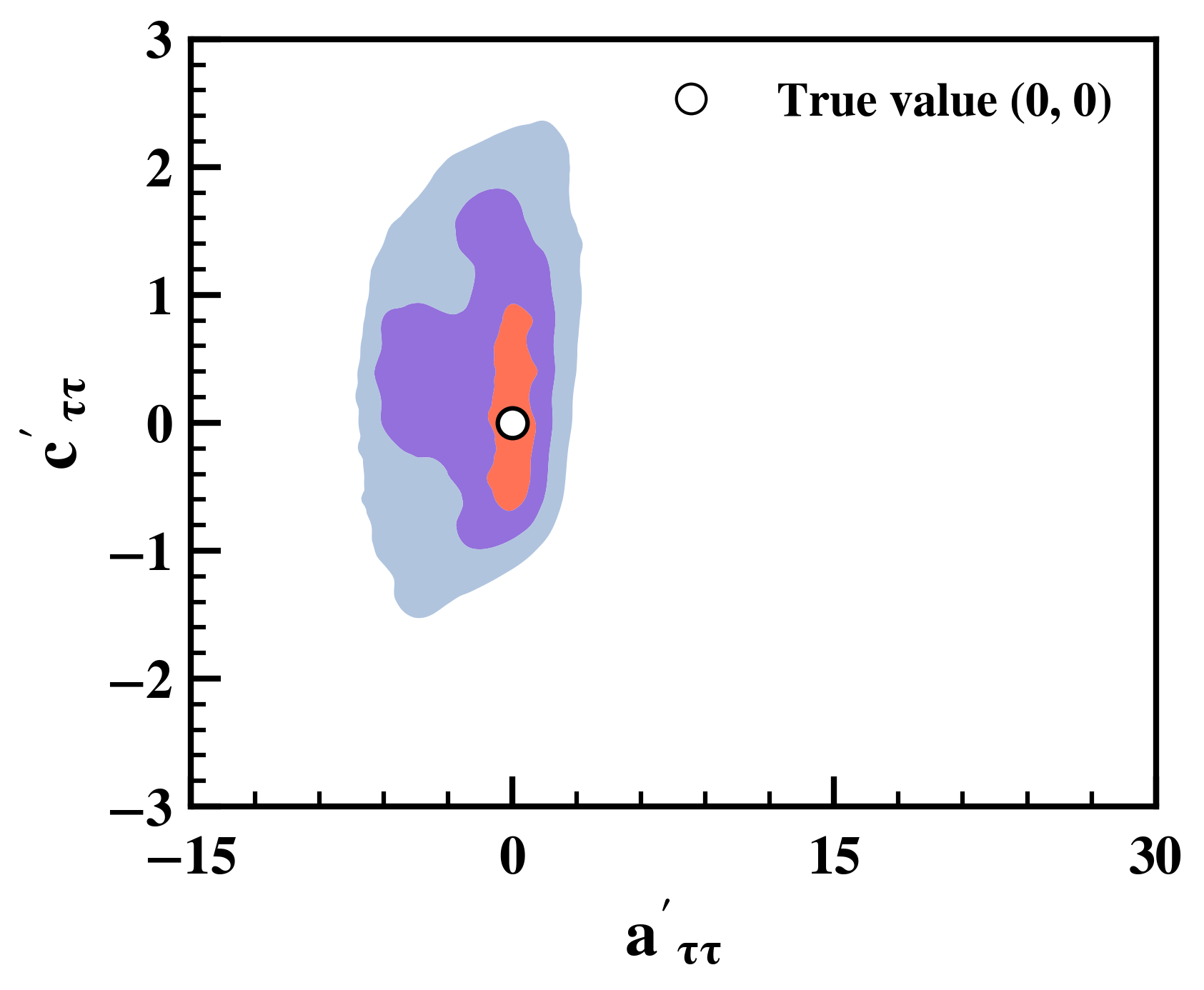}

\includegraphics[width=0.325\linewidth, height=5cm]{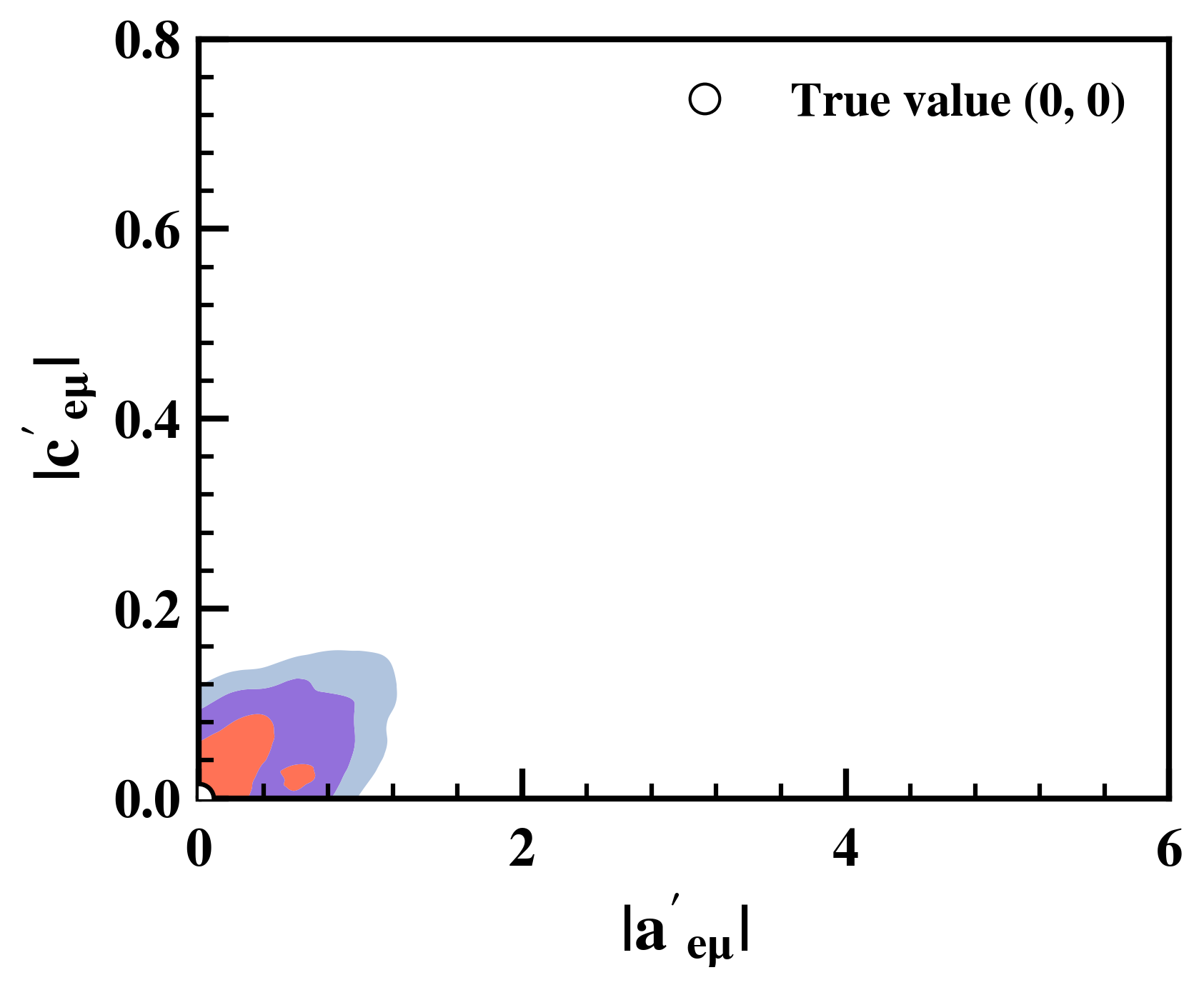}
\includegraphics[width=0.325\linewidth, height=5cm]{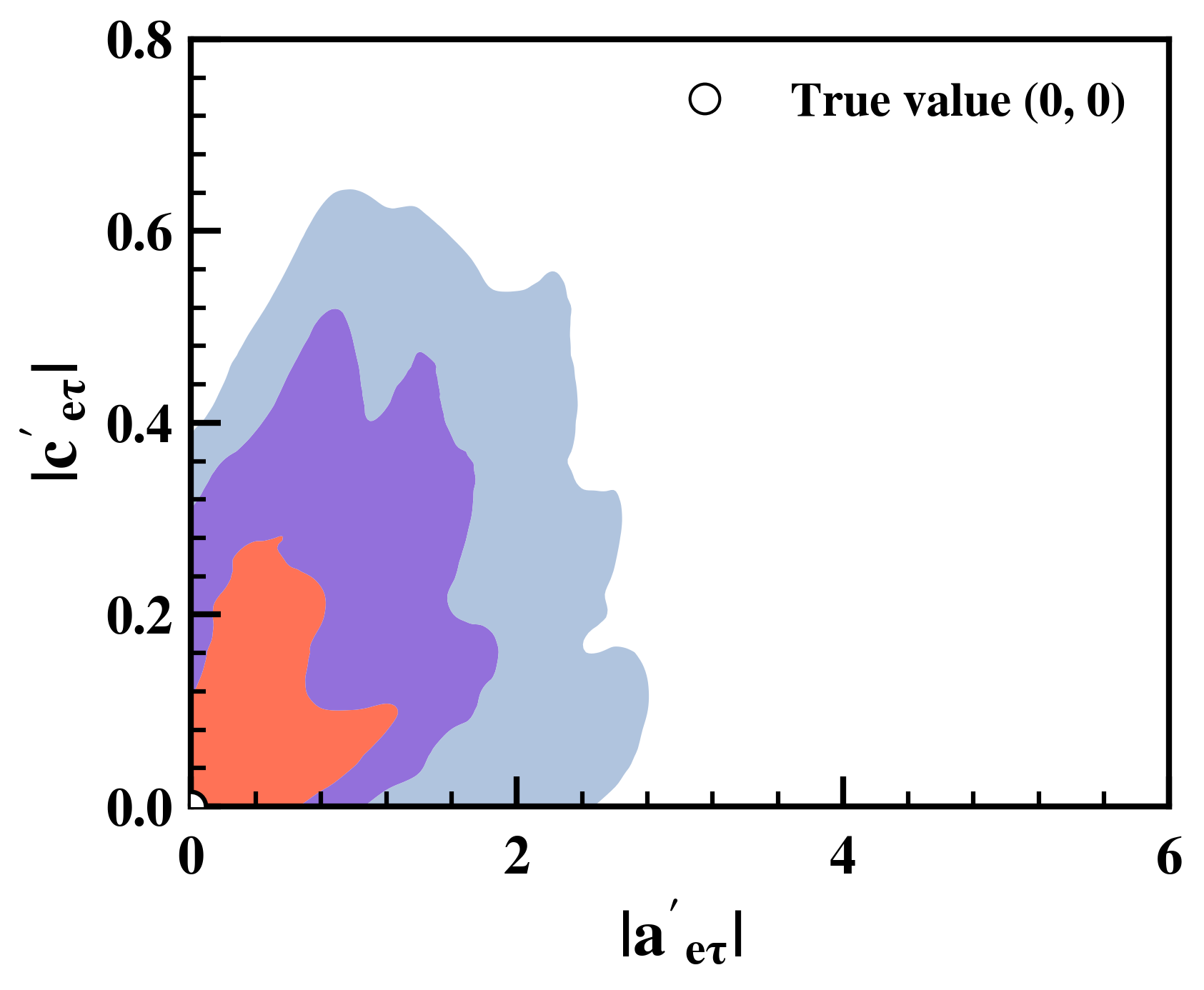}
\includegraphics[width=0.325\linewidth, height=5cm]{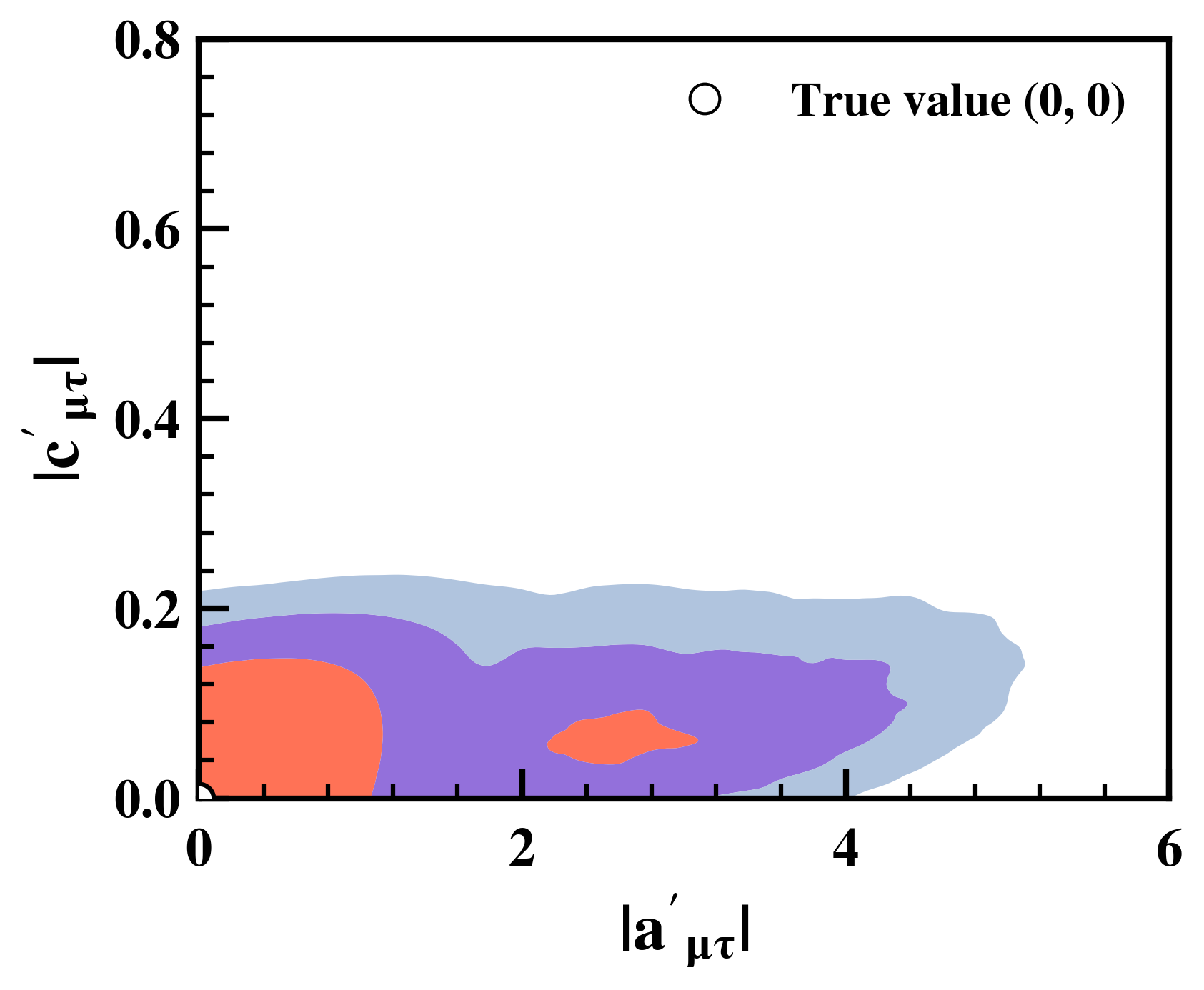}
\caption{Correlation between the LIV parameters $a'_{\alpha\beta}$ and $c'_{\alpha\beta}$ for the diagonal (top panel) and off-diagonal (bottom panel) elements. The orange, violet, and gray regions correspond to the $1\sigma$, $2\sigma$, and $3\sigma$ allowed regions, respectively. The white solid circle indicates the assumed true value.}
\label{fig:chi2_2d_allowed}
\end{figure}

In this section, we present the allowed regions in the LIV parameter space obtained using a $\Delta\chi^{2}$ analysis. For each panel, an $a_{\alpha\beta}$ parameter is varied along the $x$-axis and the corresponding $c_{\alpha\beta}$ parameter along the $y$-axis. The true values of all LIV parameters are taken to be zero, corresponding to the absence of Lorentz invariance violation. The oscillation parameters $\delta_{CP}$, $\Delta m^{2}_{31}$ and $\theta_{23}$ are marginalized over by minimizing the $\chi^{2}$ within their allowed ranges given in table~\ref{tab:bestfit}. In the case of off-diagonal LIV parameters, the corresponding phases $\phi_{\alpha\beta}^{a}$ and $\phi_{\alpha\beta}^{c}$ are additionally marginalized over the full interval $[-180^{\circ},\,180^{\circ})$. In figure~\ref{fig:chi2_2d_allowed}, we show the $1\sigma$, $2\sigma$, and $3\sigma$ allowed regions in the two-dimensional parameter planes, corresponding to $\Delta\chi^{2} = 2.30$, $6.18$, and $11.83$, respectively, for two degrees of freedom~\cite{ParticleDataGroup:2024cfk}. The main observations are summarized below

\begin{itemize}
    \item \textit{Diagonal LIV pairs (top panel)}:
     The sensitivities exhibit distinct behaviors for the different diagonal LIV parameter pairs. For the $a_{ee}$--$c_{ee}$ pair, the allowed regions are relatively broad. The irregular and elongated contour shape may indicate the presence of parameter degeneracies and a comparatively weaker constraint on $a'_{ee}$. In contrast, the $a_{\mu\mu}$--$c_{\mu\mu}$ contours are more localized around the true value and display a comparatively regular shape, suggesting improved parameter reconstruction. The $a_{\tau\tau}$--$c_{\tau\tau}$ contours are likewise compact, with no evidence of disconnected allowed regions. The contour sizes for the $a_{\mu\mu}$--$c_{\mu\mu}$ and $a_{\tau\tau}$--$c_{\tau\tau}$ pairs are broadly comparable, indicating similar reconstruction capabilities, while both are considerably better constrained than the $a_{ee}$--$c_{ee}$ pair. Our bounds are compared to the existing bounds in table~\ref{tab_bounds}.

    \item \textit{Off-diagonal LIV pairs (bottom panel)}:
     The off-diagonal LIV parameter pairs exhibit a richer correlation structure than the diagonal pairs. For the $a_{e\mu}$--$c_{e\mu}$ pair, the allowed regions are compact and localized near the true value, indicating comparatively strong constraints on both parameters. A small disconnected region appears at the $1\sigma$ level, suggesting the presence of a local secondary solution. However, the higher-confidence contours remain connected, implying that this degeneracy is relatively weak. In contrast, the $a_{e\tau}$--$c_{e\tau}$ pair exhibits the broadest allowed regions among the off-diagonal sectors. The contours extend over a large range of both $a'_{e\tau}$ and $c'_{e\tau}$, indicating weaker constraints and the presence of substantial parameter degeneracies. For the $a_{\mu\tau}$--$c_{\mu\tau}$ pair, the contours are elongated along the $a'_{\mu\tau}$ direction and develop a disconnected allowed region at larger values of $a'_{\mu\tau}$. The appearance of this isolated island suggests the existence of a residual secondary solution, pointing to a nontrivial degeneracy structure in the $\mu\tau$ sector. A common feature of all the lower panels is that, the allowed contours intersect the individual parameter axes at smaller values, compared to their full allowed range of values when both parameters are varied simultaneously. This behavior is because of the correlations between the paired LIV parameters, which broaden the joint allowed region in the two-parameter plane. This underscores the importance of considering both, CPT-conserving and CPT-violating parameters simultaneously in LIV analyses.
\end{itemize}

\begin{table}[!t]
\centering
\renewcommand{\arraystretch}{1.2}

\begin{tabular}{ccc}
\toprule
Experiments & Parameters [Unit] & Limits at $95\%$ CL \\
\midrule
SK~\cite{Super-Kamiokande:2014exs} & $a_{e\mu}$ [$10^{-23}$ GeV] & $|\mathrm{Re}(a_{e\mu})|<1.8, |\mathrm{Im}(a_{e\mu})|< 1.8$  \\   & $a_{e\tau}$ [$10^{-23}$ GeV] & $|\mathrm{Re}(a_{e\tau})|<4.1, |\mathrm{Im}(a_{e\tau})|< 2.8$ \\
  & $a_{\mu\tau}$ [$10^{-24}$ GeV] & $|\mathrm{Re}(a_{\mu\tau})|< 6.5, |\mathrm{Im}(a_{\mu\tau})|< 5.1$ \\
  & $c_{e\mu}$ $[10^{-27}]$ & $|\mathrm{Re}(c_{e\mu})|< 8.0, |\mathrm{Im}(c_{e\mu})|< 8.0$ \\
  & $c_{e\tau}$ $[10^{-25}]$ & $|\mathrm{Re}(c_{e\tau})|< 9.3, |\mathrm{Im}(c_{e\tau})|< 10$ \\
  & $c_{\mu\tau}$ $[10^{-27}]$ & $|\mathrm{Re}(c_{\mu\tau})|< 4.4, |\mathrm{Im}(c_{\mu\tau})|< 4.2$ \\
\midrule
IceCube~\cite{IceCube:2017qyp} & $a_{\mu\tau}$ [$10^{-24}$ GeV] & $|\mathrm{Re}(a_{\mu\tau})|< 2.0, |\mathrm{Im}(a_{\mu\tau})|< 2.0$ (90\% CL)  \\
  & $c_{\mu\tau}$ [$10^{-28}$] & $|\mathrm{Re}(c_{\mu\tau})|< 2.7, |\mathrm{Im}(c_{\mu\tau})|< 2.7$ (90\% CL) \\
\midrule
KM3NeT~\cite{KM3NeT:2026kuj} & $a_{e\mu}$ [$10^{-23}$ GeV] & $<2.53$  \\
 & $a_{e\tau}$ [$10^{-23}$ GeV] & $<4.70$  \\
  & $a_{\mu\tau}$ [$10^{-23}$ GeV] & $<0.48$  \\
  & $a_{\mu\mu}-a_{\tau\tau}$ [$10^{-23}$ GeV] & $<1.56$  \\
  & $c_{\mu\mu}-c_{\tau\tau}$ [$10^{-24}$] & $<1.22$  \\
\midrule
DUNE & $(a_{e\mu}, c_{e\mu})$ [$10^{-23}\,\mathrm{GeV}, 10^{-24}$] & $(<0.95, <0.12)$  \\
(This Work) & $(a_{e\tau}, c_{e\tau})$ [$10^{-23}\,\mathrm{GeV}, 10^{-24}$] & $(<1.86, <0.51)$  \\
 & $(a_{\mu\tau}, c_{\mu\tau})$ [$10^{-23}\,\mathrm{GeV}, 10^{-24}$] & $(<4.31, <0.19)$  \\
 & $(a_{ee}, c_{ee})$ [$10^{-22}\,\mathrm{GeV}, 10^{-23}$] & $([-3.08, 7.82], [-2.20, 0.85])$  \\
 & $(a_{\mu\mu}, c_{\mu\mu})$ [$10^{-22}\,\mathrm{GeV}, 10^{-23}$] & $([-4.81, 6.32], [-1.18, 1.59])$  \\
 & $(a_{\tau\tau}, c_{\tau\tau})$ [$10^{-22}\,\mathrm{GeV}, 10^{-23}$] & $([-6.39, 1.95], [-0.97, 1.81])$  \\
\bottomrule
\end{tabular}
\caption{Current experimental constraints on LIV parameters from different $\nu$-oscillation experiments, together with the projected bounds from DUNE. The experimental bounds were provided for one-parameter analysis, whereas the limits from this work are derived from a two-parameter analysis.}
\label{tab_bounds}
\end{table}

In table~\ref{tab_bounds}, we summarize the current experimental bounds on the LIV parameters from the neutrino sector. The experimental bounds were provided for one-parameter analyses. Only KM3NeT provides constraints on the diagonal parameters. In contrast, our work simultaneously considers $a_{\alpha\beta}$ and $c_{\alpha\beta}$ parameters (including phases for the off-diagonal elements), and the bounds from our analysis are presented at 95\% CL to facilitate a fair comparison. We find that DUNE can provide stronger constraints only on the off-diagonal parameters $a_{e\mu}$ and $a_{e\tau}$ than the previously reported bounds. Following the bounds derived from DUNE, we adopt a benchmark value of $(a'_{\alpha\beta}, c'_{\alpha\beta}) = (1, 0.1)$. Hereafter, we use this benchmark value to study the impact of LIV on DUNE's sensitivity to CP violation.
 
\subsection{CP violation sensitivities in presence of LIV}  \label{sec_sens}

We explore the capability of DUNE to probe CP violation in the presence of LIV parameters $a_{\alpha\beta}$ or $c_{\alpha\beta}$, as well as their combined effects. Building on the correlations and degeneracies identified in the LIV parameter space, we now assess their implications for experimental sensitivities. The CPV sensitivity of an experiment quantifies how effectively that experiment can distinguish between CP-conserving and CP-violating $\delta_{\mathrm{CP}}$ values. For sensitivity analysis, we define the test statistic as
\begin{equation}
\rm    \Delta \chi^{2}_{CPV} (\delta^{True}_{CP}) = min~[\chi^{2} (\delta^{true}_{CP}, \delta^{test}_{CP}=0), \chi^{2} (\delta^{true}_{CP}, \delta^{test}_{CP}=\pm 180^{\circ})]
\end{equation}

\noindent We minimize $\chi^{2}$ for each value of $\rm \delta_{CP}^{True}$ which constitutes the parameter space. Throughout this analysis, we assume normal ordering to be true mass ordering. All standard oscillation parameters used in the analysis, along with the marginalization ranges for $\theta_{23}$ and $\Delta m_{31}^2$, are given in table~\ref{tab:bestfit}. We additionally marginalize over the LIV parameters: for the diagonal elements, the ranges are taken as $a'_{\alpha\alpha}\in[-5,\,5]$ and $c'_{\alpha\alpha}\in[-0.5, 0.5]$; for the off-diagonal elements, we use $a'_{\alpha\beta}\in[0, 5]$ and $c'_{\alpha\beta}\in[0, 0.5]$. The corresponding LIV phases $\phi^{a}_{\alpha\beta}$ and $\phi^{c}_{\alpha\beta}$ are marginalized over the full range $[-180^{\circ},\,180^{\circ})$. The $3\sigma$ and $5\sigma$ confidence levels correspond to the $\Delta\chi^{2}=9$ and $\Delta\chi^{2}=25$, respectively.

In figure \ref{fig:CPV_sens}, we present the CPV sensitivity as a function of $\rm \delta_{CP}^{True}$. In all the panels, the standard CPV sensitivity is shown by the solid black curve, which exceeds the $5 \sigma$ level in some regions of the lower $\rm \delta_{CP}$ half plane. The red (blue) curves represent the effects of the $a_{\alpha\beta}$ ($c_{\alpha\beta}$) elements, while the magenta curves denote the combined effect of the $a_{\alpha\beta}-c_{\alpha\beta}$ pair. A detailed inspection of all panels leads to the following observations:

\begin{figure}[!b]
\includegraphics[width=0.325\linewidth, height=5cm]{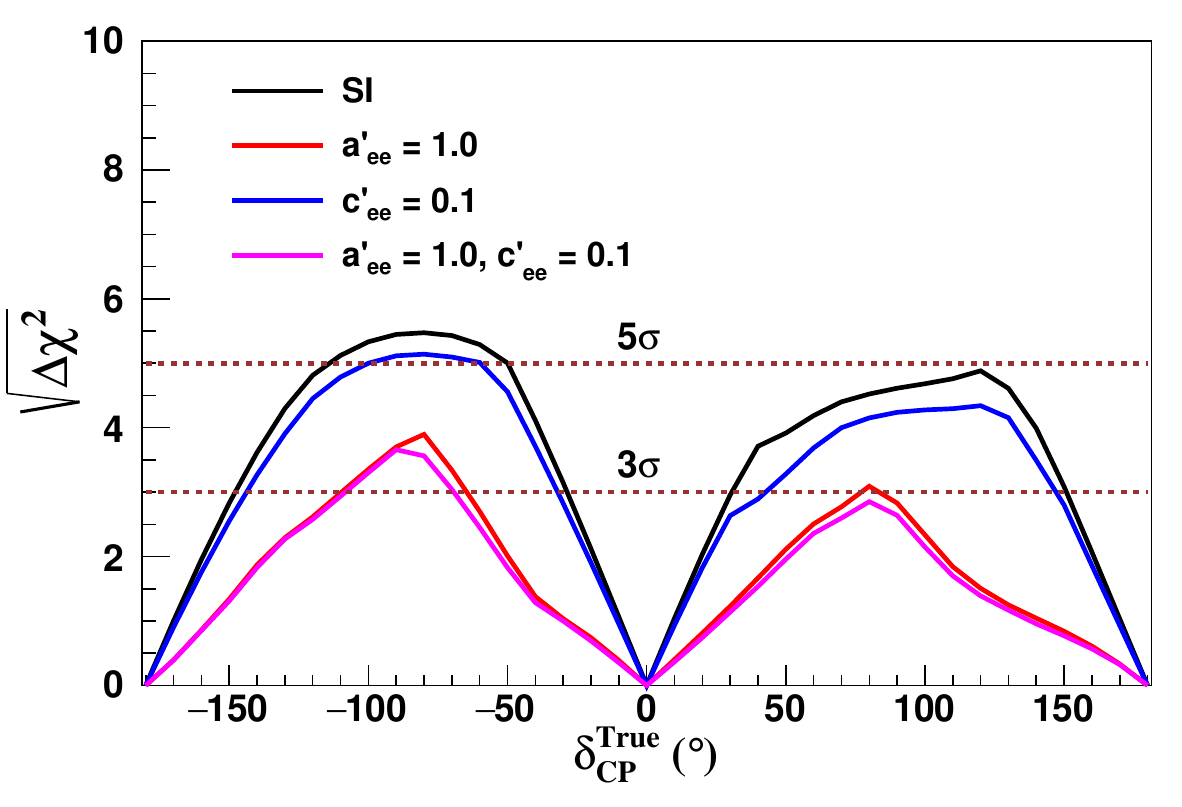}
\includegraphics[width=0.325\linewidth, height=5cm]{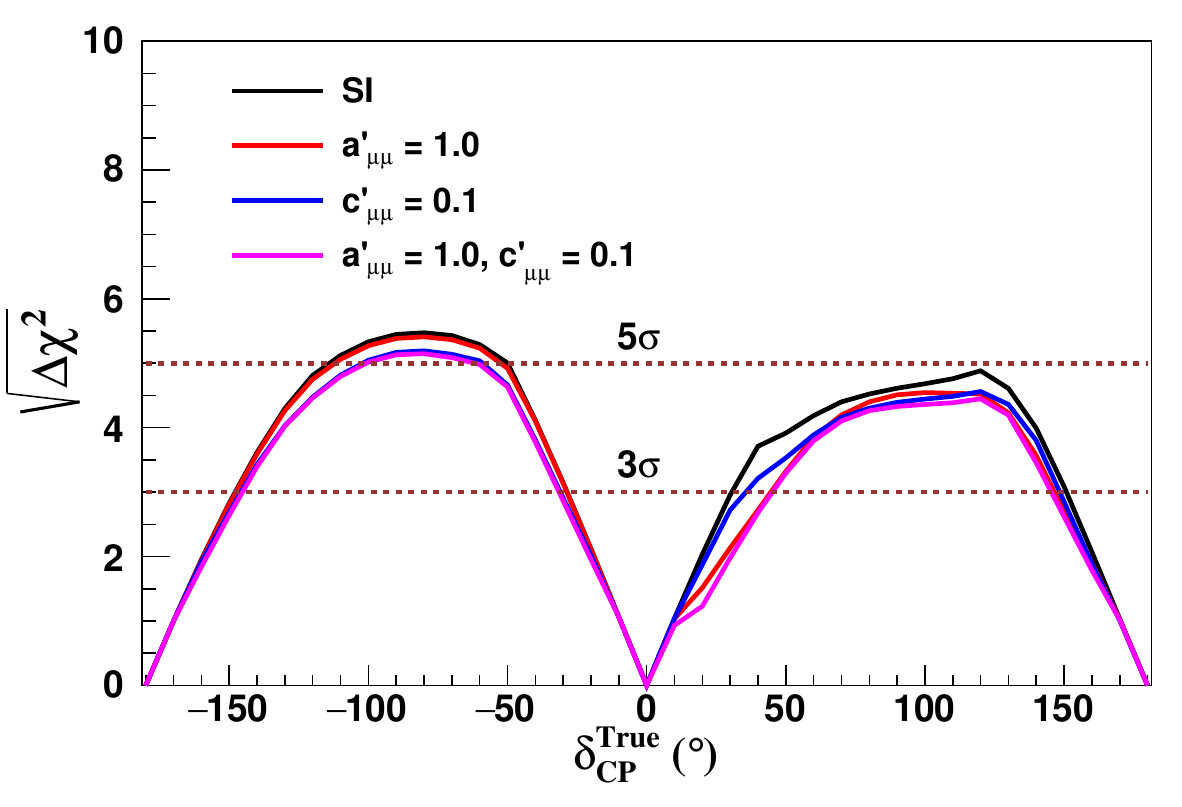}
\includegraphics[width=0.325\linewidth, height=5cm]{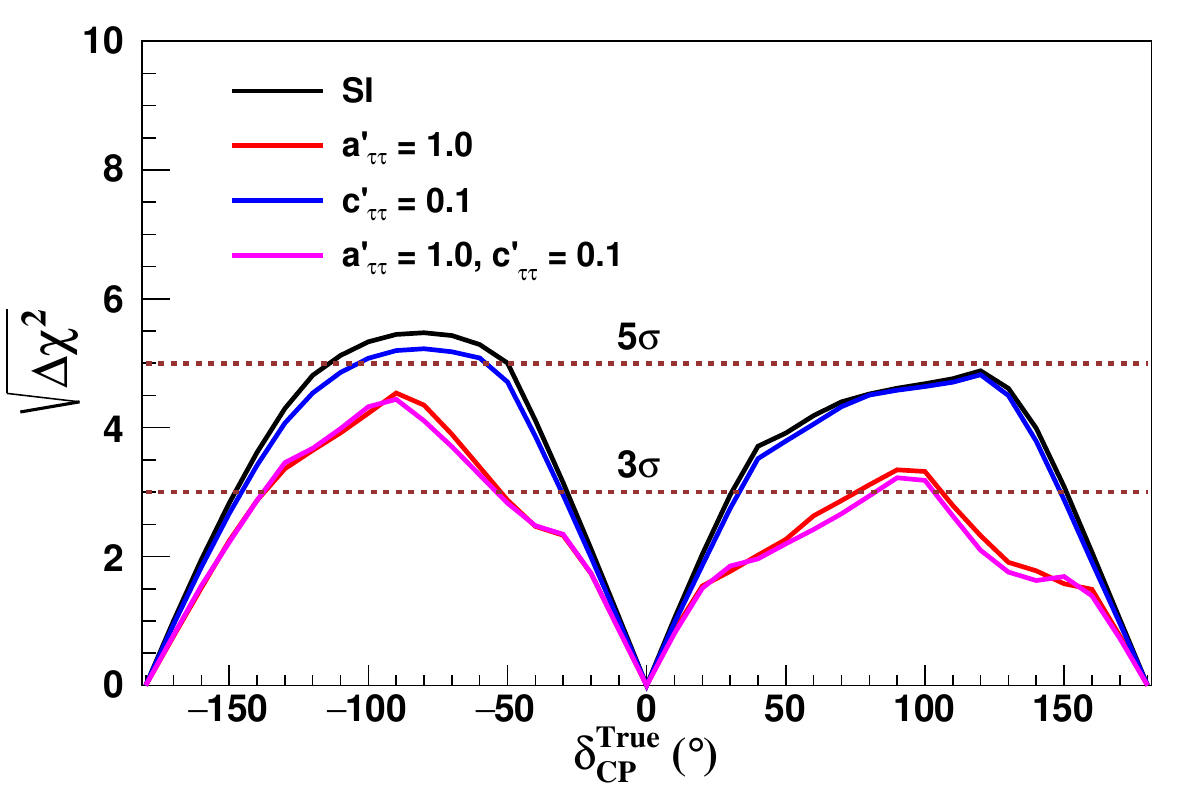}
\includegraphics[width=0.325\linewidth, height=5cm]{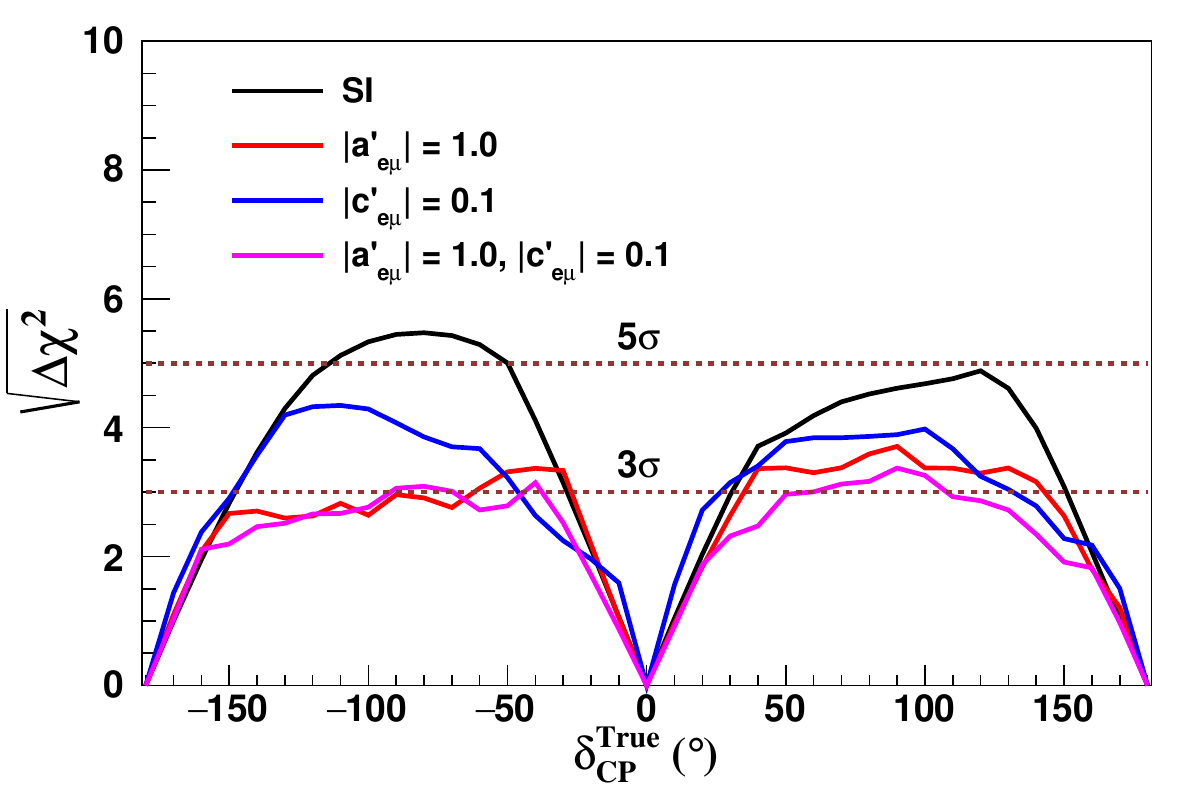}
\includegraphics[width=0.325\linewidth, height=5cm]{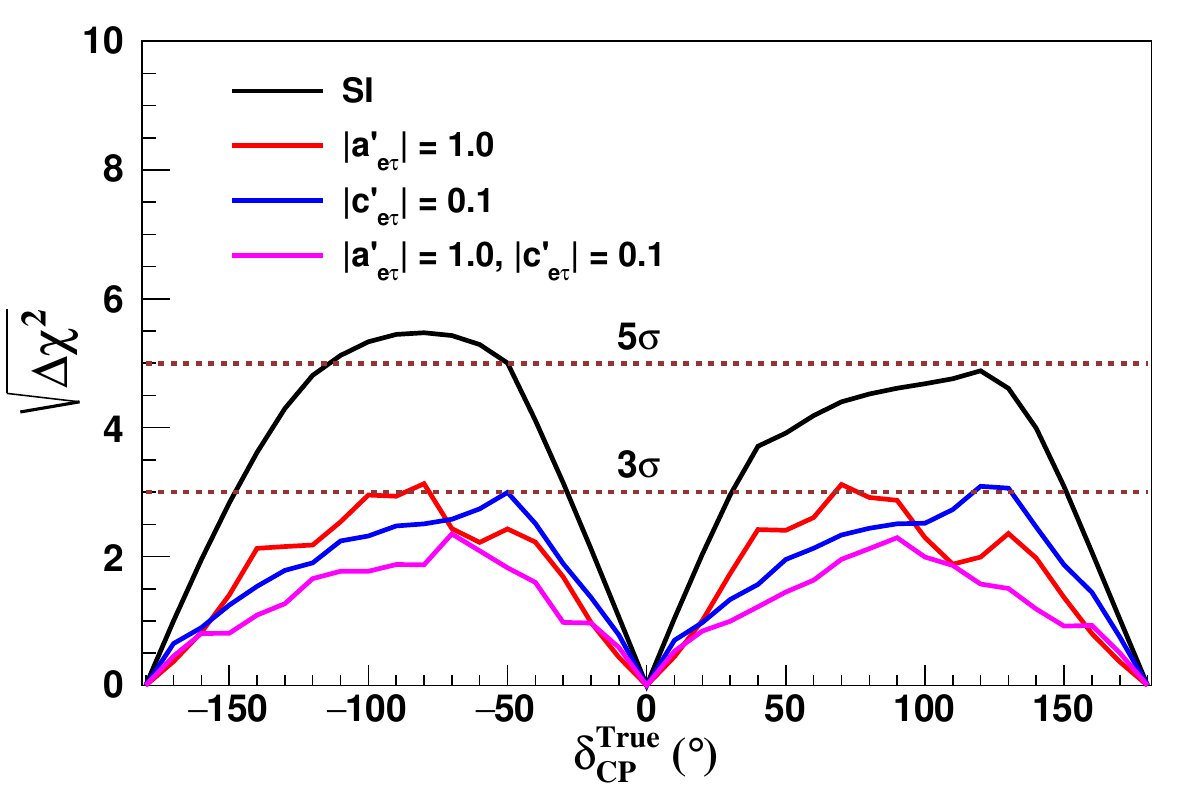}
\includegraphics[width=0.325\linewidth, height=5cm]{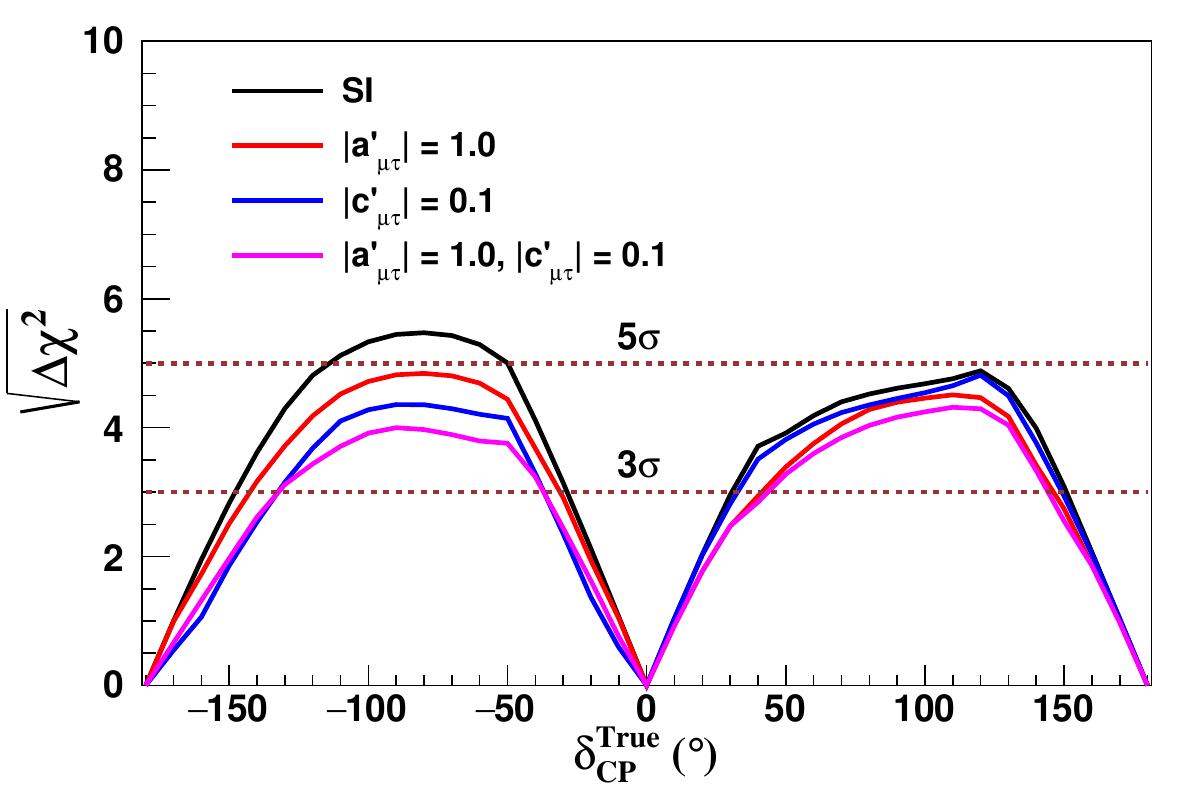}
\caption{CPV sensitivity at DUNE for $a'_{\alpha \beta}$, $c'_{\alpha \beta}$, and their combined effects. The top  (bottom) panel corresponds to diagonal (off-diagonal) elements. The true values of the off-diagonal phases $\phi_{\alpha \beta}^{a}$ and $\phi_{\alpha \beta}^{c}$ are fixed at $-90^\circ$; the test values are marginalized.}
\label{fig:CPV_sens}
\end{figure}

\begin{itemize}
    \item \textit{Diagonal elements (Top-panel)}: In the presence of $a_{ee}$, the CPV sensitivity remains suppressed below the $3\sigma$ level in the upper half-plane of $\delta_{\mathrm{CP}}$, and below $4\sigma$ in the lower half-plane. The effect of $a_{\mu\mu}$ is comparatively small, causing only a marginal suppression of the sensitivity in the upper half-plane while having a negligible impact in the lower half-plane. The behavior of $a_{\tau\tau}$ is qualitatively similar to that of $a_{ee}$, suppressing the SI sensitivity in both half-planes, with the sensitivity dropping below $3\sigma$ over most of the upper half-plane. However, all the CPT-conserving diagonal LIV elements, $c_{ee}$, $c_{\mu\mu}$, and $c_{\tau\tau}$, cause only a mild suppression throughout the $\delta_{\mathrm{CP}}$ range. The combined influence of $a_{\alpha\alpha}$ and $c_{\alpha\alpha}$ closely follows the individual effect of $a_{\alpha\alpha}$ in all three cases, leading to substantial suppression of the sensitivity for the $a_{ee}$--$c_{ee}$ and $a_{\tau\tau}$--$c_{\tau\tau}$ pairs.

    \item \textit{Off-diagonal elements (Bottom-panel)}: Among the off-diagonal elements, the contribution from \( a_{e\mu} \) leads to a significant reduction in sensitivity, dropping below \(3\sigma\) in the lower half-plane of \( \delta_{\mathrm{CP}} \). In contrast, \( a_{e\tau} \) induces a stronger suppression across the entire \( \delta_{\mathrm{CP}} \) range, pushing the sensitivity below \(3\sigma\) throughout. The effect of \( a_{\mu\tau} \) is comparatively milder, reducing the sensitivity below the \(5\sigma\) level.  The presence of \( c_{e\mu} \) results in a suppression of sensitivity in both half-cycles of \( \delta_{\mathrm{CP}} \). Similar to the \( a_{e\tau} \) case, \( c_{e\tau} \) produces a strong suppression over the full \( \delta_{\mathrm{CP}} \) range, reducing the sensitivity below \(3\sigma\). For \( c_{\mu\tau} \), a suppression is observed primarily in the lower half-plane of \( \delta_{\mathrm{CP}} \), pushing the sensitivity below \( 5\sigma \), while its effect on the upper half-plane remains minimal.  The combined effect of \( a_{\alpha\beta} \) and \( c_{\alpha\beta} \) leads to a further degradation of sensitivity. In the presence of both $a_{e\mu}$ and $c_{e\mu}$, the sensitivity falls below $3\sigma$ across most of the parameter space, except in the central regions of the lower and upper half-planes. The suppression becomes more pronounced in the presence of $a_{e\tau}$ and $c_{e\tau}$, where the sensitivity remains below \(3\sigma\) throughout. A similar trend is observed in the $a_{\mu\tau}-c_{\mu\tau}$ pair, where the combined effect produces a slightly stronger suppression than either contribution individually.
\end{itemize}

\noindent Collectively, we observe that the simultaneous presence of both $a_{\alpha\beta}$ and $c_{\alpha\beta}$ terms tends to suppress the CP sensitivity. The sensitivity behavior seen in figure~\ref{fig:CPV_sens} can be qualitatively interpreted from the probability curves in figure~\ref{fig:PvsCP}. CP discovery indicates the ability of an experiment to distinguish a CP-violating value of $\delta_{CP}$ from the CP-conserving values, $0^\circ$ and $180^\circ$. Since $a_{ee}$ has a large effect on the probability as seen in the top left panel of figure~\ref{fig:PvsCP}, a variation of this parameter results in an $a_{ee}$-$\delta_{CP}$ degeneracy. Marginalization over $a_{ee}$ therefore results in a substantial drop in the CP discovery potential, as the top left panel of figure~\ref{fig:CPV_sens} illustrates. The parameter $c_{ee}$ has a much smaller effect on the probability; therefore, marginalizing over this parameter decreases the sensitivity to a smaller extent. A similar argument applies to the effects of $a_{\tau\tau}$ and $c_{\tau\tau}$, whose impact on the oscillation probability likewise leads to a reduction in the CP discovery sensitivity. For the off-diagonal elements, referring to figure~\ref{fig:PvsCP} again, the presence of an additional phase shifts the probability curve with respect to $\delta_{CP}$, leading to a decrease in the CP discovery in all cases.

\section{Summary and Conclusion}\label{sec_clsn}
We have investigated the impact of Lorentz invariance violating parameters on $\nu$--oscillations and CPV sensitivity in an accelerator-based experiment. This work provides a systematic, analytical and numerical examination of the $a_{\alpha\beta}$ and $c_{\alpha\beta}$ LIV effects at DUNE. The analytical probability expressions were derived perturbatively, which provided insights into understanding the interplay between $\nu$--oscillation and LIV parameters.

Among the diagonal parameters, \(a_{ee}\) and \(a_{\tau\tau}\) have the strongest impact, while \(a_{e\mu}\) and \(a_{e\tau}\) dominate among the off-diagonal terms in both \(\nu\) and \(\bar{\nu}\) channels. Compared to the \(a_{\alpha\beta}\) elements, the corresponding \(c_{\alpha\beta}\) terms, owing to their smaller magnitudes, induce weaker deviations from the standard oscillations. Within the energy range relevant to DUNE, the additional energy dependence in \(c_{\alpha\beta}\) does not significantly alter the qualitative behavior. However, this energy dependence of $c_{\alpha\beta}$ may become significant in experiments that can probe higher-energy neutrinos. The parameter scans displayed as heatmaps reveal correlations among the LIV parameters and indicate regions where SI--LIV degeneracies arise. In particular, certain combinations of \(a_{\alpha\beta}\) and \(c_{\alpha\beta}\) can partially offset each other, resulting in behavior that closely resembles the SI case. The CP trajectory plots likewise show that LIV deforms the geometry of the CP ellipses, modifying the apparent CP violation and uncovering additional degeneracies.

DUNE's sensitivities to CPV show significant suppression in presence of $a_{ee}$, $a_{\tau\tau}$, $a_{e\mu}$ or $a_{e\tau}$, with the sensitivity dropping below the $3\sigma$ CL over most of the $\delta_{CP}$ range. The presence of $a_{\mu\mu}$ or $a_{\mu\tau}$ leads to only minor deviations from the standard interaction scenario. For all diagonal $c_{\alpha\alpha}$ elements, the impact on CPV sensitivity is minimal when considered individually, whereas the off-diagonal $c_{\alpha\beta}$ elements lead to a more noticeable suppression. If both $a_{\alpha\beta}$ and $c_{\alpha\beta}$ are present in nature, they can introduce non-trivial interference effects, resulting in more complex parameter dependencies than those arising from individual contributions. For the combinations - ($a_{ee}$, $c_{ee}$), ($a_{\tau\tau}$, $c_{\tau\tau}$), ($a_{e\mu}$, $c_{e\mu}$) or ($a_{e\tau}$, $c_{e\tau}$), the CPV sensitivity is pushed below the $3\sigma$ level across most of the $\delta_{CP}$ range, with the $(a_{\mu\tau},\, c_{\mu\tau})$ pair showing a slightly milder effect. We find that DUNE will be capable of providing improved bounds on the CPT-violating LIV parameters $a_{e\mu}$ and $a_{e\tau}$. A comparison of DUNE's projected bounds with current existing bounds is shown in table~\ref{tab_bounds}. Due to the correlation between $a_{\alpha\beta}$ and $c_{\alpha\beta}$ parameters, we have seen that the combined presence of LIV parameters tends to weaken the allowed bounds.

The analytical results further indicate that the off-diagonal phases $\phi^{a}_{\alpha\beta}$ and $\phi^{c}_{\alpha\beta}$ are correlated with $\delta_{CP}$ and can therefore substantially impact the measurement of true leptonic CP phase. Future studies could extend this work by exploring cross-combinations of LIV parameters and examining their implications for the determination of mass ordering and octant of $\theta_{23}$. While both \(a_{\alpha\beta}\) and \(c_{\alpha\beta}\) individually modify oscillation probabilities, the interplay between them is of particular significance, producing structures and degeneracies absent in the individual cases. Recognizing these combined effects will be vital for achieving precision measurements and for testing LIV at upcoming neutrino experiments.

\section*{Acknowledgments}
PD acknowledges the University Grants Commission (UGC), Government of India, for financial support via the UGC-NET Junior Research Fellowship (24J/03/00589). AS acknowledges financial support from the Indian Association for the Cultivation of Science in the form of Research Associateship. MMD acknowledges support from SERB, DST, Government of India, under Grant No.\ CRG/2021/002961. MMD and SR acknowledge the Workshop on High Energy Physics Phenomenology (WHEPP XVII), hosted by IIT Gandhinagar, India, in January 2024, where this work was initiated. All text and scientific content are originally written by the authors, and generative-AI tools were used only to improve the grammar, and readability of this manuscript.

\appendix
\section{Second order analytic expressions for LIV}
\label{sec_app}

The second-order components of the analytic expressions for $P_{\mu e}$ in eq.~\eqref{eq:Pac} are
\begin{align}
K_{ee}^{(2)} =\ & 
\frac{8 E^2 \left( 1 + 2 (A-1)^2 \Delta^2 - \cos{[2\Delta \,(A-1)]} + 2 (A-1)\Delta \sin{[2\Delta \,(1 - A)]} \right)}{(\Delta m_{31}^2)^2(A-1)^4} s_{13}^2 s_{23}^2
\end{align}
\begin{align}
K_{\mu\mu}^{(2)} &= 
\frac{E^2}{A^2 (A - 1)^4 (\Delta m_{31}^2)^2} \Bigg[
\bigg(
-2 (A - 1)^2 c^2_{23} \sin\delta_{CP} + A(A + (A - 2) \cos 2\theta_{23}) \sin[2\Delta + \delta_{CP}] \nonumber\\
&\quad 
+ (1 - 2A + \cos 2\theta_{23}) \sin[2 A \Delta + \delta_{CP}] + 4 (A - 1)\, A \Delta \cos[2\Delta + \delta_{CP}]\, s^2_{23}
\bigg)^2 \nonumber\\
&\quad + \bigg(
\cos[2 A \Delta + \delta_{CP}](1 - 2A + \cos 2\theta_{23}) 
+ A \cos[2\Delta + \delta_{CP}](A + (A - 2)\cos 2\theta_{23}) \nonumber\\
&\quad - 2 (A - 1) \left(
(A - 1) c^2_{23} \cos\delta_{CP}
+ 2 A \Delta \sin[2\Delta + \delta_{CP}] \sin^2\theta_{23}
\right)
\bigg)^2
\Bigg]\,s^2_{13} s^2_{23}
\end{align}

\begin{align}
K^{(2)}_{\tau\tau} =\ & 
\frac{E^2}{A^2\,(A - 1)^4 (\Delta m_{31}^2)^2}\, \Bigg[ \bigg( -2 (A - 1) A \Delta \cos{[2\Delta + \delta_{\text{CP}}]} 
+ 2 (A-2) A \cos{[\Delta + \delta_{\text{CP}}]} \sin{\Delta} \nonumber \\
& \quad + 2 \cos{[A \Delta + \delta_{\text{CP}}]} \sin{[A \Delta]} \bigg)^2 
+ \bigg( (A - 1)^2 \cos{\delta_{\text{CP}}} - (A-2) A \cos{[2\Delta + \delta_{\text{CP}}]} \nonumber \\
& \quad - \cos{[2 A \Delta + \delta_{\text{CP}}]} - 2 (A - 1) A \Delta \sin{[2\Delta + \delta_{\text{CP}}]} \bigg)^2 \Bigg] \sin^2{(2\theta_{23})} c^{2}_{23} s_{13}^2
\end{align}

\begin{align}
|K^{(2)}_{e\mu}|e^{i\phi^{a/c}_{e\mu}} &= \frac{E^2}{A^2(A-1)^2 (\Delta m_{31}^2)^2} \Bigg[
        \Big((1 - 2A + \cos(2\theta_{23}))\cos[2A\Delta - \phi_{e\mu}] + 
        2(A-1)\,c^2_{23}\cos\phi_{e\mu} \nonumber \\
        &\quad + 2A\cos[2\Delta - \phi_{e\mu}]s^2_{23}\Big)^2 
        + \Big(2A \,s^2_{23}\sin[2\Delta - \phi_{e\mu}] + (1 - 2A + \cos(2\theta_{23})) \nonumber \\
        &\quad \times \sin[2A\Delta - \phi_{e\mu}] 
        - 2(A-1)\,c^2_{23}\sin\phi_{e\mu}\Big)^2
    \Bigg]
\end{align}

\begin{align}
|K^{(2)}_{e\tau}|e^{i\phi^{a/c}_{e\tau}} &= - \frac{8E^2 \Big(A(A-1)(\cos[2\Delta]-1) -1 + A\cos[2(A -1)\Delta] - (A -1)\cos[2A\Delta]\Big) c^2_{23} s^2_{23}}{A^2(A -1)^2 (\Delta m_{31}^2)^2}
\end{align}

\begin{align}
|K^{(2)}_{\mu\tau}|\,e^{i\phi_{\mu\tau}} 
&= \frac{E^2}{A^2(A-1)^4 (\Delta m_{31}^2)^2}
\Bigg[4
    \bigg(
        \cos\delta_{\rm CP}\Big[
            \cos\phi_{\mu\tau}\Big(
                A(1 + (A-2)\cos 2\theta_{23})\sin[2\Delta] \notag\\
        &\quad
                + (-A + \cos 2\theta_{23})\sin[2A\Delta]
                + 4(A-1)A\Delta\cos[2\Delta]\sin^2\theta_{23}
            \Big) \notag\\
        &\quad
            + (A-1)\big(1-A + A\cos [2\Delta] - \cos[2A\Delta]\big)
            \sin\phi_{\mu\tau}
        \Big]
        \notag\\[1ex]
        &\quad
        + \sin\delta_{\rm CP}\Big[
            \cos\phi_{\mu\tau}\Big(
                A \big(\cos[2\Delta] - \cos [2A\Delta] 
                - 2(A-1)\Delta\sin [2\Delta] \big)
                \notag\\
        &\quad
                + \cos 2\theta_{23}\big((A-2)A\cos [2\Delta] 
                + \cos [2A\Delta] 
                + (A-1)(1-A+2A\Delta\sin [2\Delta])\big)
            \Big)
            \notag\\
        &\quad
            - (A-1)\big(A\sin[2\Delta] - \sin[2A\Delta]\big)\sin\phi_{\mu\tau}
        \Big]
    \bigg)^2 + \bigg(
        - (A-1)^2\cos[\delta_{\rm CP}-\phi_{\mu\tau}] \notag\\
        &\quad
        + \cos[2A\Delta+\delta_{\rm CP}-\phi_{\mu\tau}]
        + A\cos\phi_{\mu\tau}\Big[
            (A-2)\cos\delta_{\rm CP}
            - 2\cos[2\Delta+\delta_{\rm CP}] \notag\\
        &\quad
            + 2\cos[2A\Delta+\delta_{\rm CP}]
            + 4(A-1)\Delta\sin[2\Delta + \delta_{\rm CP}]
        \Big]
        \notag\\
        &\quad
        + 2\cos(2\theta_{23})\cos\phi_{\mu\tau}\Big[
            (A-1)^2\cos\delta_{\rm CP}
            - (A-2)A\cos[2\Delta+\delta_{\rm CP}]
            \notag\\
        &\quad
            - \cos[2A\Delta+\delta_{\rm CP}]
            - 2(A-1)A\Delta\sin[2\Delta+\delta_{\rm CP}]
        \Big]
        \notag\\
        &\quad
        + A\Big[
            -(A-2)\sin\delta_{\rm CP}
            + 2(A-1)\sin[2\Delta+\delta_{\rm CP}]
            - 2\sin[2A\Delta+\delta_{\rm CP}]
        \Big]\sin\phi_{\mu\tau}
        \notag\\
        &\quad
        + 2\sin[A\Delta]\sin[A\Delta + \delta_{\rm CP} + \phi_{\mu\tau}]
    \bigg)^2
\Bigg]\,c^2_{23}\,s^2_{13}
\end{align}

\bibliographystyle{JHEP}
\bibliography{citation}

@article{Super-Kamiokande:1998kpq,
    author = "Fukuda, Y. and others",
    collaboration = "Super-Kamiokande",
    title = "{Evidence for oscillation of atmospheric neutrinos}",
    eprint = "hep-ex/9807003",
    archivePrefix = "arXiv",
    reportNumber = "BU-98-17, ICRR-REPORT-422-98-18, UCI-98-8, KEK-PREPRINT-98-95, LSU-HEPA-5-98, UMD-98-003, SBHEP-98-5, TKU-PAP-98-06, TIT-HPE-98-09",
    doi = "10.1103/PhysRevLett.81.1562",
    journal = "Phys. Rev. Lett.",
    volume = "81",
    pages = "1562--1567",
    year = "1998"
}

@article{SNO:2001kpb,
    author = "Ahmad, Q. R. and others",
    collaboration = "SNO",
    title = "{Measurement of the rate of $\nu_e+d \to p+p+e^-$ interactions produced by $^8$B solar neutrinos at the Sudbury Neutrino Observatory}",
    eprint = "nucl-ex/0106015",
    archivePrefix = "arXiv",
    reportNumber = "UPR-0240E",
    doi = "10.1103/PhysRevLett.87.071301",
    journal = "Phys. Rev. Lett.",
    volume = "87",
    pages = "071301",
    year = "2001"
}

@article{K2K:2002icj,
    author = "Ahn, M. H. and others",
    collaboration = "K2K",
    title = "{Indications of neutrino oscillation in a 250 km long baseline experiment}",
    eprint = "hep-ex/0212007",
    archivePrefix = "arXiv",
    doi = "10.1103/PhysRevLett.90.041801",
    journal = "Phys. Rev. Lett.",
    volume = "90",
    pages = "041801",
    year = "2003"
}

@article{KamLAND:2002uet,
    author = "Eguchi, K. and others",
    collaboration = "KamLAND",
    title = "{First results from KamLAND: Evidence for reactor anti-neutrino disappearance}",
    eprint = "hep-ex/0212021",
    archivePrefix = "arXiv",
    doi = "10.1103/PhysRevLett.90.021802",
    journal = "Phys. Rev. Lett.",
    volume = "90",
    pages = "021802",
    year = "2003"
}

@article{T2K:2011ypd,
    author = "Abe, K. and others",
    collaboration = "T2K",
    title = "{Indication of Electron Neutrino Appearance from an Accelerator-produced Off-axis Muon Neutrino Beam}",
    eprint = "1106.2822",
    archivePrefix = "arXiv",
    primaryClass = "hep-ex",
    doi = "10.1103/PhysRevLett.107.041801",
    journal = "Phys. Rev. Lett.",
    volume = "107",
    pages = "041801",
    year = "2011"
}

@article{KamLAND:2008dgz,
    author = "Abe, S. and others",
    collaboration = "KamLAND",
    title = "{Precision Measurement of Neutrino Oscillation Parameters with KamLAND}",
    eprint = "0801.4589",
    archivePrefix = "arXiv",
    primaryClass = "hep-ex",
    doi = "10.1103/PhysRevLett.100.221803",
    journal = "Phys. Rev. Lett.",
    volume = "100",
    pages = "221803",
    year = "2008"
}

@article{KamLAND:2010fvi,
    author = "Gando, A. and others",
    collaboration = "KamLAND",
    title = "{Constraints on $\theta_{13}$ from A Three-Flavor Oscillation Analysis of Reactor Antineutrinos at KamLAND}",
    eprint = "1009.4771",
    archivePrefix = "arXiv",
    primaryClass = "hep-ex",
    doi = "10.1103/PhysRevD.83.052002",
    journal = "Phys. Rev. D",
    volume = "83",
    pages = "052002",
    year = "2011"
}

@article{KamLAND:2013rgu,
    author = "Gando, A. and others",
    collaboration = "KamLAND",
    title = "{Reactor On-Off Antineutrino Measurement with KamLAND}",
    eprint = "1303.4667",
    archivePrefix = "arXiv",
    primaryClass = "hep-ex",
    doi = "10.1103/PhysRevD.88.033001",
    journal = "Phys. Rev. D",
    volume = "88",
    number = "3",
    pages = "033001",
    year = "2013"
}

@article{DayaBay:2012fng,
    author = "An, F. P. and others",
    collaboration = "Daya Bay",
    title = "{Observation of electron-antineutrino disappearance at Daya Bay}",
    eprint = "1203.1669",
    archivePrefix = "arXiv",
    primaryClass = "hep-ex",
    doi = "10.1103/PhysRevLett.108.171803",
    journal = "Phys. Rev. Lett.",
    volume = "108",
    pages = "171803",
    year = "2012"
}

@article{SNO:2011hxd,
    author = "Aharmim, B. and others",
    collaboration = "SNO",
    title = "{Combined Analysis of all Three Phases of Solar Neutrino Data from the Sudbury Neutrino Observatory}",
    eprint = "1109.0763",
    archivePrefix = "arXiv",
    primaryClass = "nucl-ex",
    doi = "10.1103/PhysRevC.88.025501",
    journal = "Phys. Rev. C",
    volume = "88",
    pages = "025501",
    year = "2013"
}

@article{Burguet-Castell:2001ppm,
    author = "Burguet-Castell, J. and Gavela, M. B. and Gomez-Cadenas, J. J. and Hernandez, P. and Mena, Olga",
    title = "{On the Measurement of leptonic CP violation}",
    eprint = "hep-ph/0103258",
    archivePrefix = "arXiv",
    reportNumber = "CERN-TH-2001-081, FTUAM-01-05, IFT-UAM-CSIC-01-10, FTUV-010323, IFIC-01-10",
    doi = "10.1016/S0550-3213(01)00248-6",
    journal = "Nucl. Phys. B",
    volume = "608",
    pages = "301--318",
    year = "2001"
}

@article{DUNE:2020jqi,
    author = "Abi, B. and others",
    collaboration = "DUNE",
    title = "{Long-baseline neutrino oscillation physics potential of the DUNE experiment}",
    eprint = "2006.16043",
    archivePrefix = "arXiv",
    primaryClass = "hep-ex",
    reportNumber = "FERMILAB-PUB-20-251-E-LBNF-ND-PIP2-SCD, PUB-20-251-E-LBNF-ND-PIP2-SCD",
    doi = "10.1140/epjc/s10052-020-08456-z",
    journal = "Eur. Phys. J. C",
    volume = "80",
    number = "10",
    pages = "978",
    year = "2020"
}

@article{DUNE:2020lwj,
    author = "Abi, Babak and others",
    collaboration = "DUNE",
    title = "{Deep Underground Neutrino Experiment (DUNE), Far Detector Technical Design Report, Volume I Introduction to DUNE}",
    eprint = "2002.02967",
    archivePrefix = "arXiv",
    primaryClass = "physics.ins-det",
    reportNumber = "FERMILAB-PUB-20-024-ND, FERMILAB-DESIGN-2020-01",
    doi = "10.1088/1748-0221/15/08/T08008",
    journal = "JINST",
    volume = "15",
    number = "08",
    pages = "T08008",
    year = "2020"
}

@article{DUNE:2021tad,
    author = "Hewes, V. and others",
    collaboration = "DUNE",
    title = "{Deep Underground Neutrino Experiment (DUNE) Near Detector Conceptual Design Report}",
    eprint = "2103.13910",
    archivePrefix = "arXiv",
    primaryClass = "physics.ins-det",
    reportNumber = "FERMILAB-PUB-21-067-E-LBNF-PPD-SCD-T",
    doi = "10.3390/instruments5040031",
    journal = "Instruments",
    volume = "5",
    number = "4",
    pages = "31",
    year = "2021"
}

@article{DUNE:2021hwx,
    author = "Abud, A. Abed and others",
    collaboration = "DUNE",
    title = "{Design, construction and operation of the ProtoDUNE-SP Liquid Argon TPC}",
    eprint = "2108.01902",
    archivePrefix = "arXiv",
    primaryClass = "physics.ins-det",
    reportNumber = "FERMILAB-PUB-21-332-ND",
    doi = "10.1088/1748-0221/17/01/P01005",
    journal = "JINST",
    volume = "17",
    number = "01",
    pages = "P01005",
    year = "2022"
}

@article{DUNE:2024wvj,
    author = "Abed Abud, Adam and others",
    collaboration = "DUNE",
    title = "{DUNE Phase~II: scientific opportunities, detector concepts, technological solutions}",
    eprint = "2408.12725",
    archivePrefix = "arXiv",
    primaryClass = "physics.ins-det",
    reportNumber = "FERMILAB-TM-2833-LBNF",
    doi = "10.1088/1748-0221/19/12/P12005",
    journal = "JINST",
    volume = "19",
    number = "12",
    pages = "P12005",
    year = "2024"
}

@article{Hyper-Kamiokande:2016srs,
    author = "Abe, K. and others",
    collaboration = "Hyper-Kamiokande",
    title = "{Physics potentials with the second Hyper-Kamiokande detector in Korea}",
    eprint = "1611.06118",
    archivePrefix = "arXiv",
    primaryClass = "hep-ex",
    doi = "10.1093/ptep/pty044",
    journal = "PTEP",
    volume = "2018",
    number = "6",
    pages = "063C01",
    year = "2018"
}

@article{Hyper-Kamiokande:2018ofw,
    author = "Abe, K. and others",
    collaboration = "Hyper-Kamiokande",
    title = "{Hyper-Kamiokande Design Report}",
    eprint = "1805.04163",
    archivePrefix = "arXiv",
    primaryClass = "physics.ins-det",
    month = "5",
    year = "2018"
}

@article{BOREXINO:2014pcl,
    author = "Bellini, G. and others",
    collaboration = "BOREXINO",
    title = "{Neutrinos from the primary proton\textendash{}proton fusion process in the Sun}",
    doi = "10.1038/nature13702",
    journal = "Nature",
    volume = "512",
    number = "7515",
    pages = "383--386",
    year = "2014"
}

@article{Kostelecky:1988zi,
    author = "Kostelecky, V. Alan and Samuel, Stuart",
    title = "{Spontaneous Breaking of Lorentz Symmetry in String Theory}",
    reportNumber = "IUHET-139, CCNY-HEP-88/4",
    doi = "10.1103/PhysRevD.39.683",
    journal = "Phys. Rev. D",
    volume = "39",
    pages = "683",
    year = "1989"
}

@article{Super-Kamiokande:2019gzr,
    author = "Jiang, M. and others",
    collaboration = "Super-Kamiokande",
    title = "{Atmospheric Neutrino Oscillation Analysis with Improved Event Reconstruction in Super-Kamiokande IV}",
    eprint = "1901.03230",
    archivePrefix = "arXiv",
    primaryClass = "hep-ex",
    doi = "10.1093/ptep/ptz015",
    journal = "PTEP",
    volume = "2019",
    number = "5",
    pages = "053F01",
    year = "2019"
}

@article{IceCube:2017lak,
    author = "Aartsen, M. G. and others",
    collaboration = "IceCube",
    title = "{Measurement of Atmospheric Neutrino Oscillations at 6\textendash{}56 GeV with IceCube DeepCore}",
    eprint = "1707.07081",
    archivePrefix = "arXiv",
    primaryClass = "hep-ex",
    doi = "10.1103/PhysRevLett.120.071801",
    journal = "Phys. Rev. Lett.",
    volume = "120",
    number = "7",
    pages = "071801",
    year = "2018"
}

@article{DayaBay:2018yms,
    author = "Adey, D. and others",
    collaboration = "Daya Bay",
    title = "{Measurement of the Electron Antineutrino Oscillation with 1958 Days of Operation at Daya Bay}",
    eprint = "1809.02261",
    archivePrefix = "arXiv",
    primaryClass = "hep-ex",
    doi = "10.1103/PhysRevLett.121.241805",
    journal = "Phys. Rev. Lett.",
    volume = "121",
    number = "24",
    pages = "241805",
    year = "2018"
}

@article{RENO:2018dro,
    author = "Bak, G. and others",
    collaboration = "RENO",
    title = "{Measurement of Reactor Antineutrino Oscillation Amplitude and Frequency at RENO}",
    eprint = "1806.00248",
    archivePrefix = "arXiv",
    primaryClass = "hep-ex",
    doi = "10.1103/PhysRevLett.121.201801",
    journal = "Phys. Rev. Lett.",
    volume = "121",
    number = "20",
    pages = "201801",
    year = "2018"
}

@article{DoubleChooz:2019qbj,
    author = "de Kerret, H. and others",
    collaboration = "Double Chooz",
    title = "{Double Chooz $\theta_{13}$ measurement via total neutron capture detection}",
    eprint = "1901.09445",
    archivePrefix = "arXiv",
    primaryClass = "hep-ex",
    doi = "10.1038/s41567-020-0831-y",
    journal = "Nature Phys.",
    volume = "16",
    number = "5",
    pages = "558--564",
    year = "2020"
}

@article{MINOS:2013utc,
    author = "Adamson, P. and others",
    collaboration = "MINOS",
    title = "{Measurement of Neutrino and Antineutrino Oscillations Using Beam and Atmospheric Data in MINOS}",
    eprint = "1304.6335",
    archivePrefix = "arXiv",
    primaryClass = "hep-ex",
    reportNumber = "FERMILAB-PUB-13-102-PPD",
    doi = "10.1103/PhysRevLett.110.251801",
    journal = "Phys. Rev. Lett.",
    volume = "110",
    number = "25",
    pages = "251801",
    year = "2013"
}

@article{T2K:2019bcf,
    author = "Abe, K. and others",
    collaboration = "T2K",
    title = "{Constraint on the matter\textendash{}antimatter symmetry-violating phase in neutrino oscillations}",
    eprint = "1910.03887",
    archivePrefix = "arXiv",
    primaryClass = "hep-ex",
    doi = "10.1038/s41586-020-2177-0",
    journal = "Nature",
    volume = "580",
    number = "7803",
    pages = "339--344",
    year = "2020",
    note = "[Erratum: Nature 583, E16 (2020)]"
}

@article{NOvA:2019cyt,
    author = "Acero, M. A. and others",
    collaboration = "NOvA",
    title = "{First Measurement of Neutrino Oscillation Parameters using Neutrinos and Antineutrinos by NOvA}",
    eprint = "1906.04907",
    archivePrefix = "arXiv",
    primaryClass = "hep-ex",
    reportNumber = "FERMILAB-PUB-19-272-ND",
    doi = "10.1103/PhysRevLett.123.151803",
    journal = "Phys. Rev. Lett.",
    volume = "123",
    number = "15",
    pages = "151803",
    year = "2019"
}

@article{MiniBooNE:2020pnu,
    author = "Aguilar-Arevalo, A. A. and others",
    collaboration = "MiniBooNE",
    title = "{Updated MiniBooNE neutrino oscillation results with increased data and new background studies}",
    eprint = "2006.16883",
    archivePrefix = "arXiv",
    primaryClass = "hep-ex",
    reportNumber = "LA-UR-20-29235, LA-UR-20-24753, FERMILAB-PUB-20-288-AD-ND",
    doi = "10.1103/PhysRevD.103.052002",
    journal = "Phys. Rev. D",
    volume = "103",
    number = "5",
    pages = "052002",
    year = "2021"
}

@article{Rahaman:2021zzm,
    author = "Rahaman, Ushak and Raut, Sushant K.",
    title = "{On the tension between the latest NO$\nu $A~and T2K data}",
    eprint = "2112.13186",
    archivePrefix = "arXiv",
    primaryClass = "hep-ph",
    doi = "10.1140/epjc/s10052-022-10808-w",
    journal = "Eur. Phys. J. C",
    volume = "82",
    number = "10",
    pages = "910",
    year = "2022"
}

@article{Kostelecky:1989jp,
    author = "Kostelecky, V. Alan and Samuel, Stuart",
    title = "{Phenomenological Gravitational Constraints on Strings and Higher Dimensional Theories}",
    reportNumber = "IUHET-155",
    doi = "10.1103/PhysRevLett.63.224",
    journal = "Phys. Rev. Lett.",
    volume = "63",
    pages = "224",
    year = "1989"
}

@article{Kostelecky:1991ak,
    author = "Kostelecky, V. Alan and Potting, Robertus",
    title = "{CPT and strings}",
    reportNumber = "IUHET-197",
    doi = "10.1016/0550-3213(91)90071-5",
    journal = "Nucl. Phys. B",
    volume = "359",
    pages = "545--570",
    year = "1991"
}

@article{ParticleDataGroup:2024cfk,
    author = "Navas, S. and others",
    collaboration = "Particle Data Group",
    title = "{Review of particle physics}",
    doi = "10.1103/PhysRevD.110.030001",
    journal = "Phys. Rev. D",
    volume = "110",
    number = "3",
    pages = "030001",
    year = "2024"
}

@article{T2K:2023smv,
    author = "Abe, K. and others",
    collaboration = "T2K",
    title = "{Measurements of neutrino oscillation parameters from the T2K experiment using $3.6\times 10^{21}$ protons on target}",
    eprint = "2303.03222",
    archivePrefix = "arXiv",
    primaryClass = "hep-ex",
    doi = "10.1140/epjc/s10052-023-11819-x",
    journal = "Eur. Phys. J. C",
    volume = "83",
    number = "9",
    pages = "782",
    year = "2023"
}

@article{NOvA:2021nfi,
    author = "Acero, M. A. and others",
    collaboration = "NOvA",
    title = "{Improved measurement of neutrino oscillation parameters by the NOvA experiment}",
    eprint = "2108.08219",
    archivePrefix = "arXiv",
    primaryClass = "hep-ex",
    reportNumber = "FERMILAB-PUB-21-373-ND",
    doi = "10.1103/PhysRevD.106.032004",
    journal = "Phys. Rev. D",
    volume = "106",
    number = "3",
    pages = "032004",
    year = "2022"
}

@article{Colladay:1998fq,
    author = "Colladay, Don and Kostelecky, V. Alan",
    title = "{Lorentz violating extension of the standard model}",
    eprint = "hep-ph/9809521",
    archivePrefix = "arXiv",
    reportNumber = "IUHET-359",
    doi = "10.1103/PhysRevD.58.116002",
    journal = "Phys. Rev. D",
    volume = "58",
    pages = "116002",
    year = "1998"
}

@article{Greenberg:2002uu,
    author = "Greenberg, O. W.",
    title = "{CPT violation implies violation of Lorentz invariance}",
    eprint = "hep-ph/0201258",
    archivePrefix = "arXiv",
    reportNumber = "PP-02-027",
    doi = "10.1103/PhysRevLett.89.231602",
    journal = "Phys. Rev. Lett.",
    volume = "89",
    pages = "231602",
    year = "2002"
}

@article{MINOS:2010kat,
    author = "Adamson, P. and others",
    collaboration = "MINOS",
    title = "{A Search for Lorentz Invariance and CPT Violation with the MINOS Far Detector}",
    eprint = "1007.2791",
    archivePrefix = "arXiv",
    primaryClass = "hep-ex",
    reportNumber = "FERMILAB-PUB-10-240-E",
    doi = "10.1103/PhysRevLett.105.151601",
    journal = "Phys. Rev. Lett.",
    volume = "105",
    pages = "151601",
    year = "2010"
}

@article{MINOS:2012ozn,
    author = "Adamson, P. and others",
    collaboration = "MINOS",
    title = "{Search for Lorentz invariance and CPT violation with muon antineutrinos in the MINOS Near Detector}",
    eprint = "1201.2631",
    archivePrefix = "arXiv",
    primaryClass = "hep-ex",
    reportNumber = "FERMILAB-PUB-11-676-E",
    doi = "10.1103/PhysRevD.85.031101",
    journal = "Phys. Rev. D",
    volume = "85",
    pages = "031101",
    year = "2012"
}

@article{MiniBooNE:2011pix,
    author = "Aguilar-Arevalo, A. A. and others",
    collaboration = "MiniBooNE",
    title = "{Test of Lorentz and CPT violation with Short Baseline Neutrino Oscillation Excesses}",
    eprint = "1109.3480",
    archivePrefix = "arXiv",
    primaryClass = "hep-ex",
    reportNumber = "FERMILAB-PUB-11-834-E",
    doi = "10.1016/j.physletb.2012.12.020",
    journal = "Phys. Lett. B",
    volume = "718",
    pages = "1303--1308",
    year = "2013"
}

@article{DoubleChooz:2012eiq,
    author = "Abe, Y. and others",
    collaboration = "Double Chooz",
    title = "{First Test of Lorentz Violation with a Reactor-based Antineutrino Experiment}",
    eprint = "1209.5810",
    archivePrefix = "arXiv",
    primaryClass = "hep-ex",
    doi = "10.1103/PhysRevD.86.112009",
    journal = "Phys. Rev. D",
    volume = "86",
    pages = "112009",
    year = "2012"
}

@article{Kostelecky:2003cr,
    author = "Kostelecky, V. Alan and Mewes, Matthew",
    title = "{Lorentz and CPT violation in neutrinos}",
    eprint = "hep-ph/0309025",
    archivePrefix = "arXiv",
    reportNumber = "IUHET-459",
    doi = "10.1103/PhysRevD.69.016005",
    journal = "Phys. Rev. D",
    volume = "69",
    pages = "016005",
    year = "2004"
}

@article{Kostelecky:2011gq,
    author = "Kostelecky, Alan and Mewes, Matthew",
    title = "{Neutrinos with Lorentz-violating operators of arbitrary dimension}",
    eprint = "1112.6395",
    archivePrefix = "arXiv",
    primaryClass = "hep-ph",
    reportNumber = "IUHET-567",
    doi = "10.1103/PhysRevD.85.096005",
    journal = "Phys. Rev. D",
    volume = "85",
    pages = "096005",
    year = "2012"
}

@article{MINOS:2008fnv,
    author = "Adamson, P. and others",
    collaboration = "MINOS",
    title = "{Testing Lorentz Invariance and CPT Conservation with NuMI Neutrinos in the MINOS Near Detector}",
    eprint = "0806.4945",
    archivePrefix = "arXiv",
    primaryClass = "hep-ex",
    reportNumber = "FERMILAB-PUB-08-212-E",
    doi = "10.1103/PhysRevLett.101.151601",
    journal = "Phys. Rev. Lett.",
    volume = "101",
    pages = "151601",
    year = "2008"
}

@article{IceCube:2017qyp,
    author = "Aartsen, M. G. and others",
    collaboration = "IceCube",
    title = "{Neutrino Interferometry for High-Precision Tests of Lorentz Symmetry with IceCube}",
    eprint = "1709.03434",
    archivePrefix = "arXiv",
    primaryClass = "hep-ex",
    doi = "10.1038/s41567-018-0172-2",
    journal = "Nature Phys.",
    volume = "14",
    number = "9",
    pages = "961--966",
    year = "2018"
}

@article{T2K:2017ega,
    author = "Abe, Ko and others",
    collaboration = "T2K",
    title = "{Search for Lorentz and CPT violation using sidereal time dependence of neutrino flavor transitions over a short baseline}",
    eprint = "1703.01361",
    archivePrefix = "arXiv",
    primaryClass = "hep-ex",
    doi = "10.1103/PhysRevD.95.111101",
    journal = "Phys. Rev. D",
    volume = "95",
    number = "11",
    pages = "111101",
    year = "2017"
}

@article{Super-Kamiokande:2014exs,
    author = "Abe, K. and others",
    collaboration = "Super-Kamiokande",
    title = "{Test of Lorentz invariance with atmospheric neutrinos}",
    eprint = "1410.4267",
    archivePrefix = "arXiv",
    primaryClass = "hep-ex",
    doi = "10.1103/PhysRevD.91.052003",
    journal = "Phys. Rev. D",
    volume = "91",
    number = "5",
    pages = "052003",
    year = "2015"
}

@article{Agarwalla:2023wft,
    author = "Agarwalla, Sanjib Kumar and Das, Sudipta and Sahoo, Sadashiv and Swain, Pragyanprasu",
    title = "{Constraining Lorentz invariance violation with next-generation long-baseline experiments}",
    eprint = "2302.12005",
    archivePrefix = "arXiv",
    primaryClass = "hep-ph",
    reportNumber = "IP/BBSR/2023-01",
    doi = "10.1007/JHEP07(2023)216",
    journal = "JHEP",
    volume = "07",
    pages = "216",
    year = "2023"
}

@article{Sarker:2023mlz,
    author = "Sarker, Arnab and Medhi, Abinash and Devi, Moon Moon",
    title = "{Investigating the effects of Lorentz Invariance Violation on the CP-sensitivities of the Deep Underground Neutrino Experiment}",
    eprint = "2302.10456",
    archivePrefix = "arXiv",
    primaryClass = "hep-ph",
    doi = "10.1140/epjc/s10052-023-11785-4",
    journal = "Eur. Phys. J. C",
    volume = "83",
    number = "7",
    pages = "592",
    year = "2023"
}

@article{Majhi:2019tfi,
    author = "Majhi, Rudra and Chembra, Soumya and Mohanta, Rukmani",
    title = "{Exploring the effect of Lorentz invariance violation with the currently running long-baseline experiments}",
    eprint = "1907.09145",
    archivePrefix = "arXiv",
    primaryClass = "hep-ph",
    doi = "10.1140/epjc/s10052-020-7963-1",
    journal = "Eur. Phys. J. C",
    volume = "80",
    number = "5",
    pages = "364",
    year = "2020"
}

@article{Pan:2023qln,
    author = "Pan, Supriya and Chakraborty, Kaustav and Goswami, Srubabati",
    title = "{Sensitivity to CP discovery in the presence of Lorentz invariance-violating potential at T2HK/T2HKK}",
    eprint = "2308.07566",
    archivePrefix = "arXiv",
    primaryClass = "hep-ph",
    doi = "10.1140/epjc/s10052-024-12541-y",
    journal = "Eur. Phys. J. C",
    volume = "84",
    number = "4",
    pages = "354",
    year = "2024"
}

@article{Majhi:2022fed,
    author = "Majhi, Rudra and Singha, Dinesh Kumar and Ghosh, Monojit and Mohanta, Rukmani",
    title = "{Distinguishing nonstandard interaction and Lorentz invariance violation at the Protvino to super-ORCA experiment}",
    eprint = "2212.07244",
    archivePrefix = "arXiv",
    primaryClass = "hep-ph",
    doi = "10.1103/PhysRevD.107.075036",
    journal = "Phys. Rev. D",
    volume = "107",
    number = "7",
    pages = "075036",
    year = "2023"
}

@article{Fiza:2022xfw,
    author = "Fiza, Nishat and Khan Chowdhury, Nafis Rezwan and Masud, Mehedi",
    title = "{Investigating Lorentz Invariance Violation with the long baseline experiment P2O}",
    eprint = "2206.14018",
    archivePrefix = "arXiv",
    primaryClass = "hep-ph",
    reportNumber = "CTPU-PTC-22-15",
    doi = "10.1007/JHEP01(2023)076",
    journal = "JHEP",
    volume = "01",
    pages = "076",
    year = "2023"
}

@article{Sahoo:2022nbu,
    author = "Sahoo, Sadashiv and Kumar, Anil and Agarwalla, Sanjib Kumar and Dighe, Amol",
    title = "{Discriminating between Lorentz violation and non-standard interactions using core-passing atmospheric neutrinos at INO-ICAL}",
    eprint = "2205.05134",
    archivePrefix = "arXiv",
    primaryClass = "hep-ph",
    reportNumber = "IP/BBSR/2022-3, TIFR/TH/22-19",
    doi = "10.1016/j.physletb.2023.137949",
    journal = "Phys. Lett. B",
    volume = "841",
    pages = "137949",
    year = "2023"
}

@article{Rahaman:2021leu,
    author = "Rahaman, Ushak",
    title = "{Looking for Lorentz invariance violation (LIV) in the latest long baseline accelerator neutrino oscillation data}",
    eprint = "2103.04576",
    archivePrefix = "arXiv",
    primaryClass = "hep-ph",
    doi = "10.1140/epjc/s10052-021-09598-4",
    journal = "Eur. Phys. J. C",
    volume = "81",
    number = "9",
    pages = "792",
    year = "2021"
}

@article{Delgadillo:2024vqu,
    author = "Delgadillo, Luis A. and Miranda, O. G. and Moreno-Granados, G. and Moura, C. A.",
    title = "{Constraining the isotropic CPT-odd coefficients of the standard model extension by a combined DUNE and ESSnuSB analysis}",
    eprint = "2409.03716",
    archivePrefix = "arXiv",
    primaryClass = "hep-ph",
    doi = "10.1103/PhysRevD.111.095027",
    journal = "Phys. Rev. D",
    volume = "111",
    number = "9",
    pages = "095027",
    year = "2025"
}

@article{Cordero:2024hjr,
    author = "Cordero, Rub\'en and Delgadillo, Luis A. and Miranda, O. G. and Moura, C. A.",
    title = "{Neutrino Lorentz invariance violation from the $\textit{CPT}$-even SME coefficients through a tensor interaction with cosmological scalar fields}",
    eprint = "2407.18513",
    archivePrefix = "arXiv",
    primaryClass = "hep-ph",
    doi = "10.1140/epjc/s10052-024-13719-0",
    journal = "Eur. Phys. J. C",
    volume = "85",
    number = "1",
    pages = "6",
    year = "2025"
}

@article{Mishra:2023tdj,
    author = "Mishra, Shashank and Shukla, Saurabh and Singh, Lakhwinder and Singh, Venktesh",
    title = "{Search for Lorentz violations through the sidereal effect at the NO\ensuremath{\nu}A experiment}",
    eprint = "2309.01756",
    archivePrefix = "arXiv",
    primaryClass = "hep-ph",
    doi = "10.1103/PhysRevD.109.075042",
    journal = "Phys. Rev. D",
    volume = "109",
    number = "7",
    pages = "075042",
    year = "2024"
}

@article{Barenboim:2018ctx,
    author = "Barenboim, Gabriela and Masud, Mehedi and Ternes, Christoph A. and T\'ortola, Mariam",
    title = "{Exploring the intrinsic Lorentz-violating parameters at DUNE}",
    eprint = "1805.11094",
    archivePrefix = "arXiv",
    primaryClass = "hep-ph",
    reportNumber = "FTUV-18-05-31, IFIC-18-23",
    doi = "10.1016/j.physletb.2018.11.040",
    journal = "Phys. Lett. B",
    volume = "788",
    pages = "308--315",
    year = "2019"
}

@article{Giarnetti:2024mdt,
    author = "Giarnetti, Alessio and Marciano, Simone and Meloni, Davide",
    title = "{Exploring New Physics with Deep Underground Neutrino Experiment High-Energy Flux: The Case of Lorentz Invariance Violation, Large Extra Dimensions and Long-Range Forces}",
    eprint = "2407.17247",
    archivePrefix = "arXiv",
    primaryClass = "hep-ph",
    doi = "10.3390/universe10090357",
    journal = "Universe",
    volume = "10",
    number = "9",
    pages = "357",
    year = "2024"
}

@article{Sahoo:2021dit,
    author = "Sahoo, Sadashiv and Kumar, Anil and Agarwalla, Sanjib Kumar",
    title = "{Probing Lorentz Invariance Violation with atmospheric neutrinos at INO-ICAL}",
    eprint = "2110.13207",
    archivePrefix = "arXiv",
    primaryClass = "hep-ph",
    reportNumber = "IP/BBSR/2021-06",
    doi = "10.1007/JHEP03(2022)050",
    journal = "JHEP",
    volume = "03",
    pages = "050",
    year = "2022"
}

@article{Akhmedov:2004ny,
    author = "Akhmedov, Evgeny K. and Johansson, Robert and Lindner, Manfred and Ohlsson, Tommy and Schwetz, Thomas",
    title = "{Series expansions for three flavor neutrino oscillation probabilities in matter}",
    eprint = "hep-ph/0402175",
    archivePrefix = "arXiv",
    reportNumber = "TUM-HEP-542-04",
    doi = "10.1088/1126-6708/2004/04/078",
    journal = "JHEP",
    volume = "04",
    pages = "078",
    year = "2004"
}

@article{Chattopadhyay:2022ftv,
    author = "Chattopadhyay, Dibya S. and Chakraborty, Kaustav and Dighe, Amol and Goswami, Srubabati",
    title = "{Analytic treatment of 3-flavor neutrino oscillation and decay in matter}",
    eprint = "2204.05803",
    archivePrefix = "arXiv",
    primaryClass = "hep-ph",
    reportNumber = "TIFR/TH/22-17",
    doi = "10.1007/JHEP01(2023)051",
    journal = "JHEP",
    volume = "01",
    pages = "051",
    year = "2023"
}

@article{Bezboruah:2024yhk,
    author = "Bezboruah, Dharitree and Chattopadhyay, Dibya S. and Medhi, Abinash and Sarker, Arnab and Devi, Moon Moon",
    title = "{Neutrino oscillations in presence of diagonal elements of scalar NSI: an analytic approach}",
    eprint = "2410.05250",
    archivePrefix = "arXiv",
    primaryClass = "hep-ph",
    doi = "10.1007/JHEP12(2024)222",
    journal = "JHEP",
    volume = "12",
    pages = "222",
    year = "2025"
}

@article{Raikwal:2023lzk,
    author = "Raikwal, Deepak and Choubey, Sandhya and Ghosh, Monojit",
    title = "{Comprehensive study of Lorentz invariance violation in atmospheric and long-baseline experiments}",
    eprint = "2303.10892",
    archivePrefix = "arXiv",
    primaryClass = "hep-ph",
    doi = "10.1103/PhysRevD.107.115032",
    journal = "Phys. Rev. D",
    volume = "107",
    number = "11",
    pages = "115032",
    year = "2023"
}

@article{Huber:2004ka,
    author = "Huber, Patrick and Lindner, M. and Winter, W.",
    title = "{Simulation of long-baseline neutrino oscillation experiments with GLoBES (General Long Baseline Experiment Simulator)}",
    eprint = "hep-ph/0407333",
    archivePrefix = "arXiv",
    reportNumber = "TUM-HEP-553-04",
    doi = "10.1016/j.cpc.2005.01.003",
    journal = "Comput. Phys. Commun.",
    volume = "167",
    pages = "195",
    year = "2005"
}

@article{Kopp:2006wp,
    author = "Kopp, Joachim",
    title = "{Efficient numerical diagonalization of hermitian 3 x 3 matrices}",
    eprint = "physics/0610206",
    archivePrefix = "arXiv",
    doi = "10.1142/S0129183108012303",
    journal = "Int. J. Mod. Phys. C",
    volume = "19",
    pages = "523--548",
    year = "2008"
}

@article{Huber:2007ji,
    author = "Huber, Patrick and Kopp, Joachim and Lindner, Manfred and Rolinec, Mark and Winter, Walter",
    title = "{New features in the simulation of neutrino oscillation experiments with GLoBES 3.0: General Long Baseline Experiment Simulator}",
    eprint = "hep-ph/0701187",
    archivePrefix = "arXiv",
    reportNumber = "TUM-HEP-656-07",
    doi = "10.1016/j.cpc.2007.05.004",
    journal = "Comput. Phys. Commun.",
    volume = "177",
    pages = "432--438",
    year = "2007"
}

@article{Esteban:2024eli,
    author = "Esteban, Ivan and Gonzalez-Garcia, M. C. and Maltoni, Michele and Martinez-Soler, Ivan and Pinheiro, Jo{\~a}o Paulo and Schwetz, Thomas",
    title = "{NuFit-6.0: updated global analysis of three-flavor neutrino oscillations}",
    eprint = "2410.05380",
    archivePrefix = "arXiv",
    primaryClass = "hep-ph",
    reportNumber = "IFT-UAM/CSIC-24-140, YITP-SB-2024-24, IPPP/24/64, IPPP/24/64, IFT-UAM/CSIC-24-140, YITP-SB-2024-24",
    doi = "10.1007/JHEP12(2024)216",
    journal = "JHEP",
    volume = "12",
    pages = "216",
    year = "2024"
}

@article{Cervera:2000kp,
    author = "Cervera, A. and Donini, A. and Gavela, M. B. and Gomez Cadenas, J. J. and Hernandez, P. and Mena, Olga and Rigolin, S.",
    title = "{Golden measurements at a neutrino factory}",
    eprint = "hep-ph/0002108",
    archivePrefix = "arXiv",
    reportNumber = "CERN-TH-2000-040, FTUAM-00-03, IFT-UAM-CSIC-00-04, FTUV-00-12, IFIC-00-13",
    doi = "10.1016/S0550-3213(00)00221-2",
    journal = "Nucl. Phys. B",
    volume = "579",
    pages = "17--55",
    year = "2000",
    note = "[Erratum: Nucl.Phys.B 593, 731--732 (2001)]"
}

@article{Freund:2001pn,
    author = "Freund, Martin",
    title = "{Analytic approximations for three neutrino oscillation parameters and probabilities in matter}",
    eprint = "hep-ph/0103300",
    archivePrefix = "arXiv",
    reportNumber = "TUM-HEP-404-01",
    doi = "10.1103/PhysRevD.64.053003",
    journal = "Phys. Rev. D",
    volume = "64",
    pages = "053003",
    year = "2001"
}

@article{Denton:2016wmg,
    author = "Denton, Peter B. and Minakata, Hisakazu and Parke, Stephen J.",
    title = "{Compact Perturbative Expressions For Neutrino Oscillations in Matter}",
    eprint = "1604.08167",
    archivePrefix = "arXiv",
    primaryClass = "hep-ph",
    reportNumber = "FERMILAB-PUB-16-126-T, YACHAY-PUB-16-01-PN",
    doi = "10.1007/JHEP06(2016)051",
    journal = "JHEP",
    volume = "06",
    pages = "051",
    year = "2016"
}

@article{Minakata:2001qm,
    author = "Minakata, Hisakazu and Nunokawa, Hiroshi",
    title = "{Exploring neutrino mixing with low-energy superbeams}",
    eprint = "hep-ph/0108085",
    archivePrefix = "arXiv",
    doi = "10.1088/1126-6708/2001/10/001",
    journal = "JHEP",
    volume = "10",
    pages = "001",
    year = "2001"
}

@article{Hyde:2018tqt,
    author = "Hyde, Jeffrey M.",
    title = "{Biprobability approach to CP phase degeneracy from non-standard neutrino interactions}",
    eprint = "1806.09221",
    archivePrefix = "arXiv",
    primaryClass = "hep-ph",
    doi = "10.1016/j.nuclphysb.2019.114804",
    journal = "Nucl. Phys. B",
    volume = "949",
    pages = "114804",
    year = "2019"
}

@article{Minakata:1998bf,
    author = "Minakata, Hisakazu and Nunokawa, Hiroshi",
    title = "{CP violation versus matter effect in long baseline neutrino oscillation experiments}",
    eprint = "hep-ph/9705208",
    archivePrefix = "arXiv",
    reportNumber = "TMUP-HEL-9704, FTUV-97-21, IFIC-97-21",
    doi = "10.1103/PhysRevD.57.4403",
    journal = "Phys. Rev. D",
    volume = "57",
    pages = "4403--4417",
    year = "1998"
}

@article{DUNE:2016ymp,
    author = "Alion, T. and others",
    collaboration = "DUNE",
    title = "{Experiment Simulation Configurations Used in DUNE CDR}",
    eprint = "1606.09550",
    archivePrefix = "arXiv",
    primaryClass = "physics.ins-det",
    reportNumber = "FERMILAB-FN-1020-ND",
    doi = "10.2172/1414949",
    month = "6",
    year = "2016"
}

@article{JUNO:2025gmd,
    author = "Abusleme, Angel and others",
    collaboration = "JUNO",
    title = "{First measurement of reactor neutrino oscillations at JUNO}",
    eprint = "2511.14593",
    archivePrefix = "arXiv",
    primaryClass = "hep-ex",
    month = "11",
    year = "2025"
}

@article{Hyper-Kamiokande:2025fci,
    author = "Abe, K. and others",
    collaboration = "Hyper-Kamiokande",
    title = "{Sensitivity of the Hyper-Kamiokande experiment to neutrino oscillation parameters using accelerator neutrinos}",
    eprint = "2505.15019",
    archivePrefix = "arXiv",
    primaryClass = "hep-ex",
    doi = "10.1140/epjc/s10052-025-14938-9",
    journal = "Eur. Phys. J. C",
    volume = "86",
    number = "2",
    pages = "170",
    year = "2026"
}

@article{KM3NeT:2026kuj,
    author = "Adriani, O. and others",
    collaboration = "KM3NeT",
    title = "{Atmospheric neutrino constraints on Lorentz invariance violation with the first six detection units of KM3NeT/ORCA}",
    eprint = "2603.04264",
    archivePrefix = "arXiv",
    primaryClass = "hep-ex",
    month = "3",
    year = "2026"
}

@article{JUNO:2015zny,
    author = "An, Fengpeng and others",
    collaboration = "JUNO",
    title = "{Neutrino Physics with JUNO}",
    eprint = "1507.05613",
    archivePrefix = "arXiv",
    primaryClass = "physics.ins-det",
    doi = "10.1088/0954-3899/43/3/030401",
    journal = "J. Phys. G",
    volume = "43",
    number = "3",
    pages = "030401",
    year = "2016"
}

@article{Gonzalez-Garcia:2007dlo,
    author = "Gonzalez-Garcia, M. C. and Maltoni, Michele",
    title = "{Phenomenology with Massive Neutrinos}",
    eprint = "0704.1800",
    archivePrefix = "arXiv",
    primaryClass = "hep-ph",
    reportNumber = "YITP-SB-07-14, UB-ECM-PF-07-08",
    doi = "10.1016/j.physrep.2007.12.004",
    journal = "Phys. Rept.",
    volume = "460",
    pages = "1--129",
    year = "2008"
}

@article{Colladay:1996iz,
    author = "Colladay, Don and Kostelecky, V. Alan",
    title = "{CPT violation and the standard model}",
    eprint = "hep-ph/9703464",
    archivePrefix = "arXiv",
    reportNumber = "IUHET-354",
    doi = "10.1103/PhysRevD.55.6760",
    journal = "Phys. Rev. D",
    volume = "55",
    pages = "6760--6774",
    year = "1997"
}

@article{Kostelecky:2003fs,
    author = "Kostelecky, V. Alan",
    title = "{Gravity, Lorentz violation, and the standard model}",
    eprint = "hep-th/0312310",
    archivePrefix = "arXiv",
    reportNumber = "IUHET-461",
    doi = "10.1103/PhysRevD.69.105009",
    journal = "Phys. Rev. D",
    volume = "69",
    pages = "105009",
    year = "2004"
}

@article{KM3NeT:2025mfl,
    author = "Adriani, O. and others",
    collaboration = "KM3NeT",
    title = "{KM3NeT constraint on Lorentz-violating superluminal neutrino velocity}",
    eprint = "2502.12070",
    archivePrefix = "arXiv",
    primaryClass = "astro-ph.HE",
    doi = "10.1038/s42005-025-02347-z",
    journal = "Commun. Phys.",
    volume = "8",
    number = "1",
    pages = "457",
    year = "2025"
}

@article{Miranda:2015dra,
    author = "Miranda, O. G. and Nunokawa, H.",
    title = "{Non standard neutrino interactions: current status and future prospects}",
    eprint = "1505.06254",
    archivePrefix = "arXiv",
    primaryClass = "hep-ph",
    doi = "10.1088/1367-2630/17/9/095002",
    journal = "New J. Phys.",
    volume = "17",
    number = "9",
    pages = "095002",
    year = "2015"
}

@article{Farzan:2017xzy,
    author = "Farzan, Y. and Tortola, M.",
    title = "{Neutrino oscillations and Non-Standard Interactions}",
    eprint = "1710.09360",
    archivePrefix = "arXiv",
    primaryClass = "hep-ph",
    doi = "10.3389/fphy.2018.00010",
    journal = "Front. in Phys.",
    volume = "6",
    pages = "10",
    year = "2018"
}

@article{Medhi:2021wxj,
    author = "Medhi, Abinash and Dutta, Debajyoti and Devi, Moon Moon",
    title = "{Exploring the effects of scalar non standard interactions on the CP violation sensitivity at DUNE}",
    eprint = "2111.12943",
    archivePrefix = "arXiv",
    primaryClass = "hep-ph",
    doi = "10.1007/JHEP06(2022)129",
    journal = "JHEP",
    volume = "06",
    pages = "129",
    year = "2022"
}

@article{Sarker:2024ytu,
    author = "Sarker, Arnab and Bezboruah, Dharitree and Medhi, Abinash and Devi, Moon Moon",
    title = "{Sensitivity of DUNE in the presence of off-diagonal scalar NSI parameters}",
    eprint = "2406.15307",
    archivePrefix = "arXiv",
    primaryClass = "hep-ph",
    doi = "10.1103/dj83-rw9c",
    journal = "Phys. Rev. D",
    volume = "112",
    number = "3",
    pages = "035042",
    year = "2025"
}

@article{Gonzalez-Garcia:2004pka,
    author = "Gonzalez-Garcia, M. C. and Maltoni, Michele",
    title = "{Atmospheric neutrino oscillations and new physics}",
    eprint = "hep-ph/0404085",
    archivePrefix = "arXiv",
    reportNumber = "YITP-SB-18-04",
    doi = "10.1103/PhysRevD.70.033010",
    journal = "Phys. Rev. D",
    volume = "70",
    pages = "033010",
    year = "2004"
}

@article{Fogli:2002pt,
    author = "Fogli, G. L. and Lisi, E. and Marrone, A. and Montanino, D. and Palazzo, A.",
    title = "{Getting the most from the statistical analysis of solar neutrino oscillations}",
    eprint = "hep-ph/0206162",
    archivePrefix = "arXiv",
    doi = "10.1103/PhysRevD.66.053010",
    journal = "Phys. Rev. D",
    volume = "66",
    pages = "053010",
    year = "2002"
}

@article{Fogli:2003th,
    author = "Fogli, G. L. and Lisi, E. and Marrone, A. and Montanino, D.",
    title = "{Status of atmospheric nu(mu) ---{\ensuremath{>}} nu(tau) oscillations and decoherence after the first K2K spectral data}",
    eprint = "hep-ph/0303064",
    archivePrefix = "arXiv",
    doi = "10.1103/PhysRevD.67.093006",
    journal = "Phys. Rev. D",
    volume = "67",
    pages = "093006",
    year = "2003"
}

@article{Araya-Santander:2025jfd,
    author = "Araya-Santander, Tatiana and Bonilla, Cesar and Pan, Supriya",
    title = "{JUNO's Impact on the Neutrino Mass Ordering from Lorentz Invariance Violation}",
    eprint = "2512.11285",
    archivePrefix = "arXiv",
    primaryClass = "hep-ph",
    month = "12",
    year = "2025"
}

@article{Hilding-Norkjaer:2026fhp,
    author = "Hilding-N{\o}rkj{\ae}r, Simon and Ioannou-Nikolaides, Johann and Koskinen, D. Jason and Stuttard, Thomas",
    title = "{Interplay of Lorentz Invariance Violation and Earth's Matter Potential in High-Energy Neutrinos}",
    eprint = "2602.08076",
    archivePrefix = "arXiv",
    primaryClass = "hep-ph",
    month = "2",
    year = "2026"
}

@article{Brdar:2026jbu,
    author = "Brdar, Vedran and Mir, Samiur R.",
    title = "{Ultra-High-Energy Tau Neutrinos as Probes of Lorentz Invariance}",
    eprint = "2604.19880",
    archivePrefix = "arXiv",
    primaryClass = "hep-ph",
    month = "4",
    year = "2026"
}

@article{Bora:2025xfj,
    author = "Bora, Himanshu and Dutta, Debajyoti and Medhi, Abinash",
    title = "{Constraining and Resolving Lorentz-Violating New Physics at ESSnuSB Using Complementarity with DUNE}",
    eprint = "2512.06953",
    archivePrefix = "arXiv",
    primaryClass = "hep-ph",
    month = "12",
    year = "2025"
}

\end{document}